\documentclass[a4paper,11pt]{article}
\pdfoutput=1
\usepackage{jcappub}
\usepackage{float}

\usepackage{amsmath, amssymb, amsthm, graphicx, epsfig, fancyhdr,epsfig, slashed}

\usepackage{tikzsymbols}
\usepackage{natbib}
\usepackage{float}

\usepackage{tikz,xcolor,hyperref}

\usepackage{bm}

\usepackage{url}
\usepackage{booktabs}
\usepackage{multirow}
\usepackage{changepage}

\usepackage{natbib}
\usepackage{subcaption}
\usepackage{lmodern}

\usepackage{orcidlink}

\newcommand{\be}{\begin{equation}}
\newcommand{\ee}{\end{equation}}
\newcommand{\bea}{\begin{eqnarray}}
\newcommand{\eea}{\end{eqnarray}}

\definecolor{lime}{HTML}{A6CE39}
\DeclareRobustCommand{\orcidicon}{
	\begin{tikzpicture}
	\draw[lime, fill=lime] (0,0) 
	circle [radius=0.2] 
	node[white] {{\fontfamily{qag}\selectfont \tiny ID}};
	\draw[white, fill=white] (-0.0625,0.095) 
	circle [radius=0.007];
	\end{tikzpicture}
	\hspace{-2mm}
}

\foreach \x in {A, ..., Z}{\expandafter\xdef\csname orcid\x\endcsname{\noexpand\href{https://orcid.org/\csname orcidauthor\x\endcsname}
			{\noexpand\orcidicon}}
}

\usepackage{textcomp}
\usepackage{lscape}
\usepackage{placeins}
\usepackage{microtype}
\usepackage{xspace}

\makeatletter
\gdef\@fpheader{}
\makeatother

\begin{document}

\title{Gaussian Process Reconstruction of Cosmological Parameters with Gravitational Wave Sirens using Machine Learning}

\author[a]{Gourab Nandi\orcidlink{0009-0008-1450-7837}}
\author[b]{Anish Ghoshal\orcidlink{0000-0001-7045-302X}}
\author[c]{David F. Mota\orcidlink{0000-0003-3141-142X}}

\affiliation[a]{Department of Physics, Indian Institute of Technology Bombay, Mumbai 400076, India}
\affiliation[b]{Department of Physics and Astronomy, University of Sussex,
\\ Brighton, BN1 9RH, United Kingdom}
\affiliation[c]{Institute of Theoretical Astrophysics, University of Oslo, \\\ P.O. Box 1029 Blindern, N-0315 Oslo, Norway}

\emailAdd{gourabsony87@gmail.com}
\emailAdd{A.Ghoshal@sussex.ac.uk}
\emailAdd{d.f.mota@astro.uio.no}

\date{\today}

\abstract{Future gravitational wave (GW) standard siren catalogues will probe the late-time expansion history of the Universe across redshift ranges largely inaccessible to traditional electromagnetic observations. To determine how effectively this background distance information can distinguish between viable cosmological models, we introduce a model-independent reconstruction framework utilizing Gaussian Process Regression (GPR). Analyzing mock LISA and Einstein Telescope (ET) catalogues across six fiducial backgrounds-$\Lambda$CDM, CPL, CPL+$\Lambda$, interacting dark matter, interacting dark energy and axion inspired early dark energy. We reconstruct the comoving distance and its derivatives. Crucially, we propagated the full GP covariance, including derivative cross-covariances, to robustly evaluate diagnostics such as $H(z)$, $q(z)$ and $\mathcal{O}_{m}(z)$. While our analysis demonstrates that GW bright standard sirens faithfully recover fiducial expansion histories, applying pointwise marginal Hellinger distance reveals that background measurements alone do not provide decisive statistical separation among models. Instead, derivative sensitive diagnostics pinpoint specific redshift windows (e.g., $z\simeq1.6-1.8$ for ET and $z\simeq2.6-2.9$ for LISA) where future catalogues will maximize their discriminatory power. As machine learning methodologies become increasingly integral to astrophysics and cosmology, this Bayesian GPR pipeline offers a principled, nonparametric approach to precisely identifying where the most valuable cosmological information lies.}


\maketitle

\section{Introduction}
\label{sec:intro}

The advent of gravitational wave (GW) astronomy, inaugurated by the
landmark detections of the LIGO Scientific and Virgo Collaborations
\citep{Abbott2016, Abbott2017a, Abbott2017b}
has opened an entirely new observational window onto the Universe.
Gravitational wave standard sirens are highly compelling cosmological probes. Because the luminosity distance to a binary merger is encoded directly in the waveform, they provide self-calibrated distance measurements that require no external calibration against an electromagnetic distance ladder
\citep[e.g.,][]{Schutz1986, Holz2005, Abbott2017NatureH0}.
This property makes standard sirens very suitable to test the
expansion history of the Universe. This is complementary
and orthogonal in its systematic uncertainties to conventional distance
indicator pathway \citep{ChenFishbachHolz2018, Borhanian2020}.
Standard siren cosmology has already been demonstrated with both bright
and dark siren analyses of the present LIGO--Virgo events
\citep{Fishbach2019DarkSiren, SoaresSantos2019DarkSiren,
Abbott2021H0O2, Abbott2023GWTC3Cosmology}. These methods, along with their variants, have also been implemented to extract expansion-history information from GW source-population features
features \citep{Farr2019HubbleExpansion}.

The scientific case for GW cosmology is further sharpened by the
mounting tensions afflicting the standard $\Lambda$CDM model.
Most prominent is the Hubble tension: a statistically significant
discrepancy between the value of $H_0$ inferred from Planck CMB
anisotropies and the value measured via Cepheid-calibrated Type~Ia
supernovae currently standing at the $4$--$5\sigma$ level
\citep[e.g.,][]{Riess2022, PlanckVI2020}. Further inconsistencies
including the $S_8$ tension between weak lensing and CMB inferred
matter clustering amplitudes \citep{Heymans2021KiDS1000, Abbott2022DESY3}
and the preference of baryon acoustic
oscillation data for nonzero dark energy evolution as hinted by recent
DESI results \citep{DESI2024} suggest that the concordance model may
require revision \citep[see][for a review of proposed
solutions]{DiValentino2021}. Against this observational backdrop,
robust and model-independent probes of the late-time expansion history
are indispensable. 

Future GW observatories are poised to deliver precisely such probes.
The Einstein Telescope (ET), a proposed third-generation ground-based
interferometer, is expected to detect $\mathcal{O}(10^5)$ binary
neutron star (BNS) mergers per year out to $z\sim 2$
\citep{ET2020, Sathyaprakash2010}, while the Laser Interferometer
Space Antenna (LISA) will observe massive black hole binary (MBHB)
mergers extending to $z\sim 5$
\citep{LISA2017, Klein2016, Tamanini2016}. Together these two
detectors span a wide and complementary redshift baseline. Forecasts
for future standard-siren cosmology show that such data can provide
competitive constraints on $H_0$ and dark energy parameters
\citep{MessengerRead2012, Feeney2019StandardSirens,
Mukherjee2021RedshiftUnknown, DiazMukherjee2022ExpansionHistory,
Jin2023MultibandSirens, AfrozMukherjeeTasinato2025DarkEnergy}.
Cross-correlation methods further extend this programme to dark sirens
and to tests of GW propagation with galaxy surveys
\citep{AfrozMukherjee2024CrossCorrelation}. The next question is not
only whether $H_0$
or $H(z)$ can be recovered, but whether the same bright standard-siren data
contain enough information to distinguish cosmological models through
diagnostics that depend on derivatives of the expansion history.
Previous GPR analyses of future GW bright standard sirens, including
Shah et al.~\citep{Shah2023} and Mukherjee
et al.~\citep{MukherjeeShah2024}, demonstrated that a
machine-learning reconstruction can recover $H(z)$ and $H_0$ from
realistically generated ET and LISA catalogues. Their emphasis,
however was on the Hubble history itself. We extend prior work by testing whether GPR can accurately reconstruct cosmological parameters that depend on higher derivatives of the luminosity distance. Furthermore, we assess the statistical power of these reconstructed parameters to distinguish between competing cosmologies when full reconstruction uncertainties are propagated. Exploiting such catalogues for precision
cosmography therefore requires analysis methods capable of
reconstructing continuous cosmological functions and their derivatives
directly from discrete noisy distance measurements, while avoiding
restrictive parametric assumptions on the underlying expansion history.

Gaussian Process Regression (GPR) has emerged as the method of choice
for this task. In this context GPR is a
Bayesian machine-learning method: a Gaussian Process defines a prior
over smooth functions given training data, produces a statistically
consistent posterior over the function and its derivatives, and returns
a principled nondiagonal covariance that characterizes the full
reconstruction uncertainty \citep[e.g.,][]{Seikel2012, Holsclaw2010,
Busti2014, Yahya2014, Shafieloo2012}. GPR has already been applied
successfully to reconstruct $H(z)$ from cosmic chronometers and
supernovae to test the consistency of $\Lambda$CDM and to explore
modified gravity and dark sector models
\citep[e.g.,][]{Holsclaw2011, BilickiSeikel2012,
GomezValentAmendola2018, MukherjeeShah2024}. More broadly,
nonparametric and machine-learning reconstructions have been pursued
using complementary tools such as principal components and genetic
algorithms \citep{IshidaDeSouza2011, ArjonaNesseris2020}. In all these
applications a key methodological requirement is the retention of the
full predictive covariance matrix including cross-covariances between
the function and its derivatives when propagating uncertainties to
derived observables.
Recent Bayesian and nonparametric reconstruction studies in related
gravitational-wave and cosmological-inference settings provide
complementary examples of this broader programme
\citep{GhalebMalhotraTasinatoZavala2025, RuchikaMukherjeeFavale2025}.

In this work we apply GPR to mock gravitational wave standard siren
catalogues to perform a model-independent reconstruction of late-time
cosmological diagnostics. Using realistically generated mock
luminosity distance data for six fiducial cosmological models---the
concordance $\Lambda$CDM, two CPL dark energy variants
\citep{Chevallier2001, Linder2003} calibrated against current CMB
\citep{PlanckVI2020, PlanckV2020}, DESI \citep{DESI2024} and
PantheonPlus \citep{Brout2022} data, interacting dark matter and dark
energy scenarios and an axion-inspired early dark energy model
implemented via \texttt{AxiCLASS} \citep{Lesgourgues2011,
Blas2011, AxiCLASSGitHub, Smith2020EDE}---we reconstruct the
comoving distance and its first two
derivatives and derive $H(z)$, $q(z)$, $\mathcal{O}_m(z)$, $w_{\rm tot}(z)$
and the ratio-based diagnostic $\kappa(z)=E'(z)/E(z)$. Our mock
catalog generation methodology closely follows the approach of
Mukherjee et al.~\citep{MukherjeeShah2024} and
Shah et al.~\citep{Shah2023} to which we refer the
reader for further details. The analysis is performed separately for
the ET and LISA detector configurations each with physically motivated
redshift-dependent noise prescriptions. Building on the Bayesian GP
strategy of Ref.~\citep{MukherjeeShah2024}, we marginalize over the
kernel hyperparameters $(\sigma_f,\ell)$ with the \texttt{emcee}
ensemble sampler \citep{ForemanMackey2013} and propagate this
uncertainty through the derived diagnostics. Rather than focusing
exclusively on constraining $H_0$ our primary goal is to quantify the
ability of next-generation GW observations to discriminate between
competing cosmological models through higher-order kinematical
diagnostics.

An important outcome is not only whether the models can be ruled apart
but also where the information begins to appear. The reconstructed
diagnostics show redshift-dependent hints of discrimination: model
differences become most visible in selected redshift ranges and in
derivative-sensitive quantities, even when the final statistical
separation remains limited. After full covariance propagation these
features generally fall short of robust model separation, revealing a
practical limitation of background-expansion observables alone rather
than a failure of the reconstruction procedure. To make this statement
quantitative, we assess model separation with the pointwise marginal
Hellinger distance, a symmetric, bounded measure that tests whether the
reconstructed one-point marginal distributions of different models are
actually separated without relying on bin-dependent $\chi^2$ summaries.
Throughout, we retain the relevant joint GP predictive covariance blocks
for the reconstructed comoving-distance derivatives; this is essential
because neglecting derivative cross-correlations leads to a systematic
underestimate of the confidence regions for the derived diagnostics.
Thus, the machine-learning element in this work is the Bayesian,
nonparametric GPR framework itself, together with MCMC sampling of its
kernel hyperparameters, rather than a black-box classifier.

\textit{The paper is structured as follows:} Section~\ref{sec:reconstruction}
describes the fiducial cosmological models and the procedure for
generating mock GW standard siren catalogues. Section~\ref{sec:GPR}
presents the GPR framework including the kernel choice, hyperparameter
inference, derivative reconstruction, and covariance structure.
Reconstruction results for all diagnostics are presented and discussed
in Sec.~\ref{sec:results}. The pointwise marginal Hellinger distance analysis is presented
in Sec.~\ref{sec:hellinger}. We summarize our conclusions and outline
future directions in Sec.~\ref{sec:conclusions}.

\section{Analysis pipeline and mock standard-siren catalogues}
\label{sec:reconstruction}
Our analysis is structured around a sequence of modular steps designed
to minimize biases from specific modeling assumptions or individual mock
realizations. This systematic approach enables a transparent comparison
of reconstructed diagnostics across different fiducial cosmologies.

\begin{enumerate}

\item We generate realistic mock catalogs of gravitational wave bright
siren events based on fiducial background cosmological models. These
catalogs incorporate appropriate instrumental uncertainties and
observational errors to provide luminosity distance measurements over a
broad redshift range. This is essential for probing the late-time
expansion history of the Universe. We consider two distinct detector
configurations for this purpose: the space-based LISA detector and the
ground-based Einstein Telescope.

\item To ensure the reconstruction results do not depend solely on a
single assumed background we consider multiple fiducial cosmological
scenarios. These include the concordance $\Lambda$CDM model as well as
several extensions involving evolving dark energy, dark sector
interactions and axion-inspired modifications. This approach allows us
to assess how sensitive the reconstruction is to different underlying
cosmological dynamics.

\item We construct and analyze multiple independent realizations for
each fiducial cosmological model to avoid biases tied to any single mock
catalog. The reconstruction is performed separately for each instance so
the resulting ensemble can be used to characterize the mean behavior and
its associated uncertainties.

\item Starting from the mock luminosity distance data, we reconstruct the
comoving distance and its first two derivatives using GPR and derive the
Hubble parameter alongside higher-order model-agnostic diagnostics. The
full reconstruction framework is presented in Sec.~\ref{sec:GPR} and
the resulting diagnostic reconstructions are discussed in
Sec.~\ref{sec:results}.

\item Particular emphasis is placed on consistent uncertainty propagation
throughout the pipeline, retaining the relevant joint Gaussian Process
predictive covariance blocks rather than propagating diagonal variances
alone. The methodological importance of this choice is demonstrated in
Sec.~\ref{sec:covariance_structure}.

\item Finally, we quantify the pointwise statistical distinguishability of
reconstructed diagnostics across different fiducial models using a
Gaussian approximation to the pointwise marginal Hellinger distance at each
redshift. This moves the analysis beyond qualitative visual comparisons,
while avoiding the stronger claim that the full functional posterior
covariance across redshift has been reduced to a single separation
measure.

\end{enumerate}

This overall strategy renders the analysis largely insensitive to
specific modeling assumptions or individual mock realizations. The
framework thereby allows a robust comparison of reconstructed
diagnostics across different fiducial models and provides a transparent
method to assess the discriminatory power of gravitational wave standard
sirens in testing late-time cosmology.

\subsection{Fiducial cosmological models}
\label{sec:fiducial_models}

To generate mock gravitational wave standard siren catalogs we must
assume an underlying background cosmological model that governs the
expansion history of the Universe. There is currently no unique or
observationally preferred model for late-time cosmic acceleration,
especially in light of existing tensions and anomalies. We therefore
consider a set of representative fiducial cosmological models that span
a broad range of physically motivated scenarios. This selection allows
us to assess the robustness of our reconstruction procedure against
variations in the assumed background cosmology.

\noindent\textbf{Six fiducial cosmologies and why they are used.}
We selected these six specific backgrounds because they represent distinct, physically motivated departures from the concordance expansion history. The set contains a baseline cosmology, time varying dark energy,
a mixed CPL plus vacuum sector, two interacting dark sector cases and an
early scalar field model. This gives a controlled test of whether the
Gaussian Process reconstruction responds mainly to changes in the
background shape, to low redshift normalization or to high redshift
evolution.

\noindent\textbf{$\Lambda$CDM, CPL and CPL$+\Lambda$ models:}

As a baseline we adopt the standard $\Lambda$CDM cosmology since it
provides an excellent fit to a wide array of current observations. We
also consider two CPL-type dark energy fiducials. The first is the
standard CPL model in which the dark energy equation of state evolves as
\citep{Chevallier2001, Linder2003}
\begin{equation}
  w(z) = w_0 + w_a \frac{z}{1+z}.
  \label{eq:cpl}
\end{equation}
The second is a CPL$+\Lambda$ variant in which the dark energy sector
contains a CPL component together with an additional constant vacuum
term. Its effective $w_{\rm DE}(z)$ therefore differs from that of the
pure CPL case even though both belong to the same CPL family. For these
models only parameters governing the background expansion enter the
construction of gravitational wave luminosity distances. Parameters
associated with the primordial power spectrum and recombination history
are fixed to their Planck 2018 best fit values to compute a consistent
background evolution using \texttt{CLASS}
\citep{Lesgourgues2011, Blas2011}. The fiducial late-time parameter
values adopted for $\Lambda$CDM and the pure CPL case are taken from
the CSB constraints of Shah et al.~\citep{Shah2023}, as summarized in
Mukherjee et al.~\citep{MukherjeeShah2024}. They are reported in
Table~\ref{tab:lcdm_cpl_fiducial}.

\noindent\textbf{Key parameters for quick comparison.}
The parameters that most directly shape the late time distance relation
are $H_0$, $\Omega_{m0}$, $w_0$ and $w_a$. The remaining parameters keep
the \texttt{CLASS} background calculation consistent with the source
fits and are not varied independently in the reconstruction.

\begin{table}[ht]
\caption{\it Fiducial cosmological parameters adopted for the $\Lambda$CDM
  and CPL models. The late-time parameters correspond to the CSB
  compilation---Planck 2018 CMB, Pantheon Type~Ia supernovae and BAO
  measurements from 6dFGS, SDSS MGS and BOSS DR12---and are quoted from
  Shah et al.~\citep{Shah2023}, as also summarized in
  Mukherjee et al.~\citep{MukherjeeShah2024}.
  The primordial-sector parameters are fixed to the Planck 2018 best fit
  values \citep{PlanckVI2020}.}
\label{tab:lcdm_cpl_fiducial}
\centering
\small
\begin{tabular}{lcc}
\toprule
Parameter & $\Lambda$CDM & CPL \\
\midrule
$H_0\;[\mathrm{km\,s^{-1}\,Mpc^{-1}}]$
  & $67.72^{+0.42}_{-0.41}$ & $68.34^{+0.83}_{-0.88}$ \\
$\Omega_{m0}$
  & $0.3102^{+0.0054}_{-0.0057}$
  & $0.3064^{+0.0079}_{-0.0081}$ \\
$w_0$               & $-1$     & $-0.9571^{+0.078}_{-0.082}$ \\
$w_a$               & $0$      & $-0.2904^{+0.33}_{-0.28}$ \\
\midrule
$\tau_{\rm reio}$   & $0.054$  & $0.054$   \\
$n_s$               & $0.965$  & $0.965$   \\
$\ln(10^{10}A_s)$   & $3.044$  & $3.044$   \\
\bottomrule
\end{tabular}
\end{table}

\noindent\textbf{Interacting Dark Matter - Dark Energy Model}

Beyond phenomenological dark energy models we also consider scenarios
involving nongravitational interactions within the dark sector. We
specifically include an interacting dark matter (IntDM) model and an
interacting dark energy (IntDE) model. In both cases the background
evolution is governed by the modified continuity equations
\citep{Amendola2000, Wang2016InteractingDE, Pan2019Interacting}
\begin{align}
  \dot{\rho}_{\rm cdm} + 3H\rho_{\rm cdm} &= Q, \label{eq:intDM_cont}\\
  \dot{\rho}_{\rm de}  + 3H(1+w)\rho_{\rm de} &= -Q, \label{eq:intDE_cont}
\end{align}
where $Q$ denotes the energy transfer rate between the dark matter and
dark energy fluids. For the IntDM scenario the interaction is
proportional to the dark matter energy density,
$Q = \xi H \rho_{\rm cdm}$, while for the IntDE scenario it is
proportional to the dark energy density,
$Q = \xi H \rho_{\rm de}$, where $\xi$ is a dimensionless coupling
constant. A positive $\xi$ corresponds to energy transfer from dark
energy to dark matter. In the mock catalog generation we use
$\xi=-0.00050$ for IntDM and $\xi=0.067$ for IntDE, as listed in
Table~\ref{tab:intDM_CPLLambda}. The background expansion history is computed
self-consistently within \texttt{CLASS} for both interaction kernels.
The IntDM fiducial parameters are taken from the interacting dark sector
constraints of Pan et al.~\citep{Pan2019Interacting}, based on a
combined analysis of CMB, JLA Type~Ia supernovae, baryon acoustic
oscillation, cosmic chronometer and local $H_0$ prior data. The IntDE
case uses a separate fiducial CLASS input, so its background parameters
are not assumed to be identical to IntDM. The two interacting scenarios
therefore differ both in whether the interaction term is proportional to
$\rho_{\rm cdm}$ or $\rho_{\rm de}$ and in their fiducial parameter
choices. For comparison we also list the corresponding parameters for
the CPL$+\Lambda$ case from the
$w_0w_a\Lambda$CDM fit of Notari et al.~\citep{Notari2024DESI}, based on
Planck18 TTTEEE+lensing, DESI BAO and DES-SNYR5 supernova data. The
adopted values are reported in Table~\ref{tab:intDM_CPLLambda}.

\noindent\textbf{Interaction parameters.}
The new physical ingredient is the energy transfer rate $Q$. The two
interacting models use different choices for whether $Q$ follows
$\rho_{\rm cdm}$ or $\rho_{\rm de}$, with different fiducial values of
$\xi$ and different background parameter choices.

\begin{table}[ht]
\caption{\it Fiducial parameter values for the interacting
  dark matter (IntDM) model, the interacting dark energy (IntDE) model,
  and the CPL$+\Lambda$ model. The IntDM fiducial values are taken from
  the interacting dark sector analysis of Pan et al.~\citep{Pan2019Interacting},
  which combines CMB, JLA Type~Ia supernovae, baryon acoustic
  oscillation, cosmic chronometer and local $H_0$ prior data. The IntDE
  entries are fixed values used for the mock catalog generation; in that
  case $\Omega_{\rm cdm}h^2$ is derived from
  $\Omega_m h^2-\Omega_b h^2$, and no $\tau$ value is listed because the
  input uses \texttt{reio\_none}. The coupling entries are the fixed
  values used for the mock catalog generation. The
  CPL$+\Lambda$ column is taken from the $w_0w_a\Lambda$CDM fit of
  Notari et al.~\citep{Notari2024DESI}, using Planck18 TTTEEE+lensing,
  DESI BAO and DES-SNYR5 supernova data. Where uncertainties are shown
  they denote the quoted source $1\sigma$ intervals.}
\label{tab:intDM_CPLLambda}
\centering
\small
\setlength{\tabcolsep}{5pt}
\begin{tabular}{lccc}
\toprule
Parameter & IntDM & IntDE & CPL$+\Lambda$ \\
\midrule
$\Omega_{\rm cdm}h^2$
  & $0.11722^{+0.00066}_{-0.00128}$
  & $0.14608$
  & $0.1196\pm 0.00098$ \\
$\Omega_b h^2$
  & $0.02236^{+0.00016}_{-0.00032}$
  & $0.00020$
  & $0.02238\pm 0.00014$ \\
$\tau$
  & $0.0573^{+0.0078}_{-0.0154}$
  & --- 
  & $0.05442^{+0.0071}_{-0.0076}$ \\
$n_s$
  & $0.9772^{+0.0036}_{-0.0079}$
  & $0.9766$
  & $0.9654^{+0.0039}_{-0.0038}$ \\
$\ln(10^{10}A_s)$
  & $3.058^{+0.016}_{-0.032}$
  & $3.124$
  & $3.044^{+0.014}_{-0.015}$ \\
$w_0$              & ---  & ---  & $-0.4342^{+0.16}_{-0.41}$ \\
$w_a$              & ---  & ---  & $-2.359^{+1.8}_{-0.61}$ \\
$\Omega_{\Lambda}$ & ---  & ---  & $0.237^{+0.26}_{-0.054}$ \\
$\Omega_{m0}$
  & $0.2988^{+0.0044}_{-0.0087}$
  & $0.297$
  & --- \\
$\Omega_{\rm CPL}$ & --- & --- & $0.448^{+0.057}_{-0.28}$ \\
$\xi$
  & $-0.00050$
  & $0.067$
  & --- \\
$H_0\;[\mathrm{km\,s^{-1}\,Mpc^{-1}}]$
  & $68.56^{+0.37}_{-0.74}$
  & $70.18$
  & $67.32^{+0.62}_{-0.65}$ \\
\bottomrule
\end{tabular}
\end{table}

\noindent\textbf{Axion-inspired Early Dark Energy Model}

Finally, we include an axion-inspired scalar field model
\citep{Poulin2018Axion, Smith2020EDE} of early dark
energy to explore physically motivated departures from the concordance
paradigm. This model introduces additional dynamical degrees of freedom
that modify the background expansion history at intermediate redshifts.
The background evolution for this scenario is implemented using the
\texttt{AxiCLASS} extension of \texttt{CLASS}
\citep{Lesgourgues2011, Blas2011, AxiCLASSGitHub, Smith2020EDE,
Poulin2018Axion}. The fiducial
parameter values are taken from the \texttt{AxiCLASS}
\texttt{best\_fit\_paper\_Smithetal} input file, corresponding to the
$n$-free early dark energy best fit of Smith et al.~\citep{Smith2020EDE}. This fit
uses CMB, baryon acoustic oscillation, Type~Ia supernova luminosity
distance and SH0ES $H_0$ data. The values are summarized in
Table~\ref{tab:axion_fiducial}.

\noindent\textbf{Scalar field parameters.}
For this model the key additional quantities are the axion index
$n_{\rm axion}$, the critical scale factor $a_c$, the axion fraction
$f_{\rm axion}(a_c)$ and the initial displacement $\Theta_i$. These
parameters control when the scalar field becomes dynamically important
and how strongly it changes the background expansion.

\begin{table}[ht]
\caption{\it Fiducial cosmological parameters for the axion scalar field
  (early dark energy) model implemented using \texttt{AxiCLASS}. The
  fiducial entries are fixed best-fit values from the
  \texttt{best\_fit\_paper\_Smithetal} \texttt{AxiCLASS} input file,
  corresponding to the $n$-free best fit of Smith et al.~\citep{Smith2020EDE} to
  Planck CMB, BAO, Pantheon Type~Ia supernova and SH0ES $H_0$ data.
  The third column lists the corresponding posterior mean and $1\sigma$
  uncertainty quoted in Table~I of Smith et al.~\citep{Smith2020EDE}. The mock
  catalog generation itself uses only the fixed best-fit values in the
  second column.}
\label{tab:axion_fiducial}
\centering
\small
\begin{tabular}{lcc}
\toprule
Parameter & Fiducial value & Source mean and $1\sigma$ \\
\midrule
$\Omega_b h^2$                          & $0.02251$              & $0.02261\pm0.00024$ \\
$\Omega_{\rm cdm}h^2$                   & $0.1320$               & $0.1290^{+0.0041}_{-0.0045}$ \\
$H_0\;[\mathrm{km\,s^{-1}\,Mpc^{-1}}]$ & $72.81$                & $71.45^{+1.10}_{-1.40}$ \\
$\tau$                                  & $0.068$                & $0.070\pm0.014$ \\
$n_s$                                   & $0.9860$               & $0.9853^{+0.0073}_{-0.0079}$ \\
$A_s$                                   & $2.191\times10^{-9}$   & $(2.196\pm0.055)\times10^{-9}$ \\
$N_{\rm eff}$                           & $3.0328$               & fixed \\
$m_{\nu}\;[\mathrm{eV}]$                & $0.06$                 & fixed \\
\midrule
\multicolumn{3}{c}{\itshape Axion/scalar field parameters} \\
\midrule
$n_{\rm axion}$      & $2.6$    & $3.16^{+0.18}_{-1.16}$ \\
$\log_{10}(a_c)$     & $-3.531$ & $-3.558^{+0.110}_{-0.053}$ \\
$f_{\rm axion}(a_c)$ & $0.132$  & $0.103\pm0.035$ \\
$\Theta_i$           & $2.72$   & $2.49^{+0.52}_{-0.01}$ \\
\bottomrule
\end{tabular}

\end{table}

The background evolution equations are implemented self-consistently
within \texttt{CLASS} and its extensions for all fiducial cosmological
models \citep{Lesgourgues2011, Blas2011, AxiCLASSGitHub}. Using the
resulting expansion histories we compute the redshift
dependence of the Hubble parameter and the corresponding luminosity
distance to serve as theoretical inputs for generating mock gravitational
wave standard siren observations. By adopting fiducial parameters
inferred from existing observational data rather than arbitrary choices
we ensure that the mock catalogs are realistic and representative of
viable cosmological scenarios.

\subsection{Mock gravitational wave standard siren catalogs}
\label{sec:mock_catalogs}

We utilize the fiducial cosmological models described in
Sec.~\ref{sec:fiducial_models} to generate mock catalogs of
gravitational wave standard siren events for both the space-based LISA
detector and the ground-based Einstein Telescope. These two detectors
probe overlapping but distinct redshift windows and are subject to
different dominant noise sources which makes their combined data
particularly powerful for model-independent reconstruction.

The observed luminosity distance for each simulated event at redshift
$z_i$ is obtained by perturbing the fiducial value with a stochastic
error drawn from a Gaussian distribution:
\begin{equation}
  d_L^{\rm obs}(z_i) = d_L^{\rm fid}(z_i)
  + \mathcal{N}\!\left(0,\,\sigma_{d_L}^2(z_i)\right).
  \label{eq:mock_obs}
\end{equation}
Here $d_L^{\rm obs}$ is the mock observed luminosity distance,
$d_L^{\rm fid}$ is the fiducial luminosity distance predicted by the
chosen cosmological model, and $\mathcal{N}(0,\sigma_{d_L}^2)$ denotes a
Gaussian random error with zero mean and variance $\sigma_{d_L}^2$. The
total uncertainty $\sigma_{d_L}(z_i)$ in this expression is assembled
from independent noise contributions specific to each detector
configuration as detailed below. We generate
$N_{\rm mock}\equiv N_{\rm real}=50$ independent mock realizations for
each cosmological model. Here $N_{\rm mock}$ denotes the number of
independent simulated catalogs used to estimate the ensemble-level
reconstruction behavior. This choice ensures that the reported
uncertainties are not dominated by any single noise realization. All
realizations for a given detector employ the identical redshift sampling
strategy so the mock catalogs for different fiducial models differ only
in their underlying cosmological evolution rather than their
observational selection or noise prescriptions.

\subsubsection{LISA configuration}
\label{sec:lisa_mocks}

LISA is highly sensitive to massive black hole binary (MBHB) mergers,
which can provide bright standard sirens at high redshift. We adopt a
baseline catalog with $N_{\rm ev}=80$ events over the interval
$z\in(0,5]$. Their redshifts are drawn from an interpolated Beta
distribution constructed following the practical procedure of
Mukherjee et al.~\citep{MukherjeeShah2024} from redshift information
presented in Fig.~1 of Tamanini et al.~\citep{Tamanini2016}. That
forecast is based on LISA MBHB population models, including no-delay
high-mass-seed scenarios discussed by Klein et al.~\citep{Klein2016}.
curve itself as a physical bright siren probability density. It is used
only as a Monte Carlo sampling proxy on a finite redshift interval. The
choice of $N_{\rm ev}=80$ MBHB events follows the $\sim 10$ year
LISA bright-siren scenario used by
Mukherjee et al.~\citep{MukherjeeShah2024}; the original Tamanini
et al.~forecast gives tens of redshift-identified MBHB standard sirens
over a shorter mission, with the number depending on the assumed seed
formation and accretion model. In the event count study discussed later
we also extend the LISA sample to $N_{\rm ev}=160$ to test the scaling
of the reconstruction precision.
Electromagnetic counterpart identification for MBHB mergers, required
for spectroscopic host redshifts, is expected for a significant fraction
of these events via associated multi-messenger emission
\citep{Tamanini2016, Bogdanovic2022MBHCounterparts}.

The total LISA luminosity distance uncertainty combines four independent
contributions that each capture a distinct physical source of
observational error
\citep{Tamanini2016, Tamanini2017, Speri2021, Marsat2021, Ferreira2022}:
\begin{equation}
  \sigma_{d_L}^{\rm LISA}(z)
  = \sqrt{
    \sigma_{\rm inst}^2(z)
    + \sigma_{\rm lens}^2(z)
    + \sigma_{\rm pv}^2(z)
    + \sigma_{\rm ph\text{ }z}^2(z)
  }.
  \label{eq:lisa_total_err}
\end{equation}
Here $\sigma_{\rm inst}$ is the instrumental distance uncertainty,
$\sigma_{\rm lens}$ is the weak-lensing magnification uncertainty,
$\sigma_{\rm pv}$ is the host-galaxy peculiar-velocity contribution, and
$\sigma_{\rm ph\text{ }z}$ is the contribution induced by photometric
redshift errors.

The instrumental noise is modeled as follows:
\begin{equation}
  \sigma_{\rm inst}(z)
  = 0.05\,\frac{d_L(z)}{d_L^{\rm ref}}\,d_L(z).
  \label{eq:lisa_inst}
\end{equation}
The reference distance $d_L^{\rm ref}=36.6\;\mathrm{Gpc}$ corresponds
approximately to the luminosity distance at $z=4$ in the calibration
cosmology used for this LISA error prescription. This specific
formulation reflects the signal-to-noise-ratio dependent precision of
parameter estimation derived from the GW waveform.

The weak lensing contribution is modeled following an established
fitting formula \citep{Hirata2010, Shapiro2010}:
\begin{equation}
  \sigma_{\rm lens}(z)
  = \frac{0.066\,d_L(z)}{\Delta}
  \left[\frac{1-(1+z)^{-1/4}}{1/4}\right]^{1.8}.
  \label{eq:lisa_lens}
\end{equation}
The delensing factor $\Delta$ accounts for the partial removal of
lensing magnification using weak shear maps. We adopt $\Delta=2$ to
represent a 50\% reduction in the lensing noise achievable through
future photometric surveys.
This treatment uses weak lensing as an externally specified
distance-error contribution. In a full cosmological inference however
the magnification distribution depends on the cosmological parameters
and its scatter can itself constrain structure growth, especially
$\sigma_8$ \citep{Vaskonen2026WeakLensingSirens}. We therefore
interpret $\sigma_{\rm lens}$ here only as part of the mock
distance-noise model, not as a likelihood for lensing magnifications.

The peculiar velocity of the host galaxy contributes an additional
layer of uncertainty \citep{Kocsis2006, Tamanini2017}:
\begin{equation}
  \sigma_{\rm pv}(z)
  = d_L(z)
  \left[1+\frac{c(1+z)^2}{H(z)\,d_L(z)}\right]
  \frac{v_{\rm rms}}{c}.
  \label{eq:lisa_pv}
\end{equation}
Here $c$ is the speed of light, $H(z)$ is the Hubble parameter evaluated
in the fiducial cosmology, and $v_{\rm rms}$ is the root-mean-square
peculiar velocity of the host galaxy. We assume
$v_{\rm rms}=500\;\mathrm{km\,s^{-1}}$. This specific term forms the
dominant noise contribution at low redshift where peculiar motions
represent a significant fraction of the overall Hubble flow.

Finally, an additional photometric redshift uncertainty is included for
events at $z>2$ where host galaxies are identified photometrically
rather than spectroscopically \citep{Dahlen2013, Ilbert2013UltraVISTA}:
\begin{equation}
  \sigma_{\rm ph\text{ }z}(z) =
  \begin{cases}
    \left|\dfrac{d\,d_L}{dz}\right| \times 0.03\,(1+z), & z > 2, \\[6pt]
    0, & z \leq 2.
  \end{cases}
  \label{eq:lisa_photoz}
\end{equation}
Here $0.03(1+z)$ is the assumed one-sigma photometric redshift error.
Since $d_L(z)=(1+z)d_C(z)$, where $d_C(z)$ is the comoving distance, the
derivative entering this propagation is
$d d_L/dz=d_C+(1+z)d_C'$. This is the luminosity-distance uncertainty
induced by the photometric redshift error of the host.
This piecewise formulation inherently assumes that precise spectroscopic
identification remains broadly feasible at $z\leq 2$.
When electromagnetic counterparts or spectroscopic host identifications
are incomplete, galaxy-catalog and cross-correlation methods provide an
alternative route to statistical redshift information
\citep{Gray2020MockSiren, Gray2023DarkSirenCatalogues,
MukherjeeWandelt2018CrossCorrelation, Mukherjee2021RedshiftUnknown,
DiazMukherjee2022ExpansionHistory, AfrozMukherjee2024CrossCorrelation}.

\subsubsection{Einstein Telescope configuration}
\label{sec:ET_mocks}

The Einstein Telescope represents a proposed third-generation
ground-based detector capable of observing binary neutron star (BNS)
mergers out to $z\sim 2$ \citep{ET2020, Sathyaprakash2010}. We
distribute $N_{\rm ev}=1000$ events over the interval
$z\in[0.07,2.0]$ following the expected BNS merger rate density
\citep{Schneider2001}:
\begin{equation}
  R(z) =
  \begin{cases}
    (1+z)^{2.7}  & z \leq 1, \\[4pt]
    (1+z)^{-0.9} & z > 1.
  \end{cases}
  \label{eq:BNS_rate}
\end{equation}
This distribution tracks the cosmic star formation history while
incorporating a time delay correction for the interval between initial
star formation and the eventual binary merger. The corresponding
redshift probability density for detected events takes the following
form:
\begin{equation}
  p(z) \propto
  \frac{4\pi\,d_C^2(z)\,R(z)}{H(z)\,(1+z)}.
  \label{eq:ET_pz}
\end{equation}
Here $d_C(z)$ is the comoving distance, $H(z)$ is the fiducial Hubble
parameter, and the factor $(1+z)^{-1}$ converts the source-frame merger
rate to the observer frame. The omitted overall constants, including
the speed of light, are absorbed when the distribution is normalized.
This function is normalized over the detection window and evaluated on
the fiducial \texttt{CLASS} background. We draw specific events by
inverse cumulative distribution function (CDF) sampling of
Eq.~(\ref{eq:ET_pz}): the normalized probability density is integrated
to form a cumulative probability, and uniform random numbers are mapped
to redshifts by inverting that cumulative function. 
This substantially larger
event count compared to the LISA configuration reflects the high
detection rate expected for ET given its sensitivity to the much more
frequent BNS population. The choice of $N_{\rm ev}=1000$ BNS events
reflects not the total ET detection rate
($\sim\!10^5\,\mathrm{yr}^{-1}$) but the expected number of bright
sirens---events with a confirmed electromagnetic counterpart enabling a
host galaxy spectroscopic redshift. Following Zhao
et al.~\citep{Zhao2011} and Belgacem et al.~\citep{Belgacem2019}, we
adopt 1000 events as a benchmark corresponding to a multi-year ET
observing run assuming an EM counterpart
identification fraction of order $10\%$, consistent with the rates
inferred from GW170817 \citep{Abbott2017b}. This difference creates a qualitatively
different reconstruction regime characterized by denser sampling
confined to a lower redshift range.
Future catalogue construction can refine this benchmark with population
and phase-space information for low-mass compact objects and
formation-channel classification
\citep{AfrozMukherjee2025LMCOPhaseSpace}.
Future high-energy and multi-messenger facilities are expected to
improve counterpart discovery and follow-up for such bright siren
samples \citep{Amati2018THESEUS, Piro2022Athena, Bhalerao2024Daksha}.

The total ET luminosity distance uncertainty combines two independent
contributions:
\begin{equation}
  \sigma_{d_L}^{\rm ET}(z)
  = \sqrt{
    \sigma_{\rm inst}^2(z)
    + \sigma_{\rm lens}^2(z)
  }.
  \label{eq:ET_total_err}
\end{equation}
The instrumental component relies on a redshift-dependent fitting
formula calibrated to the ET sensitivity curve for BNS mergers
\citep{Zhao2011, MukherjeeShah2024, Ferreira2022}:
\begin{equation}
  \sigma_{\rm inst}(z)
  = \mathcal{F}(z)\,d_L(z),\quad
  \mathcal{F}(z) = 0.1449\,z - 0.0118\,z^2 + 0.0012\,z^3.
  \label{eq:ET_inst}
\end{equation}
This polynomial fitting formula captures the redshift dependence of the
SNR averaged instrumental uncertainty for BNS events which naturally
grows as the signal weakens at higher redshifts. This approach
supersedes the simple flat approximation of $\sigma_{\rm inst}=0.25\,d_L$
and produces a significantly more realistic redshift-dependent noise
floor. The fractional error sits at approximately $1.4\%$ near
$z\sim 0.1$ before rising to $\sim\!20\%$ at $z\sim 1.5$ and finally
reaching $\sim\!25\%$ near the upper edge of the ET detection window at
$z=2$. We model the weak lensing contribution as follows:
\begin{equation}
  \sigma_{\rm lens}(z) = 0.05\,z\,d_L(z).
  \label{eq:ET_lens}
\end{equation}
Peculiar velocity and photometric redshift contributions remain
subdominant for ET given its restricted redshift range and the assumed
spectroscopic identification of BNS host galaxies via electromagnetic
counterparts. We therefore exclude these terms from the total
uncertainty budget. The instrumental term dominates at low redshift
while the lensing contribution grows linearly with $z$ and becomes the
leading source of uncertainty beyond $z\sim 1$.

We summarize the key properties and error model differences between
these two detector configurations in Table~\ref{tab:detector_comparison}
for easy reference.

\begin{table}[ht]
\caption{\it Summary of LISA and ET configurations used for mock
  catalog generation.}
\label{tab:detector_comparison}
\centering
\small
\begin{tabular}{lcc}
\toprule
Property & LISA & Einstein Telescope \\
\midrule
Redshift range      & $(0,\,5]$ & $[0.07,\,2.0]$ \\
Events per catalog  & $80$      & $1000$ \\
Source type         & MBHB      & BNS \\
Redshift dist.      & Interpolated Beta proxy & BNS merger rate \\
$\sigma_{\rm inst}$ & $0.05(d_L/d_L^{\rm ref})d_L$
                    & $\mathcal{F}(z)\,d_L$, Eq.~(\ref{eq:ET_inst}) \\
$\sigma_{\rm lens}$ & Eq.~(\ref{eq:lisa_lens}) & $0.05\,z\,d_L$ \\
$\sigma_{\rm pv}$         & Yes & No \\
$\sigma_{\rm ph\text{-}z}$& Yes ($z>2$) & No \\
\bottomrule
\end{tabular}
\end{table}

\begin{figure}[ht]
  \centering
  \includegraphics[width=.50\columnwidth]{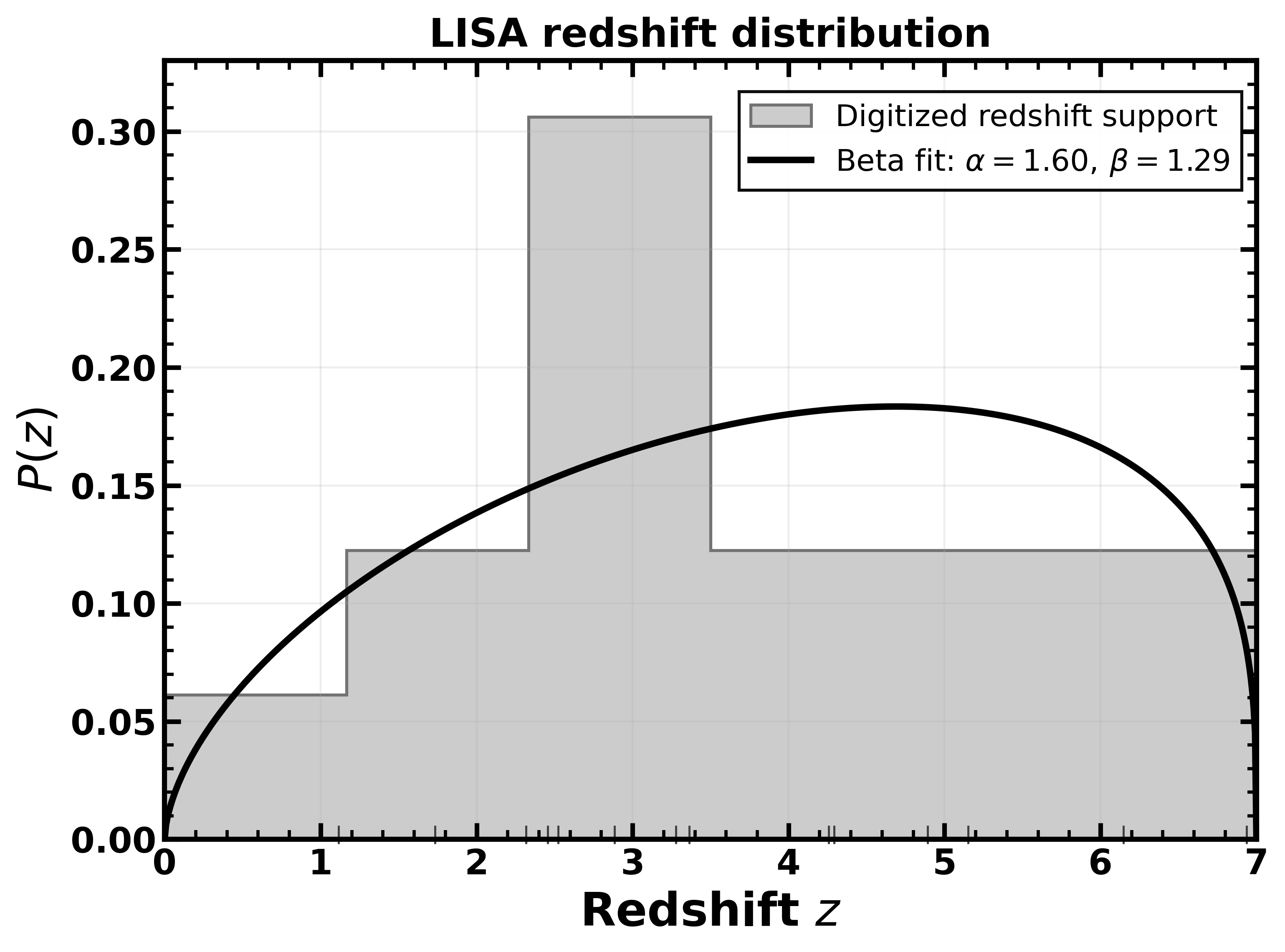}%
  \hfill%
  \includegraphics[width=.50\columnwidth]{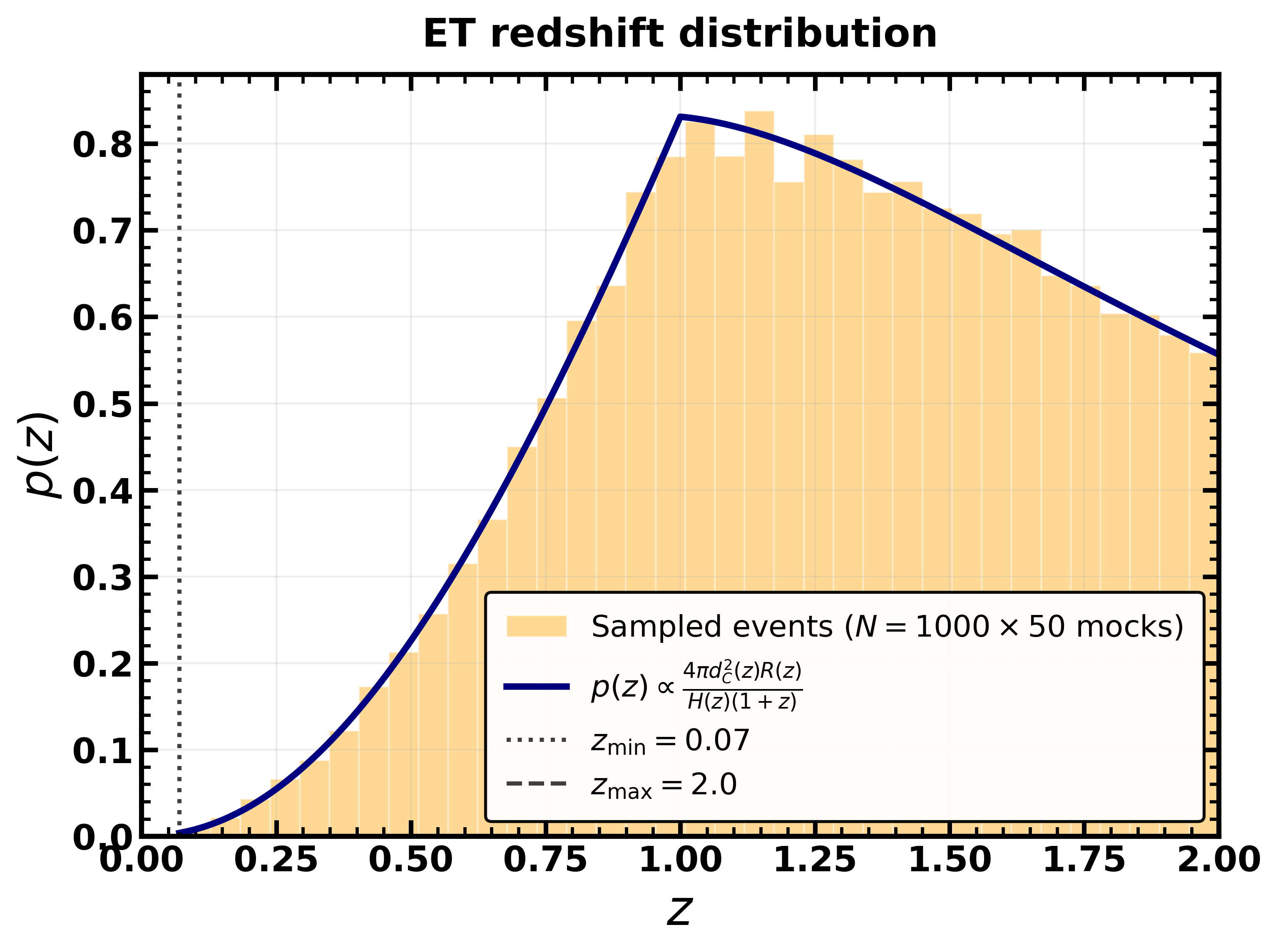}
  \caption{\it Adopted redshift distributions for mock event generation.
    \textit{Left}: Interpolated Beta distribution constructed from the
    data presented in Fig.~1 of Tamanini et al.~\citep{Tamanini2016}
    and used only as a sampling proxy for the fiducial LISA MBHB
    catalog.
    \textit{Right}: BNS merger rate redshift density $p(z)$ from
    Eq.~(\ref{eq:ET_pz}) for the ET configuration.}
  \label{fig:redshift_distributions}
\end{figure}

\FloatBarrier

The two panels show that the mock catalogues sample complementary
regions of redshift space. The LISA proxy has support over a wider
range and keeps a high redshift tail. The ET distribution is more
concentrated at low and intermediate redshift. This difference explains
why LISA is useful for testing the expansion history at larger $z$ while
ET mainly strengthens the reconstruction where the catalog is dense.

\section{Gaussian Process Reconstruction framework}
\label{sec:GPR}

In this section we present the Gaussian Process Regression (GPR)
framework adopted for the nonparametric reconstruction of cosmological
observables from gravitational wave standard siren data
\citep{Seikel2012}. Gaussian Processes provide a
probabilistic description of functions that allows for a consistent
reconstruction of a function and its derivatives directly from data
without assuming any specific functional form. This inherent flexibility
makes GPR particularly well suited for model-independent studies of the
cosmic expansion history.

\subsection{Gaussian Process formalism and kernel choice}
\label{sec:gp_formalism}

Our notation follows the standard GP regression construction
\citep{Rasmussen2005}. Closely related GP reconstructions of
cosmological distances, expansion histories and dark-energy diagnostics
have been developed in
Refs.~\citep{Seikel2012, Holsclaw2010, Holsclaw2011, Shafieloo2012,
Busti2014, Yahya2014, BilickiSeikel2012, GomezValentAmendola2018,
OColgain2021}.

A Gaussian Process (GP) is defined as a collection of random variables
where any finite subset follows a multivariate normal distribution. A
function $f(z)$ drawn from a Gaussian Process is fully specified by its
mean function $\mu(z)$ and its covariance function or kernel $k(z,z')$:
\begin{equation}
  f(z) \sim \mathcal{GP}\!\left(\mu(z),\,k(z,z')\right).
\end{equation}
Here $z$ and $z'$ denote redshifts, $f(z)$ is the continuous function to
be reconstructed from the standard-siren data, $\mu(z)$ is the prior
mean function, and $k(z,z')$ is the covariance kernel that specifies how
function values at two redshifts are correlated.
We assume a vanishing mean function $\mu(z)=0$ throughout this work to
ensure a cosmology-independent reconstruction since any nonzero mean can
always be absorbed into the data.

Given a set of training inputs $\mathbf{Z}=\{z_i\}$ with corresponding
observations $\mathbf{y}$ and noise covariance matrix $C$, the prior
covariance matrix is $k(\mathbf{Z},\mathbf{Z})$ with elements
$k(z_i,z_j)$. Here $\mathbf{y}$ denotes the measured distance data used
as GP training targets, while $C$ contains their observational
covariances. The joint prior distribution of the training data and the
function evaluated at test points $\mathbf{Z}^\star=\{z^\star_j\}$ is:
\begin{equation}
  \Sigma =
  \begin{pmatrix}
    k(\mathbf{Z},\mathbf{Z}) + C & k(\mathbf{Z},\mathbf{Z}^\star) \\[4pt]
    k(\mathbf{Z}^\star,\mathbf{Z}) & k(\mathbf{Z}^\star,\mathbf{Z}^\star)
  \end{pmatrix}.
  \label{eq:joint_prior}
\end{equation}
Conditioning on the observed data yields the predictive distribution for
the reconstructed function $f(\mathbf{Z}^\star)$, with posterior mean
$\bar f(\mathbf{Z}^\star)$:
\begin{equation}
  \bar{f}(\mathbf{Z}^\star)
  = k(\mathbf{Z}^\star,\mathbf{Z})
  \left[k(\mathbf{Z},\mathbf{Z})+C\right]^{-1}\mathbf{y},
  \label{eq:pred_mean}
\end{equation}
with the accompanying predictive covariance 
\begin{equation}
  \mathrm{cov}\!\left[f(\mathbf{Z}^\star),f(\mathbf{Z}^\star)\right]
  = k(\mathbf{Z}^\star,\mathbf{Z}^\star)
  - k(\mathbf{Z}^\star,\mathbf{Z})\left[k(\mathbf{Z},\mathbf{Z})+C\right]^{-1}k(\mathbf{Z},\mathbf{Z}^\star).
  \label{eq:pred_cov}
\end{equation}

The choice of kernel function heavily influences the generalization
properties and smoothness of the final reconstruction. The squared
exponential (SE) kernel takes the form:
\begin{equation}
  k_{\rm SE}(z,\tilde{z}) = \sigma_f^2
  \exp\!\left[-\frac{(z-\tilde{z})^2}{2\ell^2}\right].
\end{equation}
Here $\sigma_f$ is the signal-amplitude hyperparameter, $\ell$ is the
correlation length scale in redshift, and $\tilde z$ denotes a second
redshift argument. Although the SE kernel is mathematically convenient
because it is infinitely differentiable, this smoothness can be too
restrictive for cosmological reconstruction. It strongly correlates
neighboring redshifts and can therefore over-smooth real features in the
expansion history, producing confidence intervals that are artificially
tight. Cosmological distance data are sparse and noisy, especially for
standard sirens, so it is useful to choose a kernel whose differentiability
is sufficient for the required diagnostics but not unlimited.

The Mat\'ern family provides this intermediate option: it gives a
controlled degree of smoothness while retaining analytic covariance
functions for half-integer values of $\nu$ \citep{SeikClark2013,
OColgain2021}. In cosmology this is useful because one
often wants model-independent reconstructions of $H(z)$ and derivative
diagnostics without imposing the very smooth behavior built into the SE
kernel. The differentiability is governed by the parameter
$\nu=p+1/2$, where $p$ is a non-negative integer:
\begin{multline}
  k_{\nu=p+1/2}(z,\tilde{z}) =
  \sigma_f^2
  \exp\!\left(-\frac{\sqrt{2p+1}\,|z-\tilde{z}|}{\ell}\right) \\
  \times\frac{p!}{(2p)!}
  \sum_{i=0}^{p}
  \frac{(p+i)!}{i!(p-i)!}
  \left(\frac{2\sqrt{2p+1}\,|z-\tilde{z}|}{\ell}\right)^{p-i}.
  \label{eq:matern_general}
\end{multline}
In this expression $i$ is the summation index and $p$ fixes the order of
the half-integer Mat\'ern kernel. The kernel must satisfy $\nu>n$ to
reconstruct an $n$th order derivative. In the present work the strongest
requirement comes not only from reconstructing $d_C''(z)$, but from its
covariance block $\mathrm{Cov}[d_C''(z),d_C''(z')]$, which requires the
kernel derivative $k^{(2,2)}$ and hence four total differentiations of
$k(z,z')$. We therefore adopt the Mat\'ern $\nu=9/2$ kernel ($p=4$),
which supplies this fourth-order differentiability while remaining less
restrictive than the infinitely differentiable SE kernel:
\begin{multline}
  k(z,\tilde{z}) =
  \sigma_f^2
  \exp\!\left(-\frac{3|z-\tilde{z}|}{\ell}\right)
  \Biggl[
  1 + \frac{3|z-\tilde{z}|}{\ell}
  + \frac{27(z-\tilde{z})^2}{7\ell^2} \\
  + \frac{18|z-\tilde{z}|^3}{7\ell^3}
  + \frac{27(z-\tilde{z})^4}{35\ell^4}
  \Biggr].
  \label{eq:matern92}
\end{multline}
This specific selection ensures sufficient differentiability for
calculating higher-order diagnostics while successfully avoiding the
over-correlation associated with the SE kernel
\citep{SeikClark2013, OColgain2021}.

\subsection{Hyperparameter inference}
\label{sec:hyperparameters}

The Mat\'ern $\nu=9/2$ kernel~(\ref{eq:matern92}) contains two free
hyperparameters: the signal variance $\sigma_f$ and the length scale
$\ell$. They control the amplitude and correlation range of the
reconstructed function respectively. Their values are not prescribed
\textit{a priori} but are instead directly inferred from the data.

We infer the hyperparameters from the GP marginal likelihood. For fixed
$(\sigma_f,\ell)$, analytically marginalizing over the latent function
values gives the standard GP evidence
\citep{Seikel2012}: 
\begin{equation}
  \ln p(\mathbf{y}\,|\,\mathbf{Z},\,\sigma_f,\ell)
  = -\frac{1}{2}\mathbf{y}^{\top}\left[k(\mathbf{Z},\mathbf{Z})+C\right]^{-1}\mathbf{y}
  - \frac{1}{2}\ln\det\!\left[k(\mathbf{Z},\mathbf{Z})+C\right] - \frac{n}{2}\ln 2\pi,
  \label{eq:log_evidence}
\end{equation}
Here the training points are the input redshifts
$\mathbf{Z}=\{z_i\}_{i=1}^{n}$ at which the mock standard-siren catalog
provides measured distance values $\mathbf{y}$ and observational
covariance $C$; thus $n$ is the number of such training redshifts. The
first term in Eq.~(\ref{eq:log_evidence}) measures how well a chosen
kernel with hyperparameters $(\sigma_f,\ell)$ fits the data, the
log-determinant term penalizes overly flexible covariance structures
and the final term is the Gaussian normalization. 
This objective therefore balances data fit against model complexity to
penalize kernels that might overfit noisy data. We use
Eq.~(\ref{eq:log_evidence}) as the likelihood term in a Markov Chain
Monte Carlo (MCMC) marginalization over $(\sigma_f,\ell)$, thereby
propagating hyperparameter uncertainty to the reconstructed diagnostics.
We implement this with the \texttt{emcee}
ensemble sampler \citep{ForemanMackey2013} using log-uniform priors over
physically motivated ranges. We use the posterior samples to compute an
ensemble of GP reconstructions and we incorporate the resulting spread
into the reported uncertainties. We find that hyperparameter
marginalization increases the reported confidence intervals by
approximately $10$ to $30\%$ relative to holding the hyperparameters
fixed near a representative posterior value. This confirms it is a
nonnegligible yet subdominant source of uncertainty compared to the
underlying data noise.

The two hyperparameters play qualitatively distinct roles in shaping the
reconstructed precision. The signal variance $\sigma_f$ enters the kernel
as a global amplitude factor,
\begin{equation}
  k(z,\tilde{z}) = \sigma_f^2\,\mathcal{K}\!\left(\frac{|z-\tilde{z}|}{\ell}\right),
  \label{eq:kernel_factored}
\end{equation}
where $\mathcal{K}$ denotes the dimensionless Mat\'ern kernel shape as a
function of the scaled redshift separation. The signal variance
$\sigma_f$ therefore primarily changes the overall signal-to-noise
balance and the vertical scale of the posterior uncertainty. The length
scale $\ell$ by contrast governs the correlation range in redshift space and therefore
affects the \emph{shape} of the precision curve. Below the first
low-redshift events the GP has no anchoring data and derivative
uncertainties are extrapolation dominated. Once the first events are
included the reconstruction enters an interpolation regime and the
precision curve reaches its minimum slightly above this low-redshift
anchor. Thus $\sigma_f$ mainly affects how precise the reconstruction is
\emph{at} $z^*$, while $\ell$ and the location of the first events
influence \emph{where} $z^*$ occurs. Since ET ($z_{\min}=0.07$) and LISA
(effective $z_{\min}\sim 0.1$--$0.2$ from the Beta distribution) have
similar low-redshift anchoring regions, the resulting $z^*$ values are
close despite their different event counts and noise prescriptions. This
interpretation is empirical for the mock catalogues studied here and is
discussed in detail in Sec.~\ref{sec:zstar}.

\subsection{Reconstruction of derivatives}
\label{sec:derivatives}

A key property of Gaussian Processes is that their derivatives are
themselves Gaussian Processes. This mathematically allows for a
statistically consistent reconstruction of derivatives directly from the
kernel function without assuming any additional functional form.

More precisely, for a GP with kernel $k(z,z')$, the predictive mean of
the $m$-th derivative is obtained by differentiating the kernel $m$
times with respect to the first argument:
\begin{equation}
  \bar{f}^{(m)}(\mathbf{Z}^\star)
  = k^{(m,0)}(\mathbf{Z}^\star,\mathbf{Z})
  \left[k(\mathbf{Z},\mathbf{Z})+C\right]^{-1}\mathbf{y},
  \label{eq:mean_deriv_general}
\end{equation}
where $k^{(m,n)}(z,z')$ denotes the kernel differentiated $m$ times
with respect to $z$ and $n$ times with respect to $z'$. Setting $m=0$
recovers Eq.~(\ref{eq:pred_mean}). The joint predictive covariance
between the $m$-th and $n$-th derivatives is obtained by differentiating
the corresponding kernel blocks \citep{SeikClark2013}:
\begin{equation}
  \mathrm{cov}\!\left[f^{(m)},f^{(n)}\right]
  = k^{(m,n)}(\mathbf{Z}^\star,\mathbf{Z}^\star)
  - k^{(m,0)}(\mathbf{Z}^\star,\mathbf{Z})\left[k(\mathbf{Z},\mathbf{Z})+C\right]^{-1}k^{(0,n)}(\mathbf{Z},\mathbf{Z}^\star),
  \label{eq:cov_deriv_general}
\end{equation}
which reduces to Eq.~(\ref{eq:pred_cov}) for $m=n=0$. The specific
cases required in this work are $m,n\in\{0,1,2\}$, yielding the
predictive means for $d_C$, $d_C'$ and $d_C''$ and the covariance and
cross-covariance blocks of the joint predictive covariance matrix.
The kernel derivatives $k^{(m,n)}(z,z')$ for $m+n\leq 4$ are computed
analytically from the Mat\'{e}rn $\nu=9/2$ kernel of
Eq.~(\ref{eq:matern92}), which is sufficiently differentiable for this
purpose by construction. Retaining the joint covariance blocks---rather
than only the diagonal variances---is
essential for consistent uncertainty propagation to derived cosmological
observables, as we discuss in detail in
Sec.~\ref{sec:covariance_structure}.

\subsection{Cosmological observables and diagnostic quantities}
\label{sec:observables}

\noindent\textbf{Distance and expansion rate.}
The primary reconstructed quantity is the comoving distance:
\begin{equation}
  d_C(z) = \frac{d_L(z)}{1+z},
  \label{eq:comoving}
\end{equation}
where $d_L(z)$ denotes the luminosity distance inferred from
gravitational wave standard sirens. Throughout this reconstruction we
assume a spatially flat FLRW background, for which the line-of-sight
comoving distance obeys $d_C'(z)=c/H(z)$. We therefore obtain the
Hubble parameter from the first derivative of $d_C$ as follows:
\begin{equation}
  H(z) = \frac{c}{d_C'(z)}.
  \label{eq:Hz}
\end{equation}
Its first derivative encodes additional information about the underlying
expansion dynamics:
\begin{equation}
  H'(z) = -c\,\frac{d_C''(z)}{\left[d_C'(z)\right]^2}.
  \label{eq:Hprime}
\end{equation}

We use $H(z)$ and its derivatives to construct a set of kinematical and
dynamical diagnostic quantities widely used to test deviations from the
concordance $\Lambda$CDM model.

\begin{itemize}
\item \textbf{Deceleration parameter $q(z)$:}
\begin{equation}
  q(z) = -1 - (1+z)\,\frac{d_C''(z)}{d_C'(z)}.
  \label{eq:qz}
\end{equation}
A transition from $q>0$ to $q<0$ signals the onset of accelerated
expansion.

\item \textbf{Dimensionless Hubble parameter $E(z)$:}
\begin{equation}
  E(z) = \frac{H(z)}{H_0}.
  \label{eq:Ez}
\end{equation}
This quantity removes the explicit dependence on $H_0$ and provides a
model-independent characterization of the overall expansion history.

\item \textbf{Om diagnostic $\mathcal{O}_m(z)$:}
\begin{equation}
  \mathcal{O}_m(z) = \frac{E^2(z)-1}{(1+z)^3-1}.
  \label{eq:Omz}
\end{equation}
This expression is the standard $Om(z)$ null diagnostic
written in terms of $E(z)$ \citep{SahniShafielooStarobinsky2008}. It
implicitly assumes a spatially flat universe with
$\Omega_{\Lambda0}=1-\Omega_{m0}$ in the $\Lambda$CDM limit and
therefore carries a degree of model dependence. It is not the physical
matter fraction $\Omega_m(z)$; rather, it reduces to the present-day
matter density parameter $\Omega_{m0}$ only for spatially flat
$\Lambda$CDM. Any significant redshift evolution consequently signals
departures from the concordance scenario.

\item \textbf{Total effective equation of state $w_{\rm tot}(z)$:}
\begin{equation}
  w_{\rm tot}(z) = -1 - \frac{2}{3}(1+z)\,\frac{d_C''(z)}{d_C'(z)}.
  \label{eq:wtot}
\end{equation}
This quantity is the total effective equation of state of the cosmic
fluid inferred from the background expansion, not the dark-energy
equation of state. In spatially flat $\Lambda$CDM it is approximately
$-\Omega_{\Lambda0}$ today (about $-0.7$ for $\Omega_{\Lambda0}\simeq
0.7$), approaches zero at high redshift during matter domination, and
approaches $-1$ only asymptotically in the far future. Alternative
models may exhibit nontrivial redshift evolution.
\end{itemize}

\subsection{Structure of the predictive covariance matrix}
\label{sec:covariance_structure}


The covariance structure is central to the physical interpretation of
the reconstruction, not merely a numerical detail. The diagnostics used
below are nonlinear functions of the reconstructed derivatives. For
example, $H'(z)$, $q(z)$, $w_{\rm tot}(z)$ and $\kappa(z)$ depend on the
ratio $d_C''/d_C'$. Their uncertainties therefore depend not only on the
individual variances of $d_C'$ and $d_C''$, but also on how these two
quantities co-vary at the same and different redshifts. If these
off-diagonal terms are discarded, the propagated confidence regions can
be biased. This would directly affect the later model-comparison step:
underestimated uncertainties can create artificial separation between
cosmologies, while overestimated uncertainties can hide real but weak
differences.

We therefore compute the joint predictive covariance matrix of the
reconstructed comoving distance and its first two derivatives. In block
form this covariance can be written as
\[
\boldsymbol{\Sigma}_{\rm GP} =
\begin{pmatrix}
\mathbf{C}_{00} & \mathbf{C}_{01} & \mathbf{C}_{02} \\
\mathbf{C}_{10} & \mathbf{C}_{11} & \mathbf{C}_{12} \\
\mathbf{C}_{20} & \mathbf{C}_{21} & \mathbf{C}_{22}
\end{pmatrix},
\qquad
\left[\mathbf{C}_{mn}\right]_{ij}
= \mathrm{Cov}\!\left[d_C^{(m)}(z_i),d_C^{(n)}(z_j)\right],
\]
where $m,n=0,1,2$ correspond respectively to $d_C$, $d_C'$ and
$d_C''$. The blocks with $i\neq j$ encode correlations between different
redshift points, while the off-diagonal blocks with $m\neq n$ encode
correlations between different derivative orders. In the diagnostic
sampling step, the most important sub-block is
\[
\boldsymbol{\Sigma}_{12} =
\begin{pmatrix}
\mathrm{Cov}[d_C',d_C'] & \mathrm{Cov}[d_C',d_C''] \\
\mathrm{Cov}[d_C'',d_C'] & \mathrm{Cov}[d_C'',d_C'']
\end{pmatrix},
\]
which is the block structure used to jointly sample $d_C'$ and $d_C''$.
For a derived diagnostic $g(d_C',d_C'')$, the linearized propagated
variance contains the term
$2(\partial g/\partial d_C')(\partial g/\partial d_C'')
\mathrm{Cov}(d_C',d_C'')$. A diagonal-only treatment would omit this
term entirely.

\begin{figure*}[t]
  \centering
  \begin{subfigure}[t]{0.4\textwidth}
    \includegraphics[width=\linewidth]{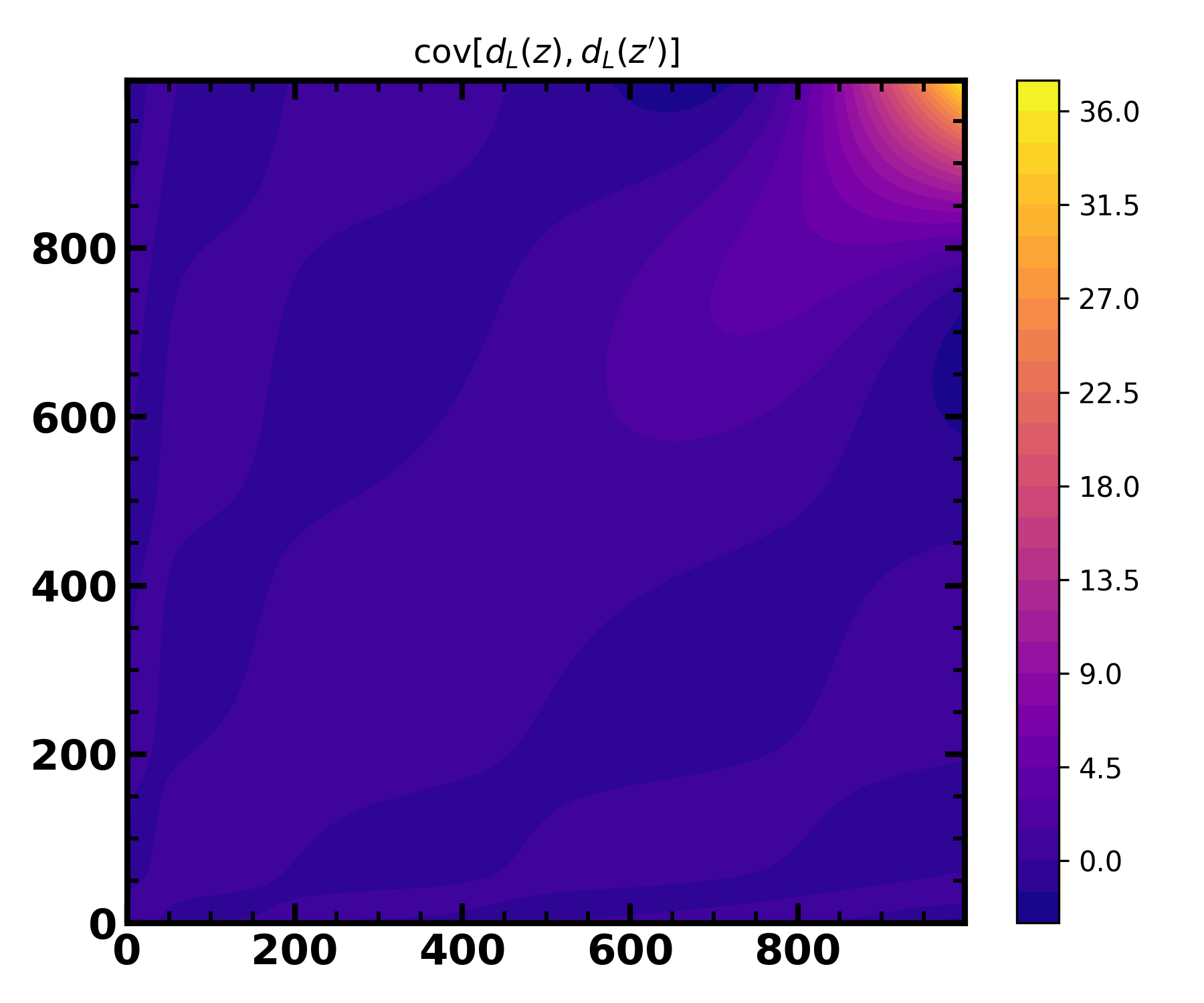}
    \caption{\it $\mathrm{Cov}[d_C(z),\,d_C(z')]$}
  \end{subfigure}
  \hfill
  \begin{subfigure}[t]{0.4\textwidth}
    \includegraphics[width=\linewidth]{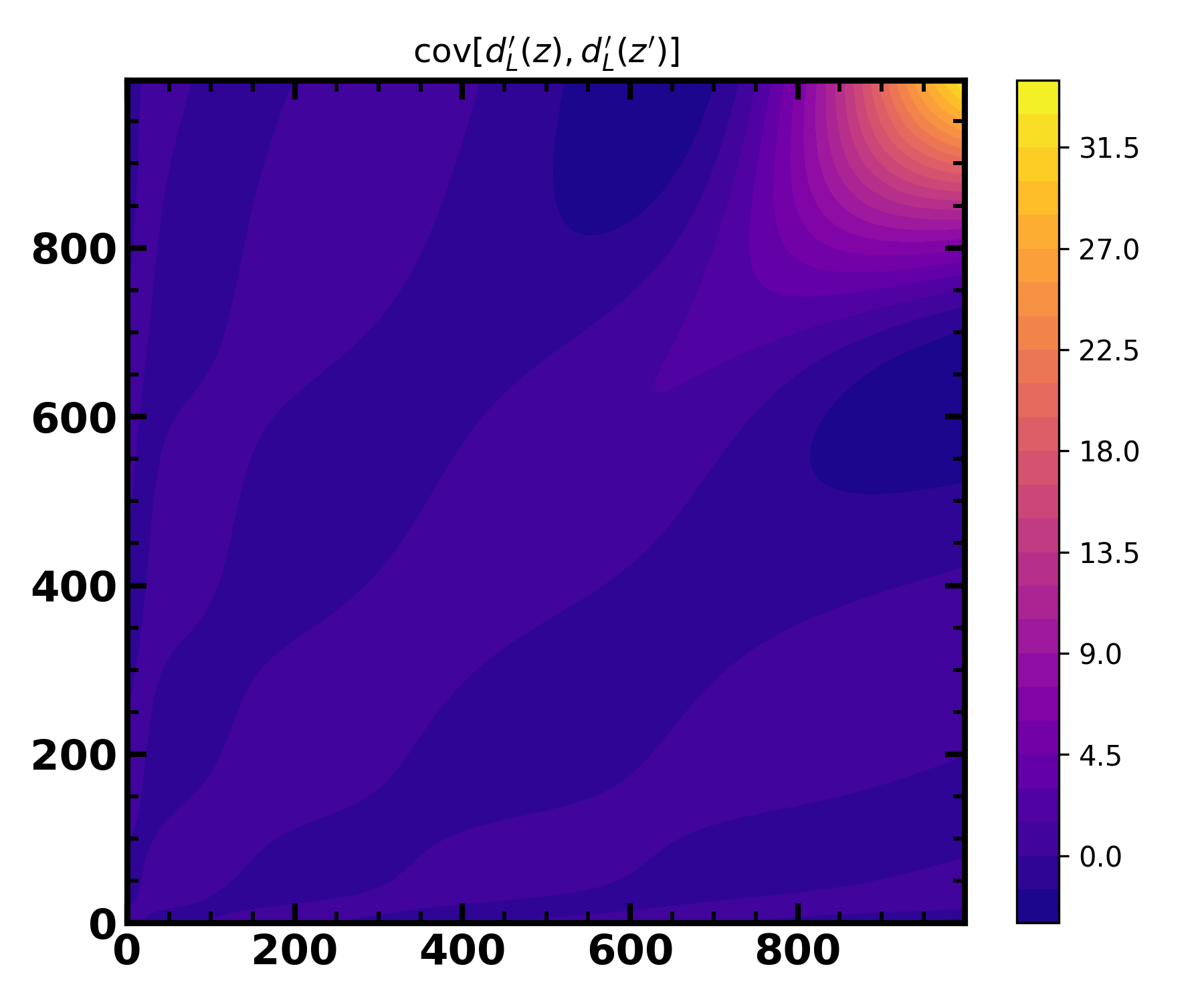}
    \caption{\it $\mathrm{Cov}[d_C'(z),\,d_C'(z')]$}
  \end{subfigure}

  \vspace{.4em}

  \begin{subfigure}[t]{0.4\textwidth}
    \includegraphics[width=\linewidth]{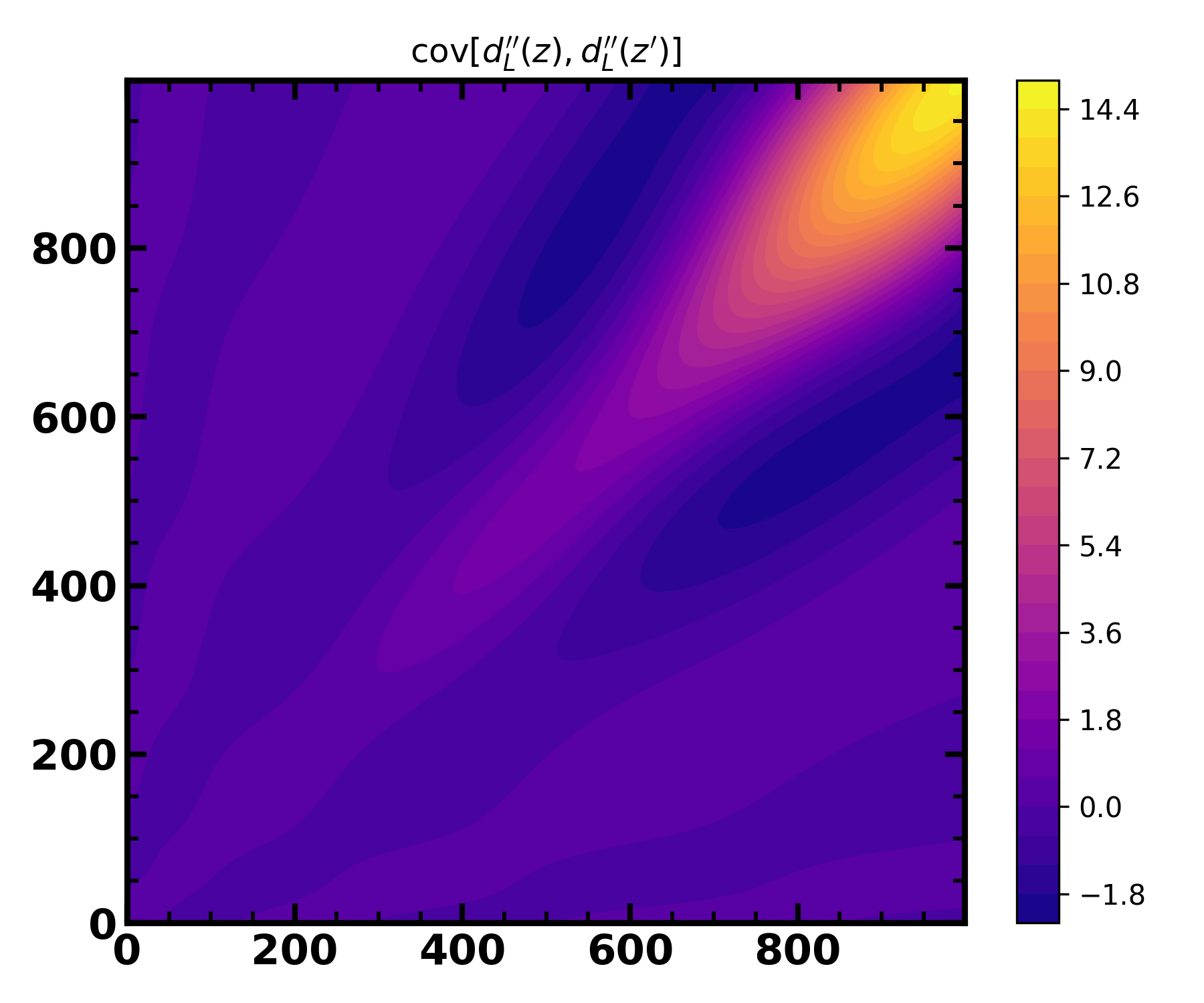}
    \caption{\it $\mathrm{Cov}[d_C''(z),\,d_C''(z')]$}
  \end{subfigure}
  \hfill
  \begin{subfigure}[t]{0.4\textwidth}
    \includegraphics[width=\linewidth]{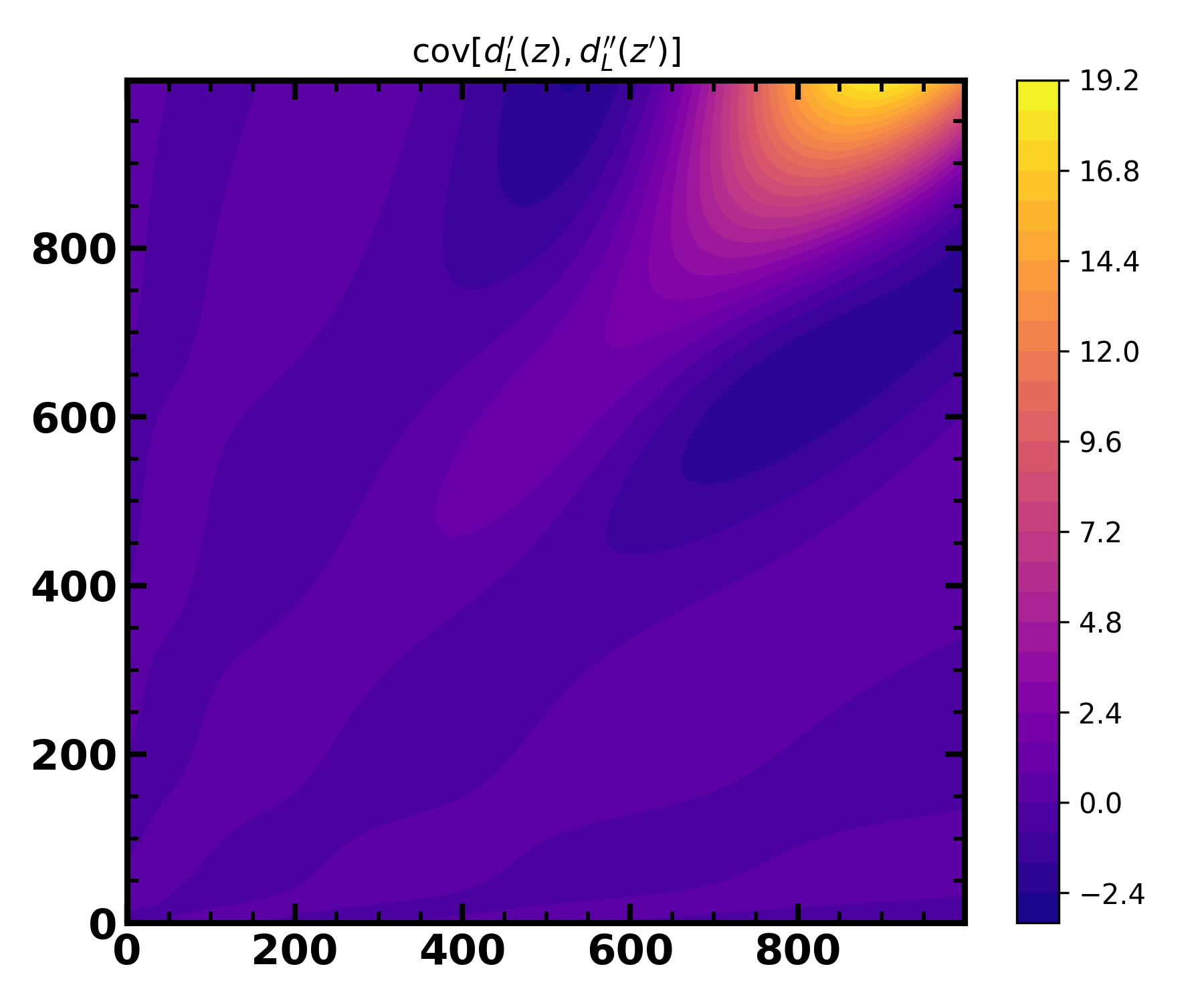}
    \caption{\it $\mathrm{Cov}[d_C'(z),\,d_C''(z')]$}
  \end{subfigure}
  \caption{\it Representative blocks of the Gaussian Process predictive
    covariance matrix for a reconstructed mock catalog.
    \textit{(a)}~$\mathrm{Cov}[d_C(z),d_C(z')]$.
    \textit{(b)}~$\mathrm{Cov}[d_C'(z),d_C'(z')]$.
    \textit{(c)}~$\mathrm{Cov}[d_C''(z),d_C''(z')]$.
    \textit{(d)}~Cross-covariance $\mathrm{Cov}[d_C'(z),d_C''(z')]$.
    The extended off-diagonal structure and increasingly oscillatory
    behavior for higher derivatives illustrate the nonlocal and strongly
    correlated nature of Gaussian Process reconstructions.}
  \label{fig:covariance_blocks}
\end{figure*}

Figure~\ref{fig:covariance_blocks} illustrates representative covariance
blocks, all evaluated at pairs of redshifts $(z,z')$:
\[
\mathrm{Cov}[d_C,d_C],\quad
\mathrm{Cov}[d_C',d_C'],\quad
\mathrm{Cov}[d_C'',d_C''],\quad
\mathrm{Cov}[d_C',d_C''].
\]
The covariance structure exhibits extended off-diagonal correlations
that reflect the nonlocal nature of Gaussian Process inference.
Derivative covariances display increasingly oscillatory behavior and
amplified amplitudes toward higher redshift, arising directly from the
repeated differentiation of the kernel function.

The auto covariance blocks are symmetric under
$z\leftrightarrow z'$, as required for any covariance between the same
quantity evaluated at two redshifts. Their largest amplitudes occur when
both arguments lie near the high redshift edge, where the reconstruction
is least anchored by the data. The cross covariance block
$\mathrm{Cov}[d_C'(z),d_C''(z')]$ is different. It need not be symmetric
by itself since exchanging the two axes also exchanges the derivative
order. Moreover, it is not expected to peak on the diagonal
$z=z'$. For a stationary even kernel such as the Mat\'ern kernel used
here, $\mathrm{Cov}[d_C'(z),d_C''(z')]$ involves an odd total derivative
of the kernel with respect to the separation $z-z'$, so the same-redshift
contribution is suppressed and the extrema occur at finite separation.
This explains the tilted positive and negative bands in panel (d): a
curvature fluctuation at one redshift changes the inferred slope most
strongly at nearby, but not identical, redshifts. The finite redshift
range, the low-redshift anchor $d_C(0)=0$, and the weaker high-redshift
data constraints then amplify and shift these lobes toward the less
constrained high-redshift region. Thus the off-diagonal maximum in the
mixed block is a feature of derivative GP inference, not a numerical
artifact. These structures also explain the rapid growth of uncertainties
observed for higher-order derivatives at large redshift and motivate our
preference for diagnostics that avoid third-order differentiation, as
discussed further in Appendix~\ref{app:third_deriv}.

\FloatBarrier

\section{Reconstructed cosmological diagnostics}
\label{sec:results}

We present the Gaussian Process reconstructions of the cosmological
diagnostic quantities defined in Sec.~\ref{sec:observables} using mock
gravitational wave standard siren catalogs. We discuss the results for
the $\Lambda$CDM fiducial model\footnote{The shorthand ``LCDM''
appearing in some figure labels and file names denotes the same
$\Lambda$CDM fiducial model.} in detail while the qualitatively similar
findings for alternative cosmologies are presented in
Appendix~\ref{app:precision}. Unless stated otherwise, the reconstruction
figures use the full ensemble of $N_{\rm mock}=50$ independent mock
realizations defined in Sec.~\ref{sec:mock_catalogs}; the shorthand
``Nmocks50'' denotes this choice. The primary goal of this section is to
examine the qualitative behavior and uncertainty properties of each
diagnostic as a function of redshift before we introduce the
quantitative distinguishability measure detailed in
Sec.~\ref{sec:hellinger}. Shaded regions throughout these plots denote
confidence intervals obtained from joint Gaussian Process covariance
propagation.

\begin{figure*}[t]
  \centering
  \includegraphics[height=.24\textheight,keepaspectratio]{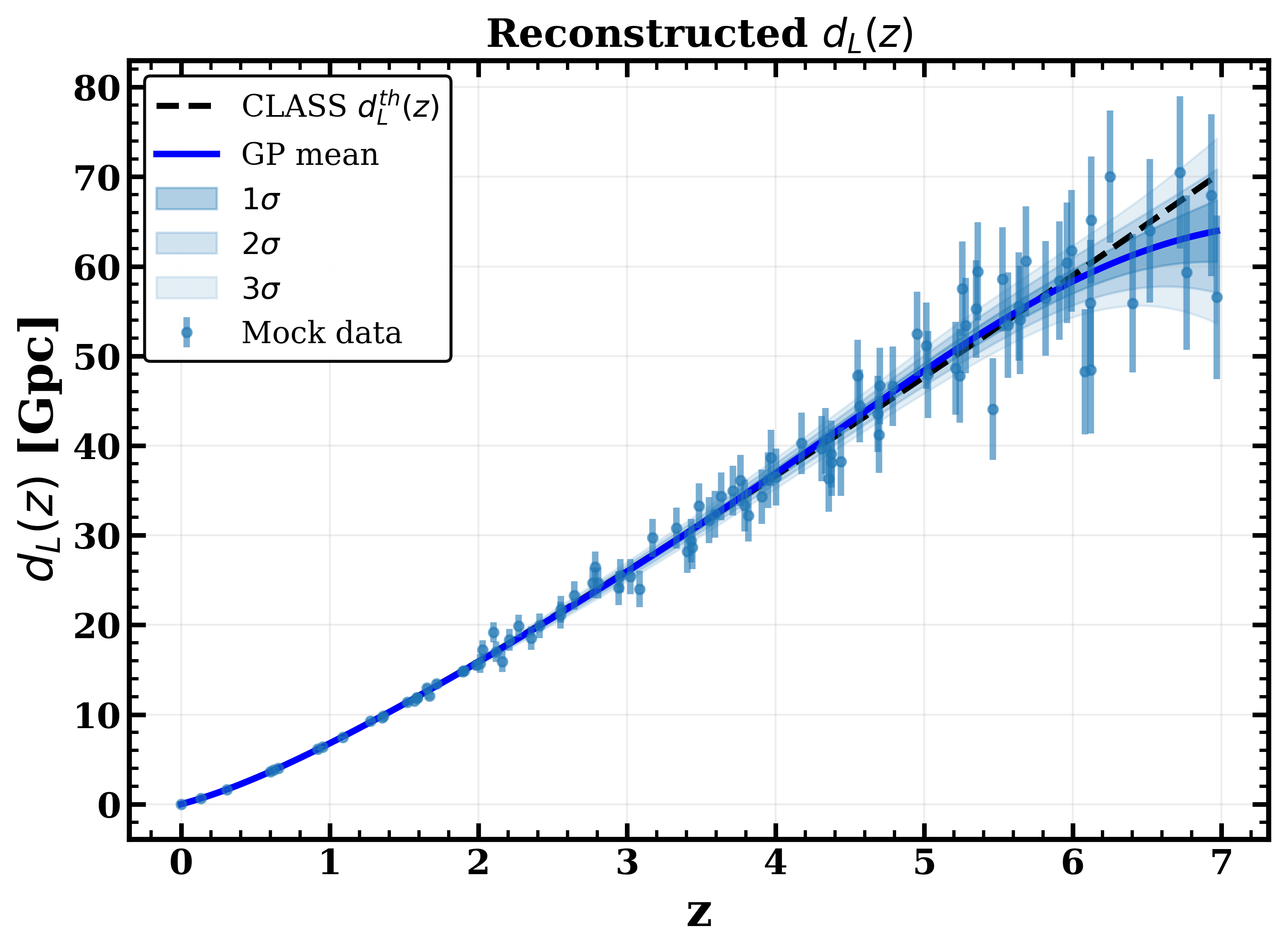}
    \includegraphics[height=.24\textheight,keepaspectratio]{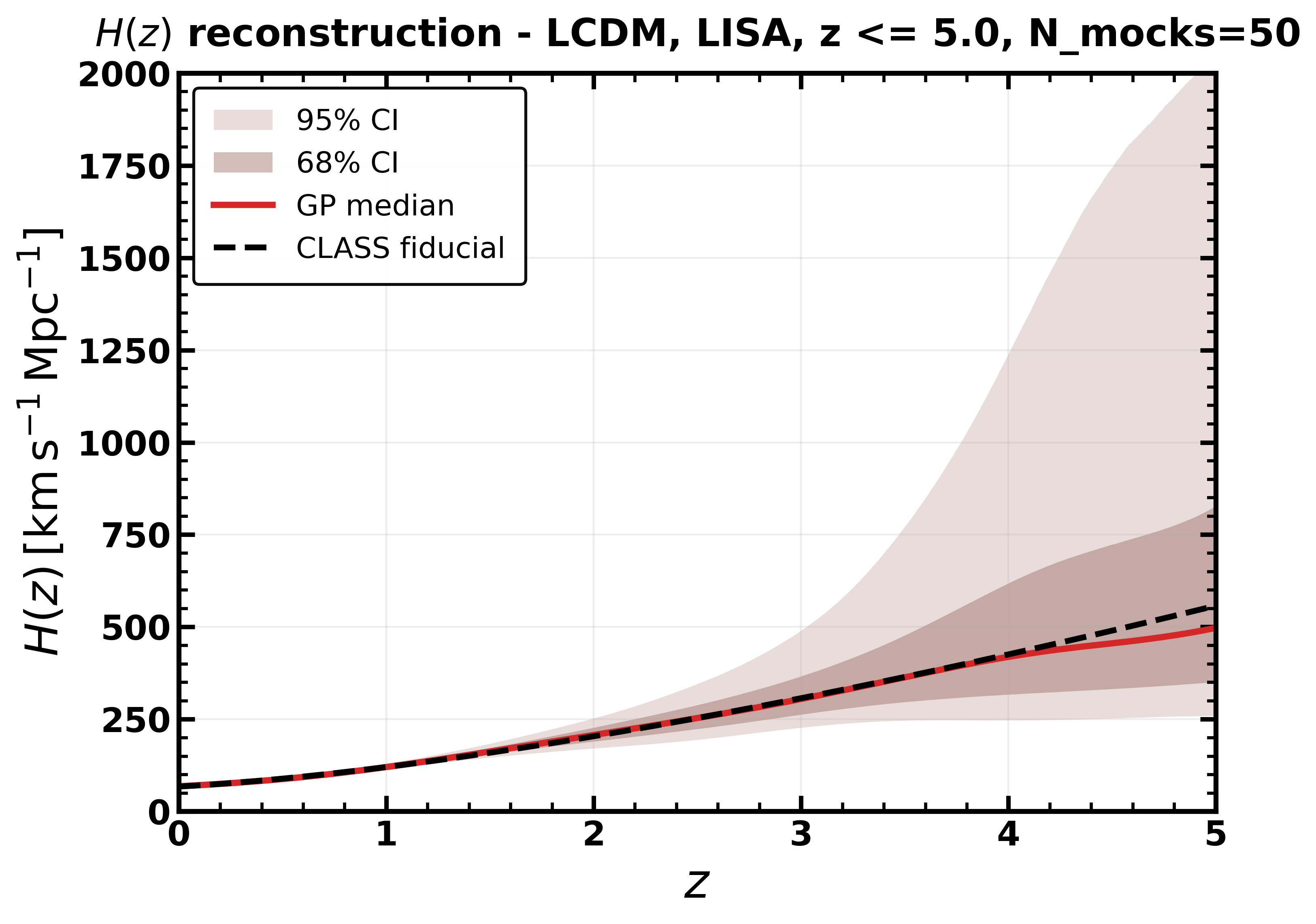}
    \includegraphics[height=.24\textheight,keepaspectratio]{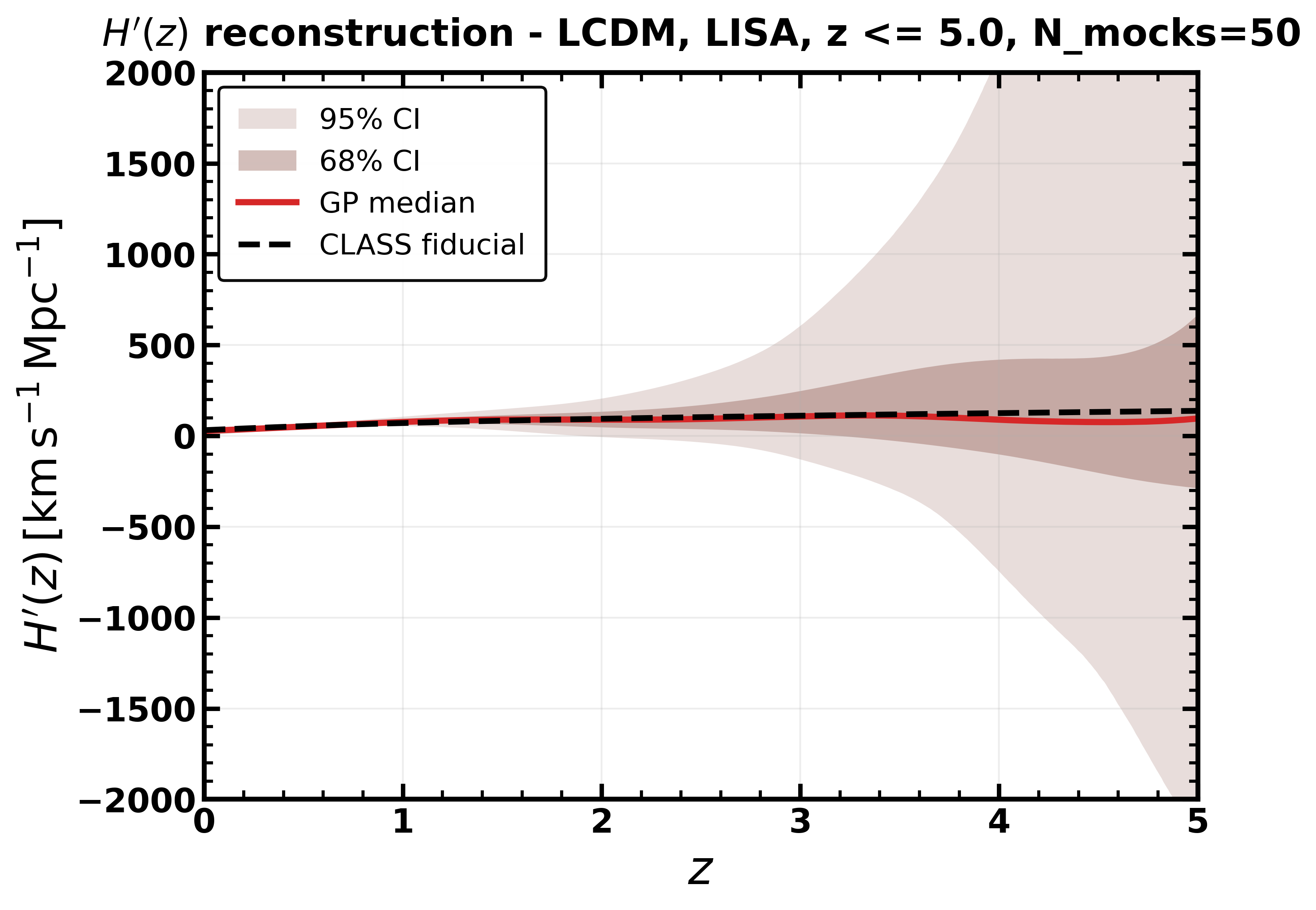}
  \caption{\it Gaussian Process reconstruction from a representative
    $\Lambda$CDM mock gravitational wave standard siren catalog. The
    upper row shows the distance reconstruction at left and the Hubble
    parameter at right. The lower panel shows the derivative of the
    Hubble parameter. \textit{Top left}: reconstructed luminosity
    distance $d_L(z)$ with the GP median shown as a solid blue line. The
    shaded bands show the $1\sigma$ through $3\sigma$ confidence
    regions. The dashed black curve is the fiducial \texttt{CLASS}
    prediction and the points are the mock data. \textit{Top right}:
    derived Hubble parameter $H(z)$ with $1\sigma$ and $2\sigma$
    confidence regions. \textit{Bottom}: $H'(z)$ with the rapid growth
    of uncertainty at high redshift.}
  \label{fig:gp_reconstruction_overview}
\end{figure*}

Figure~\ref{fig:gp_reconstruction_overview} provides an overview of the
reconstruction pipeline by illustrating the reconstructed luminosity
distance $d_L(z)$ alongside the derived Hubble parameter $H(z)$ and its
derivative $H'(z)$ for a representative $\Lambda$CDM mock catalog.
The figure shows that the GP first follows the distance data smoothly and
then transfers this information into the reconstructed expansion rate.
The median curves track the fiducial \texttt{CLASS} prediction well in
the region where the mock events provide support. The main loss of
precision appears in $H'(z)$, where the confidence band opens rapidly
after $z\sim 3$. This identifies derivative amplification as the first
major limitation of the pipeline.

\FloatBarrier

\subsection{First-order diagnostics}
\label{sec:first_order}

We begin our analysis with the Hubble parameter $H(z)$ which encodes
the overall expansion history of the Universe and is derived directly
from the first derivative of the reconstructed comoving distance. We
revisit $H(z)$ here specifically in the context of model comparison
even though it was already shown in Fig.~\ref{fig:gp_reconstruction_overview}
as a pipeline validation.

\begin{figure}[ht]
  \centering
  \includegraphics[width=.95\columnwidth]{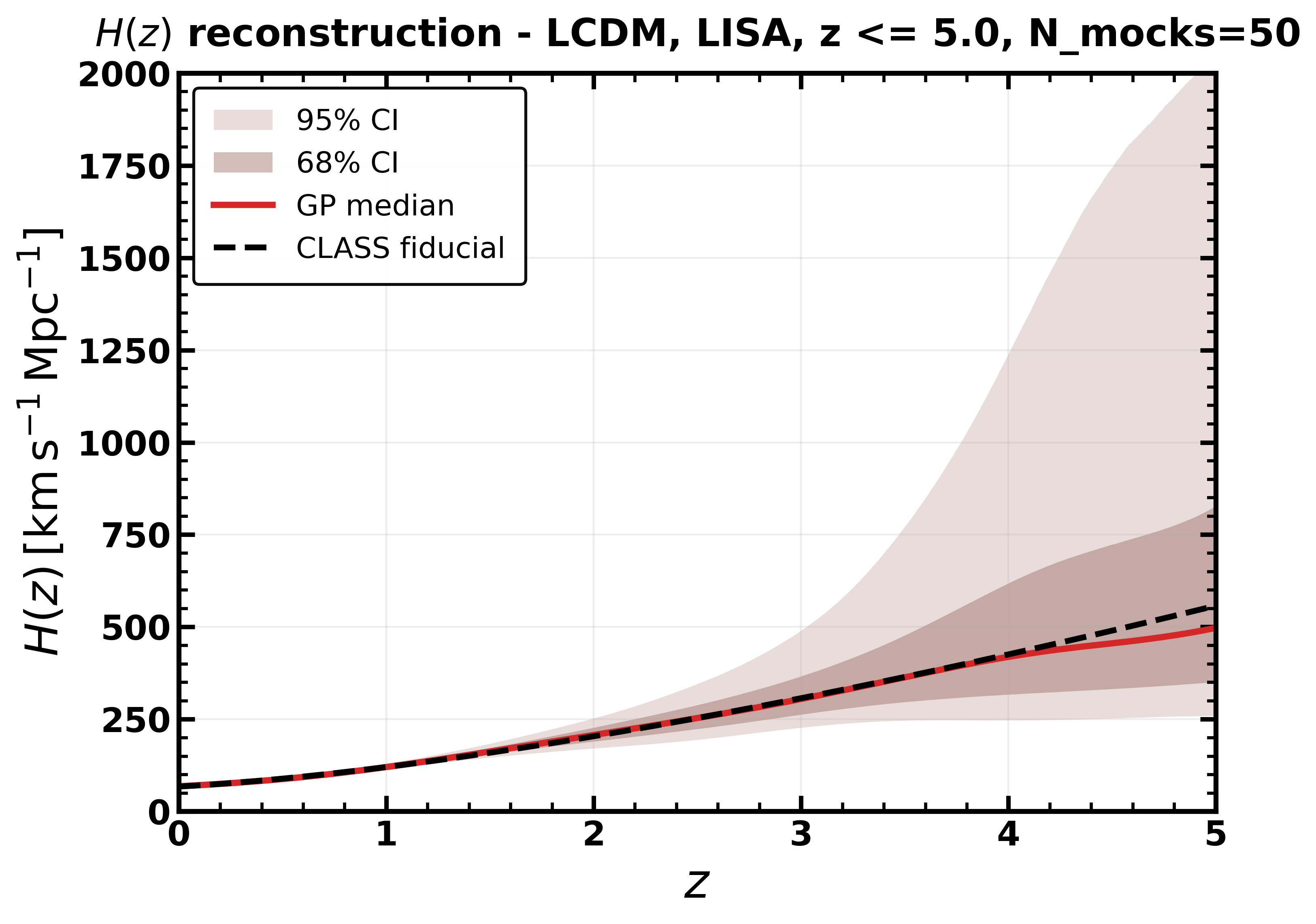}
  \caption{\it Gaussian Process reconstruction of the Hubble parameter
    $H(z)$. The solid line shows the reconstructed median while shaded
    regions correspond to $1\sigma$ and $2\sigma$ confidence intervals
    from joint covariance propagation. The dashed curve represents the
    fiducial \texttt{CLASS} evolution.}
  \label{fig:H_recon}
\end{figure}

Figure~\ref{fig:H_recon} displays the reconstructed $H(z)$ together
with $1\sigma$ and $2\sigma$ confidence regions compared directly to the
fiducial background evolution. The reconstructed median closely follows
the fiducial model over a broad redshift range to demonstrate that the
Gaussian Process framework can successfully recover the underlying
expansion history in a model-independent manner. These uncertainties naturally increase with redshift, reflecting both the sparsity of high-redshift standard siren events and the cumulative propagation of distance errors throughout the reconstruction pipeline.

The Hubble parameter provides a robust and smooth characterization of
the expansion history but its reconstructed confidence regions overlap
substantially across the different fiducial cosmological models
considered here at intermediate and high redshift. An important
caveat applies at low redshift $z\to 0$: the six fiducial cosmologies
adopt different values of $H_0$ (ranging from $67.32$ to
$72.81\;\mathrm{km\,s^{-1}\,Mpc^{-1}}$, see
Tables~\ref{tab:lcdm_cpl_fiducial}--\ref{tab:axion_fiducial}) and
since the GP reconstruction is highly precise at low redshift it
cleanly resolves these offsets. Any apparent model separation seen in
$H(z)$ near $z\sim 0$ therefore reflects the differing input $H_0$
values rather than genuinely distinct expansion dynamics and should
not be interpreted as discriminatory power over the underlying
cosmological models. This point is quantified explicitly via the
pointwise marginal Hellinger distance analysis in
Sec.~\ref{sec:hellinger_results}.
At intermediate and high redshift, where the $H_0$ offset becomes
subdominant, the reconstructed confidence regions overlap substantially
across all fiducial models, demonstrating that first-order diagnostics
lack the discriminatory power to distinguish between the cosmological
scenarios examined in this work.

\FloatBarrier

\subsection{Second-order diagnostics}
\label{sec:second_order}

We consider diagnostic quantities that depend on the second derivative
of the reconstructed comoving distance to probe more subtle features of
the late-time expansion history. These quantities include the
deceleration parameter $q(z)$ alongside the Om diagnostic
$\mathcal{O}_m(z)$ and the total effective equation of state
$w_{\rm tot}(z)$.

\begin{figure*}[t]
  \centering
  \includegraphics[height=.24\textheight,keepaspectratio]{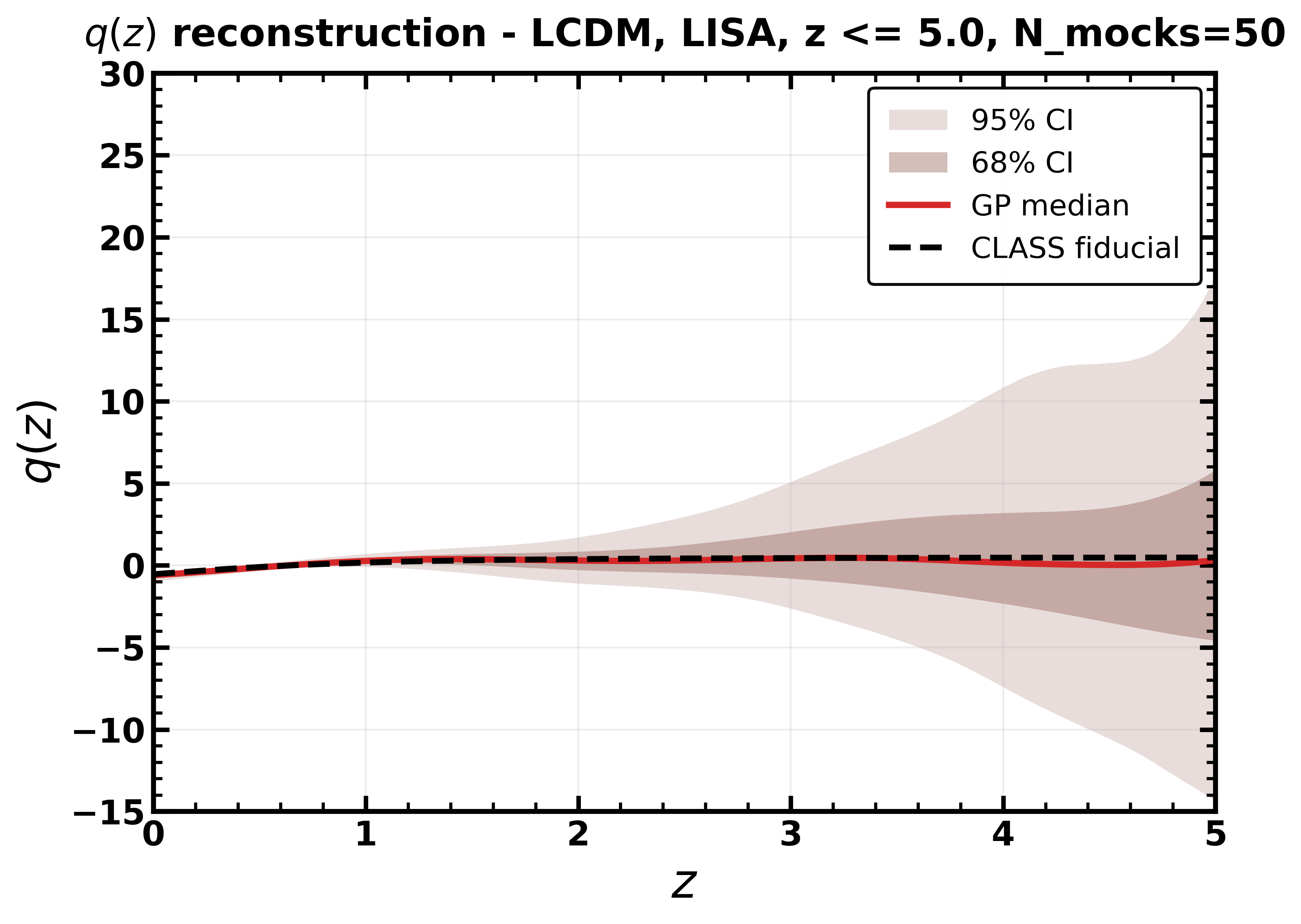}
  \includegraphics[height=.24\textheight,keepaspectratio]{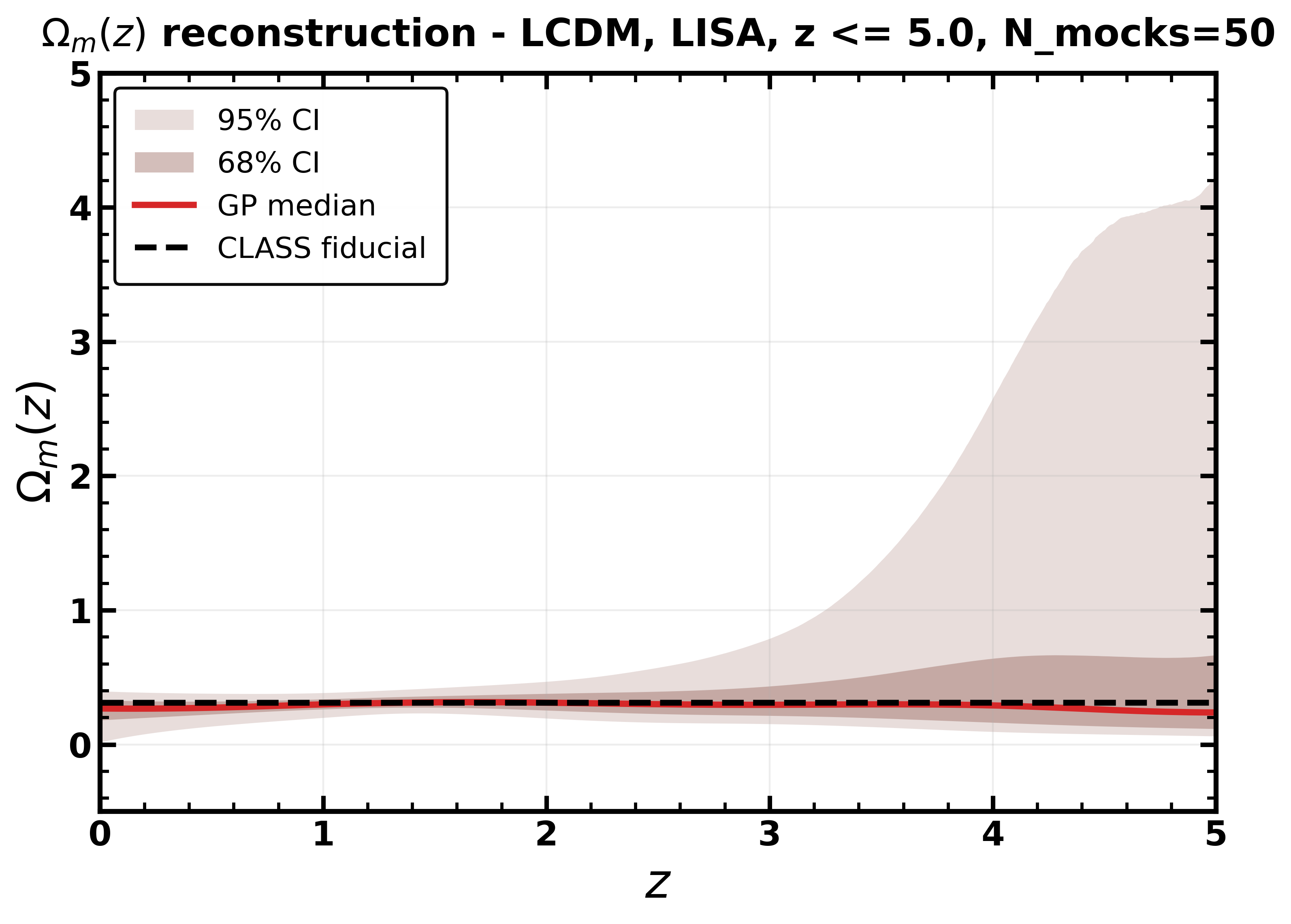}%
  \includegraphics[height=.24\textheight,keepaspectratio]{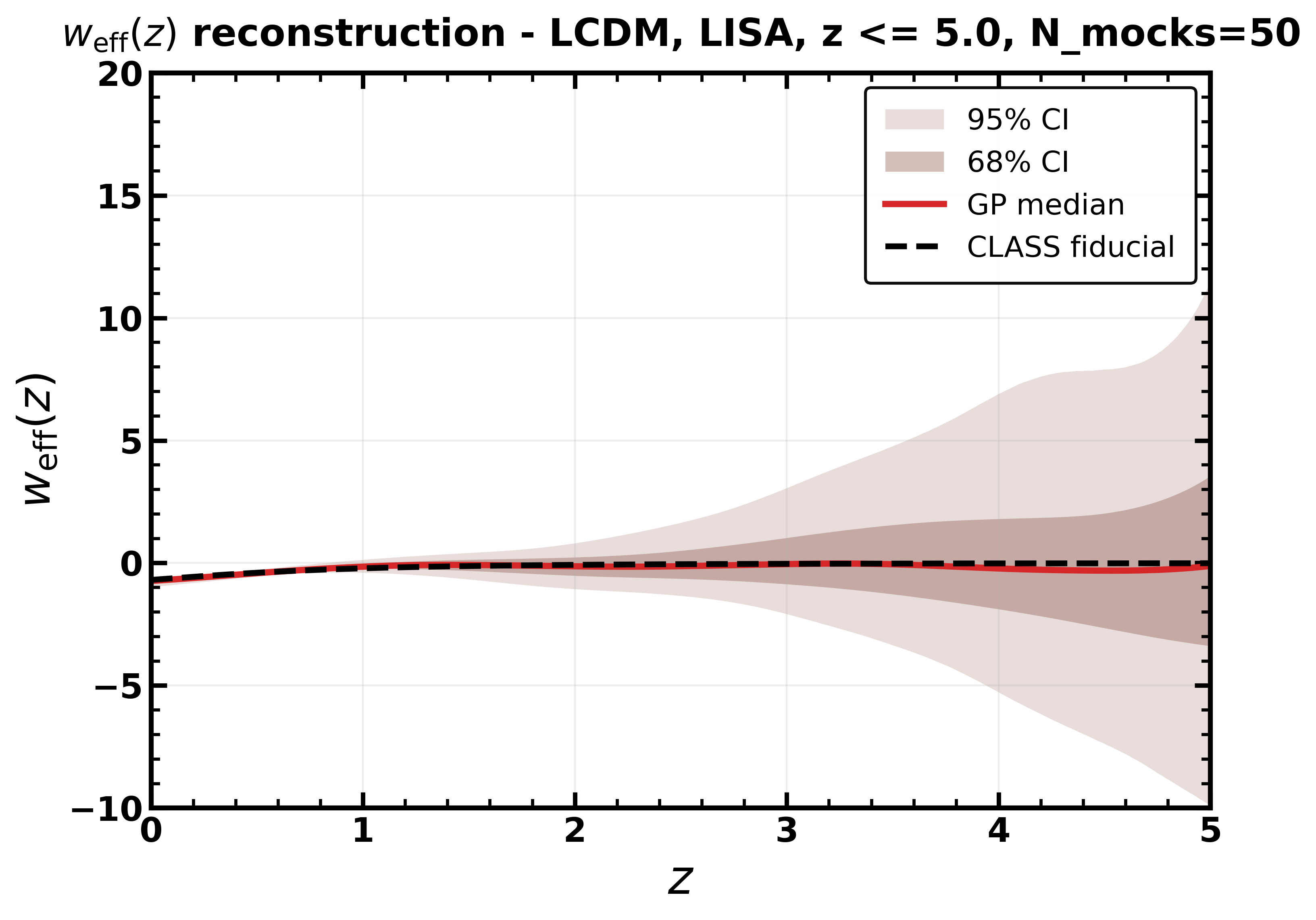}
  \caption{\it Gaussian Process reconstruction of second order cosmological
    diagnostics. The upper row shows the deceleration parameter at left
    and the Om diagnostic at right. The lower panel shows the total
    effective equation of state. \textit{Top left}: deceleration
    parameter $q(z)$. \textit{Top right}: Om diagnostic
    $\mathcal{O}_m(z)$. \textit{Bottom}: total effective equation of
    state $w_{\rm tot}(z)$. Shaded regions show the $1\sigma$ and
    $2\sigma$ confidence intervals from joint covariance propagation.}
  \label{fig:q_Om_wtot_recon}
\end{figure*}

Figure~\ref{fig:q_Om_wtot_recon} presents the detailed reconstructions
of these three specific quantities. These diagnostics exhibit enhanced
sensitivity to departures from the concordance $\Lambda$CDM model when
compared directly to $H(z)$. Deviations from a constant
$\mathcal{O}_m(z)$, or from the fiducial total equation-of-state
history $w_{\rm tot}(z)$, serve as clear signatures of nonstandard
expansion dynamics.
The panels show that the nonlinear transformations do not introduce a
visible bias in the median reconstruction. The $q(z)$ curve captures the
transition from accelerated to decelerated expansion. The
$w_{\rm tot}(z)$ curve shows the same evolution in equation of state
form and moves toward the matter dominated value near zero at larger
redshift. The $\mathcal{O}_m(z)$ panel remains close to the expected
constant value for $\Lambda$CDM over the best constrained part of the
range. At high redshift all three panels become dominated by wide
confidence regions, so median level differences are not sufficient for
model separation.

The uncertainties associated with these quantities grow
rapidly with redshift due to the inherent amplification of errors
through higher-order derivatives. The reconstructed median curves
display nontrivial redshift evolution but the corresponding confidence
regions remain broad and overlapping across all fiducial models which
strongly motivates the rigorous quantitative analysis presented in Sec.~\ref{sec:hellinger}.

\FloatBarrier

\subsection{Ratio-based diagnostic $\kappa(z)$}
\label{sec:kappa}

Higher-order kinematical diagnostics like the cosmological jerk
parameter involve third derivatives of the reconstructed comoving
distance and are therefore will be dominated by large uncertainties within a
Gaussian Process framework. We therefore use the following
ratio-based diagnostic as a numerically stable consistency check on the
second-order reconstruction:
\begin{equation}
  \kappa(z) \equiv \frac{E'(z)}{E(z)}
  = -\frac{d_C''(z)}{d_C'(z)}.
  \label{eq:kappa}
\end{equation}
This quantity is directly related to the deceleration parameter through
\begin{equation}
  \kappa(z) = \frac{1+q(z)}{1+z},
  \label{eq:kappa_q_relation}
\end{equation}
and equivalently to the total effective equation of state through
\begin{equation}
  \kappa(z) = \frac{3}{2}\,\frac{1+w_{\rm tot}(z)}{1+z}.
\end{equation}
These identities follow from Eqs.~(\ref{eq:qz}) and (\ref{eq:wtot})
together with the relation $H'/H = E'/E$. Geometrically $\kappa(z)$
measures the logarithmic slope of $E(z)$ so that
$\kappa(z)=d\ln E/dz$.
At fixed redshift it is algebraically equivalent to $q(z)$ and
$w_{\rm tot}(z)$ and therefore does not add independent background
information.

In the $\Lambda$CDM model during matter domination
$E(z)\propto(1+z)^{3/2}$ and therefore
$\kappa(z)=3/[2(1+z)]$. As dark energy begins to dominate $E(z)$ varies
more slowly and $\kappa(z)$ moves toward zero. Its practical value is
numerical rather than informational. Because $\kappa(z)$ is a ratio of
$d_C''(z)$ to $d_C'(z)$ it remains better conditioned than higher-order
kinematical diagnostics that require third-order differentiation
although it still depends on the second derivative of the reconstructed
comoving distance.

\begin{figure}[ht]
  \centering
  \includegraphics[width=.95\columnwidth]{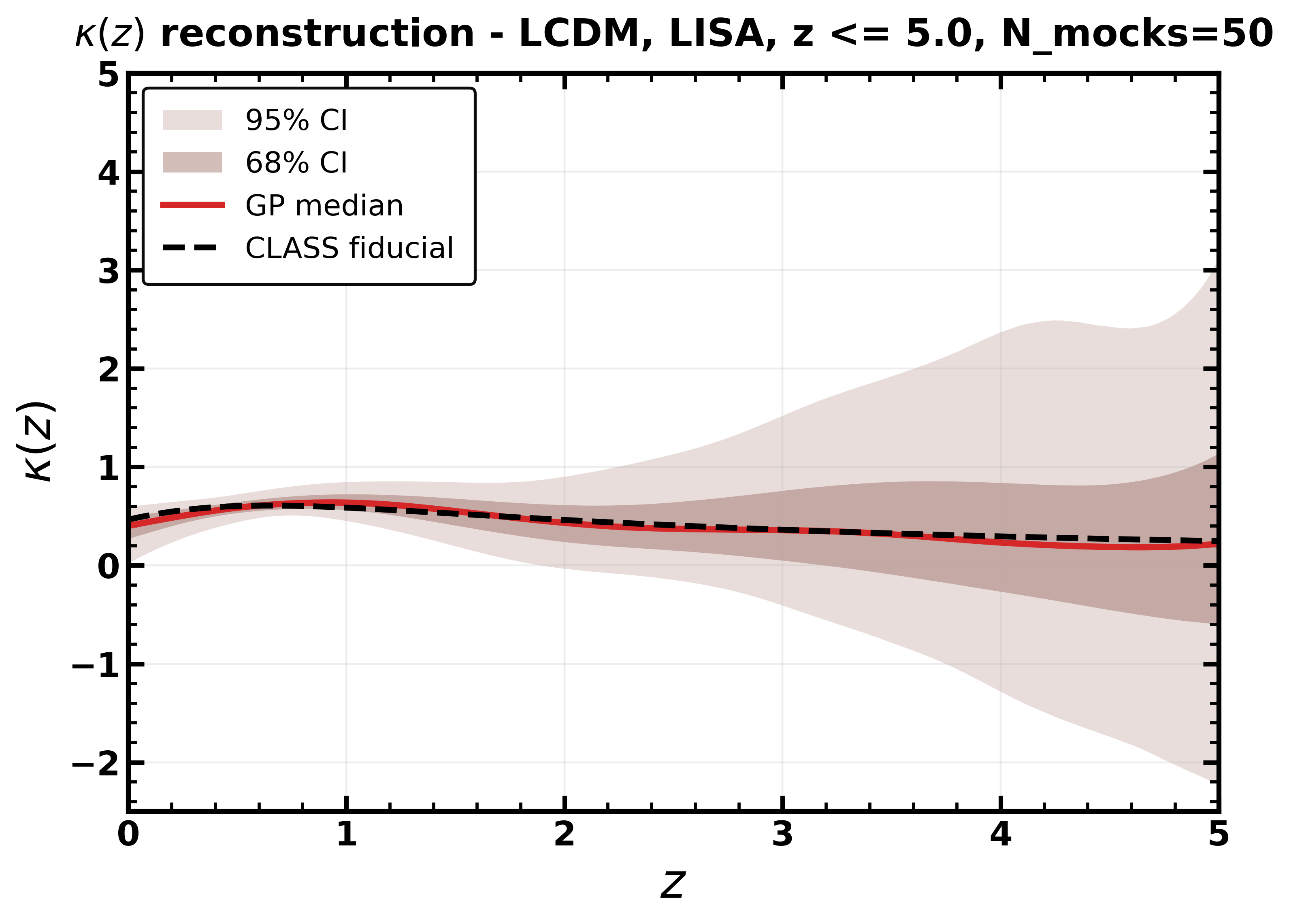}
  \caption{\it Gaussian Process reconstruction of the ratio-based diagnostic
    $\kappa(z)=E'(z)/E(z)$. The solid line denotes the reconstructed
    median and shaded regions indicate $1\sigma$ and $2\sigma$ confidence
    intervals.}
  \label{fig:kappa_recon}
\end{figure}

Figure~\ref{fig:kappa_recon} shows the detailed reconstruction of
$\kappa(z)$. The reconstructed profile follows the expected
$\Lambda$CDM behavior at low redshift and the confidence regions
remain substantially narrower than those of $H'(z)$ across the full
redshift range, confirming the numerical advantages discussed above.
The plot shows that $\kappa(z)$ remains positive and slowly decreases
after the low redshift maximum, as expected for the logarithmic slope of
$E(z)$ in $\Lambda$CDM. Its uncertainty grows more gently than the
uncertainty in $H'(z)$ because the ratio cancels part of the common
amplitude variation. At high redshift the interval is still broad, so
$\kappa(z)$ is most useful as a stable consistency check on the
second order reconstruction.

\FloatBarrier

\subsection{Redshift-dependent reconstruction precision}
\label{sec:precision}

We evaluate the redshift-dependent relative precision to appropriately
quantify the statistical stability of the reconstructed diagnostics:
\begin{equation}
  \mathcal{P}(z) \equiv \frac{\sigma(z)}{|\mathrm{median}(z)|},
  \label{eq:rel_precision}
\end{equation}
where $\sigma(z)$ denotes the $1\sigma$ uncertainty obtained from joint
covariance propagation with hyperparameter marginalization. We
compute the precision curves using the complete ensemble of mock
realizations so the reported uncertainties accurately reflect both
reconstruction variance and mock-to-mock fluctuations. We report the
absolute uncertainty $\sigma(z)$ in place of the ratio for the
zero crossing diagnostics $q(z)$, $w_{\rm tot}(z)$, and $\kappa(z)$
where the median may pass through zero.

\begin{figure*}[t]
  \centering
  \includegraphics[width=\linewidth]{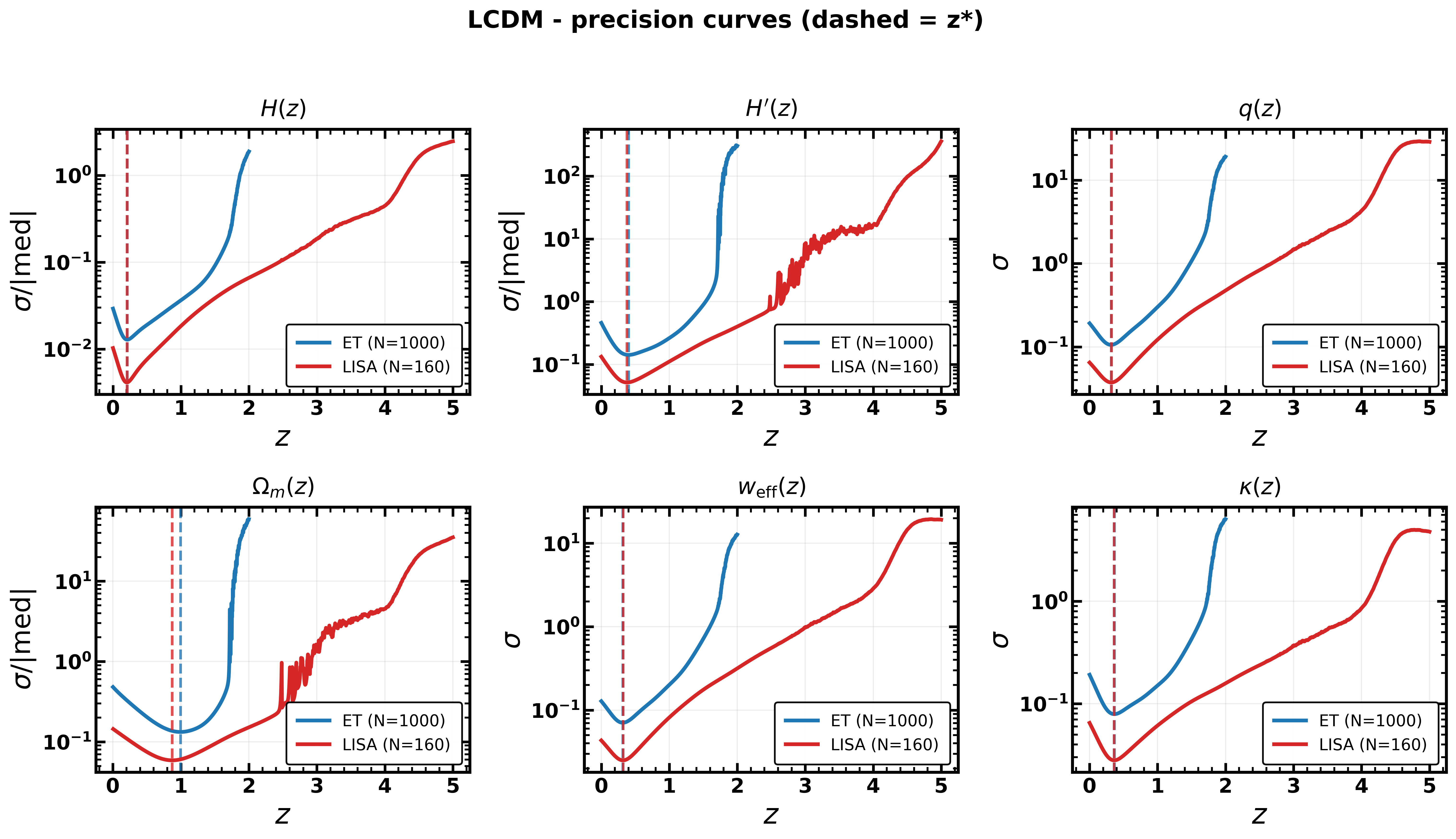}
  \caption{\it Redshift-dependent precision curves for all reconstructed
    cosmological diagnostics in the $\Lambda$CDM fiducial model showing
    both the ET (blue, $N_{\rm ev}=1000$) and LISA (red, $N_{\rm ev}=160$)
    configurations overlaid. For $H$, $H'$ and $\mathcal{O}_m$ the ordinate
    is the relative precision $\sigma/|\mathrm{median}|$; for $q$,
    $w_{\rm tot}$ and $\kappa$ the absolute uncertainty $\sigma$ is shown
    to avoid divergences near zero crossings. Vertical dashed lines mark
    the optimal redshift $z^*$ (argmin of the precision metric) for each
    detector. Uncertainties correspond to $1\sigma$ confidence regions
    from joint covariance propagation with hyperparameter marginalization.}
  \label{fig:relative_precision}
\end{figure*}

Figure~\ref{fig:relative_precision} displays the precision curves for
the $\Lambda$CDM fiducial cosmology for both detector configurations
overlaid. The baseline LISA catalogs described in
Sec.~\ref{sec:lisa_mocks} use $N_{\rm ev}=80$. Here we evaluate the
curves at the maximum event counts explored in the scaling sweep
corresponding to $N_{\rm ev}=1000$ for ET and
$N_{\rm ev}=160$ for LISA. We find qualitatively similar behavior across
all models and provide additional examples in Appendix~\ref{app:precision}.

\FloatBarrier

Several important trends emerge from this evaluation. The Hubble
parameter $H(z)$ remains the most statistically stable diagnostic
across the full redshift range for both detectors. It achieves
sub-percent precision at low redshift and degrades gradually toward
higher $z$ which remains consistent with its strict dependence on the
first derivative of $d_C(z)$ alone.

The derivative $H'(z)$ exhibits a contrasting rapid growth in relative
uncertainty beyond $z\sim 2$ for LISA and beyond $z\sim 1$ for ET
before eventually exceeding unity at high redshift. This serves as a
direct consequence of the derivative amplification inherent to Gaussian
Process differentiation as we discussed in
Sec.~\ref{sec:covariance_structure}.

Second-order diagnostics such as $q(z)$, $\mathcal{O}_m(z)$ and
$w_{\rm tot}(z)$ display intermediate behavior. They are statistically
informative at low to moderate redshift but become increasingly
noise-dominated toward the high $z$ boundary of the reconstruction
window.

The ratio-based diagnostic $\kappa(z)$ demonstrates improved stability
relative to $H'(z)$ confirming that avoiding explicit third-order
differentiation successfully mitigates uncertainty amplification. These
results clearly highlight that while reconstructed diagnostics may
exhibit visible deviations between cosmological models the confidence
regions remain broad and overlapping. This directly motivates the
quantitative approach presented in the next section.

\subsubsection{Optimal reconstruction redshift and detector comparison}
\label{sec:zstar}

We identify the optimal redshift $z^*$ for each diagnostic as the
location of the minimum precision metric within the detector redshift
window:
\begin{equation}
  z^* = \underset{z\in[z_{\min},z_{\max}]}{\arg\min}\;\mathcal{P}(z).
  \label{eq:zstar}
\end{equation}
In words, we evaluate the precision curve $\mathcal{P}(z)$ over the
allowed detector redshift range and choose the redshift where this curve
is smallest. The notation $\arg\min$ returns the location of that
minimum, not the minimum value itself. Thus $z^*$ identifies the precise
redshift where the GP reconstruction achieves its tightest constraint on
a given observable. It therefore serves as a useful figure of merit for
comparing different detector configurations.

Table~\ref{tab:zstar_lcdm} summarizes $z^*$ and the corresponding
precision metric value for all diagnostics and both detector
configurations assuming the $\Lambda$CDM fiducial cosmology. LISA
achieves a lower and thus better precision metric than ET across all
diagnostics. The LISA values are smaller by roughly a factor of two to
three, more precisely about $2.3$--$3.1$ for the $\Lambda$CDM case,
depending on the specific diagnostic. This ordering is a direct
consequence of the distinct noise prescriptions for the two detectors.
The LISA instrumental noise
$\sigma_{\rm inst}=0.05\,(d_L/d_L^{\rm ref})\,d_L$
(Eq.~\ref{eq:lisa_inst}) is quadratically suppressed relative to the
calibration scale $d_L^{\rm ref}=36.6\;\mathrm{Gpc}$. Both detectors
probe redshifts $z\lesssim 2$ where the luminosity distances remain
below this reference scale. The LISA
per-event fractional error remains far smaller than the ET value encoded
by $\mathcal{F}(z)$ in this regime even though ET observes significantly
more events. We must keep this asymmetry in mind when interpreting
comparative precision statements. These results directly reflect the
performance achievable under the specific mock error models adopted here.
The relative ranking of the two detectors remains entirely contingent on
those precise noise prescriptions.

\begin{table}[ht]
\caption{\it Optimal reconstruction redshift $z^*$ and precision metric
  value at $z^*$ for all cosmological diagnostics comparing the
  ET ($N_{\rm ev}=1000$) and LISA ($N_{\rm ev}=160$) configurations
  for the $\Lambda$CDM fiducial cosmology. For $H$, $H'$ and
  $\mathcal{O}_m$ the metric is $\sigma/|\mathrm{median}|$; for $q$,
  $w_{\rm tot}$ and $\kappa$ it is the absolute $\sigma$.}
\label{tab:zstar_lcdm}
\centering
\small
\begin{tabular}{llcccc}
\toprule
Diagnostic & Metric
  & $z^*_{\rm ET}$ & $\mathcal{P}(z^*_{\rm ET})$
  & $z^*_{\rm LISA}$ & $\mathcal{P}(z^*_{\rm LISA})$ \\
\midrule
$H$           & $\sigma/|\mathrm{med}|$ & $0.210$ & $0.01287$ & $0.205$ & $0.00414$ \\
$H'$          & $\sigma/|\mathrm{med}|$ & $0.400$ & $0.14201$ & $0.375$ & $0.05172$ \\
$q$           & $\sigma$               & $0.316$ & $0.10599$ & $0.320$ & $0.03753$ \\
$\mathcal{O}_m$    & $\sigma/|\mathrm{med}|$ & $0.993$ & $0.13275$ & $0.871$ & $0.05887$ \\
$w_{\rm tot}$ & $\sigma$               & $0.316$ & $0.07066$ & $0.320$ & $0.02502$ \\
$\kappa$      & $\sigma$               & $0.364$ & $0.07916$ & $0.360$ & $0.02799$ \\
\bottomrule
\end{tabular}
\end{table}

A striking feature is visible in both Table~\ref{tab:zstar_lcdm} and
Fig.~\ref{fig:relative_precision}. The optimal reconstruction redshifts
for ET and LISA are nearly coincident for most diagnostics. The ET
minimum is sharply defined and shifts to a significantly lower redshift when we 
compare to the flat $25\%$ noise approximation when using the
redshift-dependent ET noise model of Eq.~(\ref{eq:ET_inst}). The dashed
vertical lines marking $z^*_{\rm ET}$ and $z^*_{\rm LISA}$ essentially
overlap in Fig.~\ref{fig:relative_precision} for $H(z)$, $q(z)$ and
$w_{\rm tot}(z)$. The primary exception is $\mathcal{O}_m$ where both
detectors find their minimum pushed to $z^*\sim 0.9$ to $1.0$.
This shift occurs due to the vanishing of the denominator $(1+z)^3-1$
as $z\to 0$. Another exception is $H'(z)$ where the rapid growth of
derivative uncertainties places the ET minimum at a somewhat higher
redshift than LISA.

This near coincidence of $z^*$ between two detectors with fundamentally
different noise prescriptions, event rates and redshift windows is a
nontrivial result. It does not imply similar astrophysical redshift
populations for BNS and MBHB sources. The two distributions are visibly
different in Fig.~\ref{fig:redshift_distributions}: ET peaks at
$z\sim 0.5$--$1.0$ while the LISA proxy extends to much higher redshift.
The key point is that $z^*$ is controlled mainly by the low-redshift
boundary of the catalog. For LISA the reconstruction is additionally
stabilized at the origin by imposing the boundary condition
$d_C(0)=0$. This auxiliary anchor is not treated as an observed bright
siren event and it does not by itself determine $z^*$. The minimum
appears slightly above the boundary once the first few nonzero redshift
events allow the GP to move from extrapolation to interpolation and
thereby fix the effective local correlation scale. ET behaves in the
same way. Its catalog starts
at $z_{\min}=0.07$ while the first LISA events lie near
$z\sim 0.1$--$0.2$. Because these low-redshift anchoring regions are
similar, the effective local correlation scale inferred by the GP is
similar and $z^*$ is close for the two detectors. Changes in noise
amplitude shift the vertical level of the precision curve; in these mock
catalogues they do not strongly shift the location of its minimum. We
show in Appendix~\ref{app:precision} that the same behavior persists
across all six fiducial cosmological models.
Within the detector setups and mock catalogs studied here this makes
$z^*$ a robust and largely model-insensitive figure of merit for
gravitational wave standard siren reconstructions.

A further distinction is worth drawing between two separate aspects of
$z^*$ stability: its near coincidence between ET and LISA and its
stability across cosmological models within each detector.

For LISA the stability of $z^*$ across all six cosmological models is
largely a consequence of the mock catalog construction. The 80 LISA
events are drawn from a fixed interpolated Beta distribution
constructed from the data presented in Fig.~1 of
Tamanini et al.~\citep{Tamanini2016} following
Mukherjee et al.~\citep{MukherjeeShah2024}. As a result the
lower boundary of the LISA event distribution is
essentially identical across all six models. Since $z^*$ is anchored
near this lower boundary it converges to nearly the same value for all
cosmologies. The stability of $z^*_{\rm LISA}$ across cosmologies is
therefore expected by construction.

For ET the situation is fundamentally different and the stability is
genuinely nontrivial. The 1000 ET events are drawn from the redshift
probability density $p(z)\propto 4\pi d_C^2(z)R(z)/[H(z)(1+z)]$
(Eq.~\ref{eq:ET_pz}) which is evaluated self-consistently on the
fiducial background of each cosmological model. The event redshift
distribution is therefore genuinely distinct for each cosmology:
$\Lambda$CDM, AXI\_CLASS and IntDM each produce different $p(z)$
through their different $H(z)$ and $d_C(z)$. Despite this the range
of $z^*_{\rm ET}$ across the six cosmologies is only $0.060$--$0.084$
for $H$, $q$, $w_{\rm tot}$ and $\kappa$ (Table~\ref{tab:zstar_all}).
This tight stability is an earned emergent result. The ET lower cutoff
$z_{\min}=0.07$ is set by the detector sensitivity floor and is
identical across all cosmological models by construction. Since $z^*$
is anchored near $z_{\min}$ rather than near the bulk of $p(z)$, the
cosmology-to-cosmology differences in the event distribution at higher
redshift are subdominant for the location of $z^*$.

Taken together these results reveal a useful separation between the
scale of the reconstruction uncertainty and the location of its minimum.
The signal variance $\sigma_f$ and the detector noise prescription mainly
set the \emph{absolute value} of the precision metric at $z^*$---this is
why LISA achieves a roughly $3\times$ better precision metric than ET
across $H$, $q$, $w_{\rm tot}$ and $\kappa$. The lengthscale $\ell$ and
the low-redshift boundary of the catalogue shape the precision curve near
the transition from extrapolation to interpolation and therefore help set
the \emph{location} of $z^*$. The two effects are not mathematically
independent, but in the mock catalogues considered here varying the noise
level changes how precise the reconstruction is \emph{at} $z^*$ much
more than it changes the location of $z^*$.

\section{Pointwise marginal Hellinger distance analysis}
\label{sec:hellinger}

The reconstructed diagnostics presented in Sec.~\ref{sec:results}
demonstrate that second-order and ratio-based observables can in
principle enhance sensitivity to nonstandard cosmological dynamics.
However the substantial overlap between reconstructed confidence
regions especially at intermediate and high redshifts makes it
difficult to assess whether apparent differences between models are
statistically significant based on visual inspection alone.

To address this, we employ the pointwise marginal Hellinger distance as
a robust metric for quantifying the statistical
distinguishability of reconstructed diagnostics across different
fiducial cosmological models at fixed redshift
\citep{Hellinger1909, LeCamYang2000}.
For two continuous probability
distributions $P$ and $Q$, the Hellinger distance is defined as
\begin{equation}
  H(P,Q) = \frac{1}{\sqrt{2}}
  \left[\int_{-\infty}^{\infty}
  \left(\sqrt{p(x)}-\sqrt{q(x)}\right)^2 \mathrm{d}x
  \right]^{1/2},
  \label{eq:hellinger_def}
\end{equation}
where $x$ is the value of the random variable whose distribution is being
compared, and $p(x)$ and $q(x)$ are the probability density functions of
$P$ and $Q$ respectively. In our application, $x$ represents the value of
the reconstructed diagnostic at fixed redshift, for example $H(z)$,
$H'(z)$, $q(z)$, $\mathcal{O}_m(z)$, $w_{\rm tot}(z)$ or $\kappa(z)$.
The Hellinger distance is bounded
$H(P,Q)\in[0,1]$ with $H=0$ indicating identical distributions and
$H=1$ indicating orthogonal (fully distinguishable) distributions. It is
symmetric, satisfies the triangle inequality and is more robust to
binning choices than the commonly used $\chi^2$ statistic.

For each redshift we summarize the reconstructed marginal distribution
of a diagnostic by a Gaussian with median $\mu$ and standard deviation
$\sigma$. Under this Gaussian marginal approximation, for distributions
$\mathcal{N}(\mu_1,\sigma_1^2)$ and $\mathcal{N}(\mu_2,\sigma_2^2)$,
Eq.~(\ref{eq:hellinger_def}) reduces to the closed form expression:
\begin{equation}
  H^2(P,Q) = 1 - \sqrt{\frac{2\sigma_1\sigma_2}{\sigma_1^2+\sigma_2^2}}
  \exp\!\left(-\frac{(\mu_1-\mu_2)^2}{4(\sigma_1^2+\sigma_2^2)}\right).
  \label{eq:hellinger_gaussian}
\end{equation}
This expression is evaluated independently at each redshift $z$ where $\mu_i(z)$ and
$\sigma_i(z)$ are the reconstructed median and $1\sigma$ uncertainty for
fiducial models $i=1,2$. For nonlinear diagnostics such as $q$,
$\mathcal{O}_m$, $w_{\rm tot}$ and $\kappa$ the exact sample distributions
need not be perfectly Gaussian; the Hellinger curves should therefore be
interpreted as compact Gaussian summaries of pointwise marginal
distinguishability. They do not represent a Hellinger distance between
the full functional GP posteriors across redshift, because the
off-diagonal covariance between different redshift points is not folded
into a single distribution-level model-separation statistic.

\subsection{Results}
\label{sec:hellinger_results}

\begin{figure}[!t]
  \centering
  \begin{subfigure}[t]{0.95\textwidth}
    \centering
    \includegraphics[width=\linewidth]{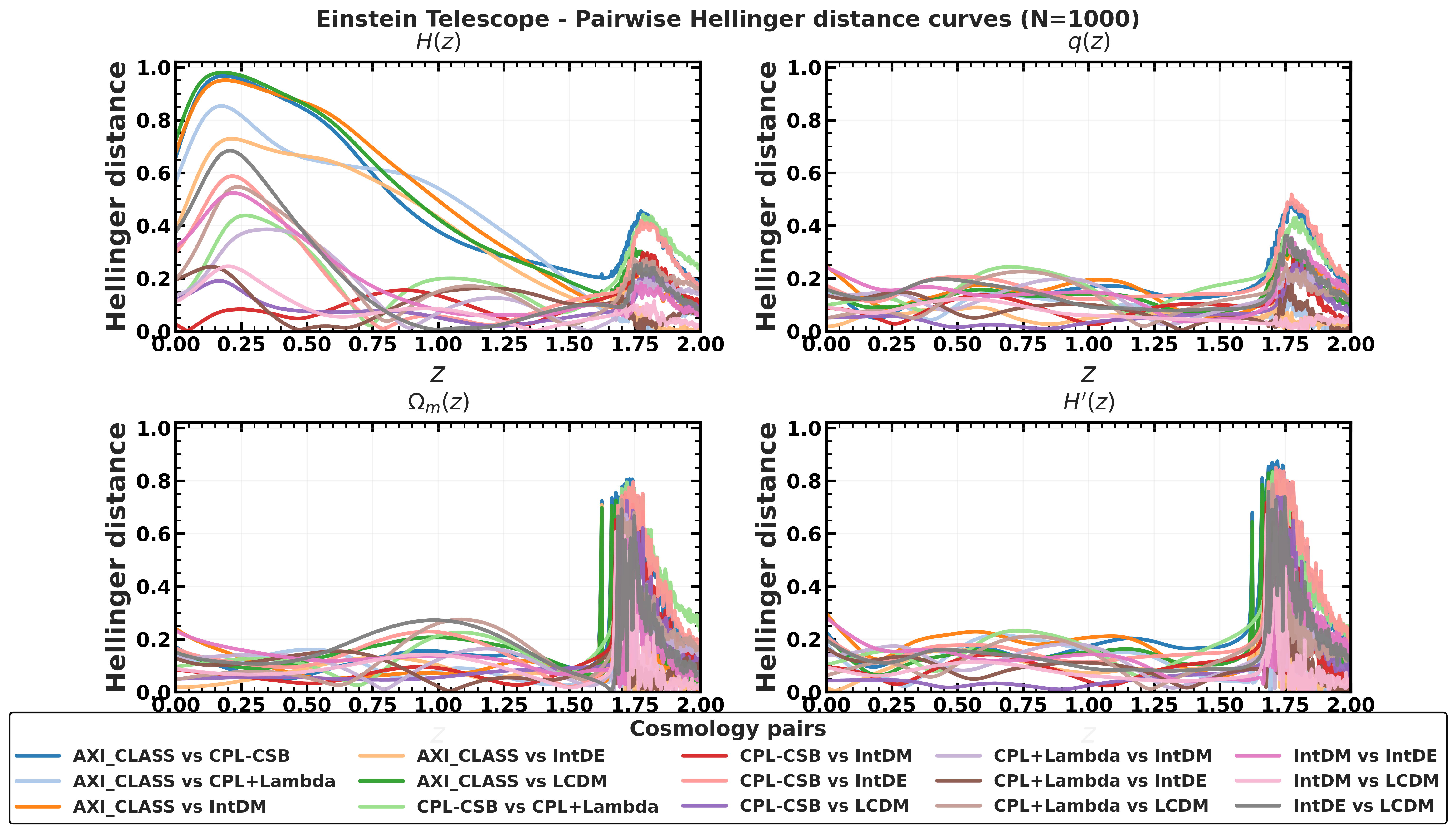}
    \caption{\it Einstein Telescope}
  \end{subfigure}

  \vspace{0.5em}

  \begin{subfigure}[t]{0.95\textwidth}
    \centering
    \includegraphics[width=\linewidth]{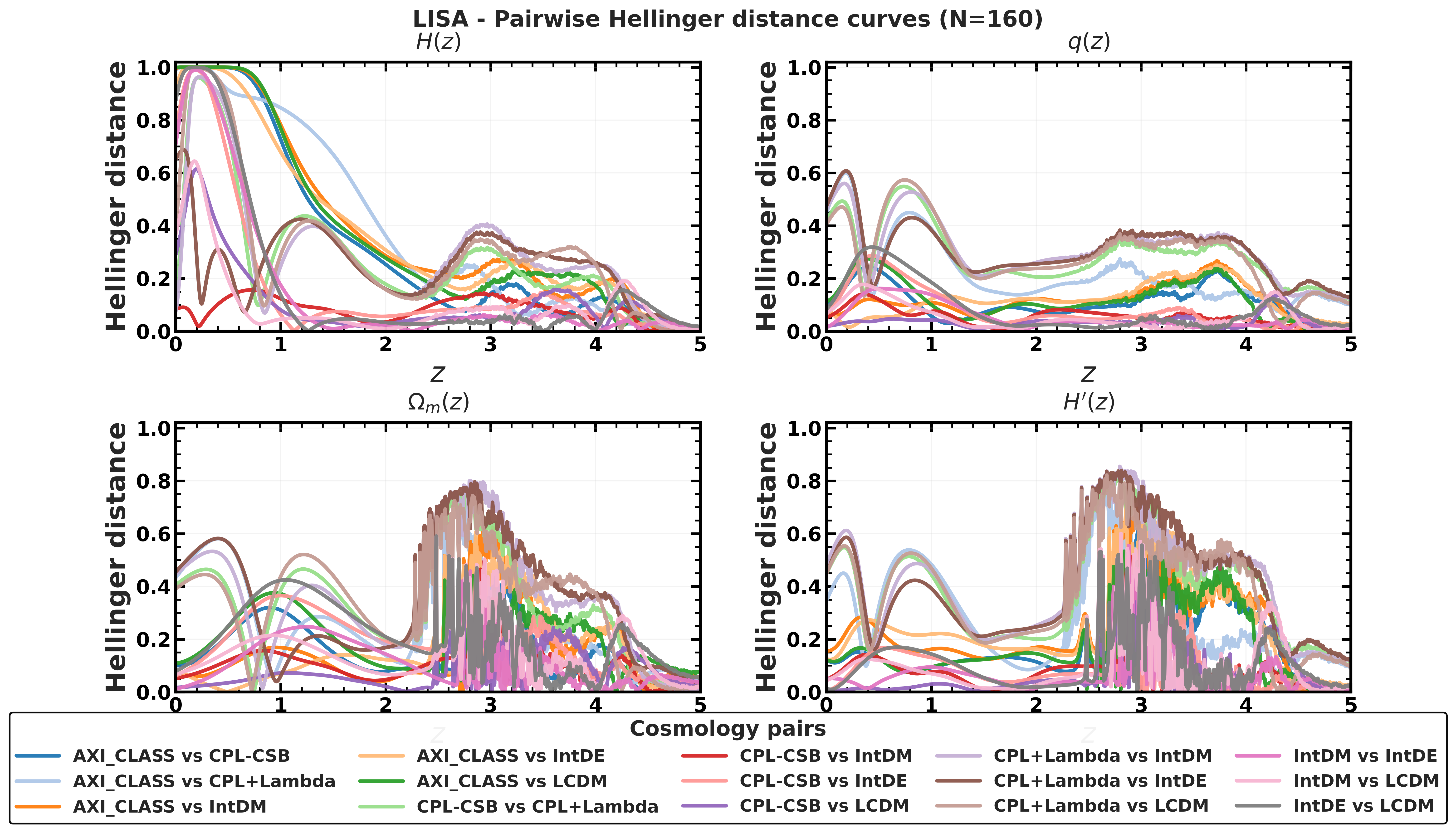}
    \caption{\it LISA}
  \end{subfigure}
  \caption{\it Summary of the pairwise pointwise marginal Hellinger distance curves for the ET
    and LISA reconstructions. The $H(z)$ panels are dominated at low
    redshift by the differing fiducial $H_0$ values. The $q(z)$,
    $w_{\rm tot}(z)$ and $\kappa(z)$ panels are nearly identical in
    practice because these diagnostics are related by simple affine
    transformations at fixed redshift. Detailed detector-specific plots
    for all diagnostics are shown in Appendix~\ref{app:hellinger}.}
  \label{fig:hellinger_summary}
\end{figure}

We compute the pointwise marginal Hellinger distance between
reconstructed diagnostics for all pairs of fiducial cosmological models
and for each detector separately. Figure~\ref{fig:hellinger_summary}
shows summary panels for ET and LISA. The full detector-specific plots are collected in
Appendix~\ref{app:hellinger} and shown in
Figs.~\ref{fig:hellinger_H} through \ref{fig:hellinger_kappa}.

The behavior of $H(z)$ is distinct from the other diagnostics. For both
detectors the pointwise marginal Hellinger distance is largest at very low redshift because
the GP reconstruction resolves the different fiducial $H_0$ values with
high precision. This produces near-unity values for several model pairs.
It does not by itself imply strong separation of the late-time
expansion dynamics. The detector-specific $H(z)$ curves are shown in
Fig.~\ref{fig:hellinger_H}.

The diagnostics $q(z)$, $w_{\rm tot}(z)$ and $\kappa(z)$ track each
other very closely and are nearly identical in practice. This is
expected because at fixed redshift they are related by the affine maps
\begin{equation}
  \kappa(z)=\frac{1+q(z)}{1+z}
  = \frac{3}{2}\,\frac{1+w_{\rm tot}(z)}{1+z}.
\end{equation}
The Gaussian pointwise marginal Hellinger distance is invariant under the same one-to-one
affine transformation applied to both reconstructed distributions. In
exact arithmetic the three curves would coincide. In practice finite
sampling and numerical smoothing leave only very small residual
differences. Their near coincidence is therefore a consistency check
rather than a new source of independent information. This behavior is
shown explicitly in Figs.~\ref{fig:hellinger_q},
\ref{fig:hellinger_wtot} and \ref{fig:hellinger_kappa}.

The detailed redshift dependence differs between the two detectors. For
ET the $q(z)$, $w_{\rm tot}(z)$ and $\kappa(z)$ curves remain modest
over most of the range and rise most strongly near $z\sim 1.7$--$1.9$
close to the upper edge of the catalog. For LISA the same diagnostics
show prominent low-redshift maxima for some model pairs, together with a
broader high-redshift shoulder from $z\sim 2.5$ to $z\sim 4$ that crests
around $z\sim 3$. The $H'(z)$ and $\mathcal{O}_m(z)$ diagnostics develop the
sharpest high-redshift structure for both detectors. In $\mathcal{O}_m(z)$
this is amplified by the denominator in Eq.~(\ref{eq:Omz}) while
$H'(z)$ is more sensitive to fluctuations in $d_C''(z)$ and to sign
changes in that quantity. The corresponding detector-specific curves are
shown in Figs.~\ref{fig:hellinger_Hp} and \ref{fig:hellinger_Om}.

Despite these detector-specific features, the overall conclusion is
unchanged. Away from the low-redshift $H(z)$ feature, all pairwise
pointwise marginal Hellinger distances remain well below unity. The axion-inspired early
dark energy model and the interacting dark matter scenario usually show
the largest separation from $\Lambda$CDM, but the reconstructed
distributions still overlap strongly. Background luminosity-distance
data alone therefore do not provide genuine statistical separation
between the six fiducial cosmologies considered here.

These results show that $\kappa(z)$ is best viewed as a numerically
stable consistency check on $q(z)$ and $w_{\rm tot}(z)$. It tests the
same second-order reconstruction without pushing the analysis to third-order
differentiation. In the pointwise marginal Hellinger analysis it does not provide
independent discriminating power. More generally no diagnostic achieves
genuine statistical separation between the six fiducial cosmologies.
GPR applied to background luminosity distance data therefore defines a
practical benchmark on what these catalogs can and cannot distinguish.

The pattern described above reflects a fundamental asymmetry between
what controls the width of the reconstructed posteriors and what
controls the separation between them. The posterior width is set by the
noise amplitude and event count, both of which improve with larger, better
catalogs. The physical model separation, however, is set by the
difference in the functional \emph{shapes} of $d_L(z)$ across
cosmologies. Increasing event counts or reducing noise narrows the
reconstructed posteriors and can increase the pointwise marginal
Hellinger distance for a
fixed nonzero separation. It does not, however, change the underlying
separation of the fiducial backgrounds themselves. When the $d_L(z)$
shapes are very similar, very large gains in precision are required
before the Gaussian pointwise marginal Hellinger measure approaches genuine
separation. In this sense the distinguishability is limited primarily by
shape, with noise controlling how sharply that shape difference can be
resolved.
Genuine improvement in model discriminability therefore requires either
accessing redshifts where the models diverge more significantly
($z\sim 3$--$5$ where LISA's reach becomes advantageous) or
incorporating observables that probe different sectors of the physics
such as the growth rate $f\sigma_8(z)$ from redshift space distortions
\citep{Alam2017}, which is sensitive to the perturbation sector and can
break degeneracies that the background expansion history cannot.

\FloatBarrier

\section{Summary and conclusions}
\label{sec:conclusions}

We have presented a model-independent reconstruction of late-time
cosmological diagnostics from mock gravitational wave standard siren
catalogs employing Gaussian Process Regression as the core inference
framework. The analysis encompasses six fiducial cosmological
models: the concordance $\Lambda$CDM, two CPL dark energy variants,
interacting dark matter and dark energy scenarios and an axion-inspired early dark energy model investigated for two complementary forthcoming GW detectors LISA and the Einstein Telescope each assigned physically motivated
redshift-dependent noise prescriptions. The adopted source redshift
distributions are shown in Fig.~\ref{fig:redshift_distributions}, while
the LISA and ET error models are defined in
Eqs.~(\ref{eq:lisa_total_err}) and (\ref{eq:ET_total_err}). The central aim is to assess the
discriminatory power of future GW standard siren observations from
next-generation detectors. Our principal findings are as follows.

\begin{enumerate}

\item \textit{Reconstruction fidelity.} The Gaussian Process framework
employing a Mat\'ern $\nu=9/2$ kernel with MCMC hyperparameter
marginalization successfully recovers the
underlying expansion history across all fiducial models and for both
detector configurations. The reconstructed median follows the fiducial
evolution closely and $\Lambda$CDM remains consistent with the
reconstructed observables within inferred confidence regions in all
cases. This behavior is illustrated by the reconstruction overview in
Fig.~\ref{fig:gp_reconstruction_overview} and by the detailed
$H(z)$ reconstruction in Fig.~\ref{fig:H_recon}. The GP kernel and
hyperparameter likelihood used for this result are given in
Eqs.~(\ref{eq:matern92}) and (\ref{eq:log_evidence}).
The comparison across fiducial cosmologies shows that changing the
background model mainly changes the median cosmological history, not the
statistical hierarchy of the reconstruction. In the LISA precision
comparison with $N_{\rm ev}=160$, the direct Hubble reconstruction is
the most robust diagnostic: the optimal relative precision for $H(z)$
lies in the narrow range $0.394\%$--$0.421\%$ across the six fiducial
models, with $z^*_{\rm LISA}=0.165$--$0.205$. The first derivative is
less tightly constrained, with the best relative precision for $H'(z)$
lying in the range $5.09\%$--$6.55\%$ at
$z^*_{\rm LISA}=0.320$--$0.385$. The derived diagnostics inherit this
loss of precision: the best absolute uncertainties are
$0.0355$--$0.0471$ for $q(z)$, $0.0237$--$0.0314$ for
$w_{\rm tot}(z)$ and $0.0267$--$0.0360$ for $\kappa(z)$, while
$\mathcal{O}_m(z)$ reaches best relative precision
$5.74\%$--$6.12\%$. Thus the reconstructions teach us that the
dominant limitation is derivative amplification and high-redshift
uncertainty growth rather than a failure of the GP to track any
particular fiducial background.

\item \textit{Uncertainty propagation.} Retaining the relevant joint
Gaussian Process predictive covariance blocks, especially the
cross-covariance between $d_C'$ and $d_C''$, is the central
methodological requirement of this analysis. Neglecting these cross
correlations leads
to a systematic underestimate of the confidence regions for $H(z)$,
$q(z)$, $\mathcal{O}_m(z)$, $w_{\rm tot}(z)$ and $\kappa(z)$. MCMC
marginalization over $(\sigma_f,\ell)$ broadens the inferred intervals
by about $10$ to $30\%$ and should also be included in realistic error
budgets.
This requirement follows from the derivative covariance expression in
Eq.~(\ref{eq:cov_deriv_general}) and is visualized directly in
Fig.~\ref{fig:covariance_blocks}. The higher derivative behavior shown
in Fig.~\ref{fig:covariance_third_deriv} explains why the main analysis
avoids diagnostics that require third order differentiation.

\item \textit{Hierarchy of diagnostics.} The Hubble parameter $H(z)$
achieves sub-percent reconstruction precision at low redshift but shows
limited discriminatory power in isolation. Second-order diagnostics
($q$, $\mathcal{O}_m$, $w_{\rm tot}$) enhance model sensitivity but suffer
from rapid uncertainty growth at high $z$. The ratio-based diagnostic
$\kappa(z)=E'(z)/E(z)$ is best viewed as a numerically stable
consistency check on $q(z)$ and $w_{\rm tot}(z)$. It repackages the
same second-order information in logarithmic slope form and is useful
because it avoids pushing the reconstruction to third-order
differentiation. It should not be interpreted as an independent source
of model separation. The definitions of these observables are given in
Eqs.~(\ref{eq:Hz}), (\ref{eq:Hprime}), (\ref{eq:qz}),
(\ref{eq:Omz}), (\ref{eq:wtot}), and (\ref{eq:kappa}). Their
reconstructed behavior is shown in
Figs.~\ref{fig:H_recon}, \ref{fig:q_Om_wtot_recon}, and
\ref{fig:kappa_recon}, while their precision hierarchy is summarized in
Fig.~\ref{fig:relative_precision}.

\item \textit{Complementarity and precision of LISA and ET.} The two
detector configurations probe the cosmic expansion history in
complementary regimes. ET provides dense sampling at $z\lesssim 2$
owing to its large BNS event count while LISA extends the
reconstruction to $z\sim 5$ with sparser but high-redshift events.
Adopting a realistic redshift-dependent ET instrumental noise
model $\sigma_{\rm inst}(z)=\mathcal{F}(z)\,d_L(z)$ with
$\mathcal{F}(z)=0.1449\,z-0.0118\,z^2+0.0012\,z^3$ yields
substantially improved ET reconstruction precision at low redshift
compared to the flat $25\%$ approximation and shifts the ET optimal
redshift $z^*$ to significantly lower values. Under the adopted error
prescriptions LISA achieves a lower precision metric than ET across all
diagnostics and cosmologies owing to LISA's quadratically suppressed
per event noise at the relevant redshifts. In the updated precision
summaries the improvement at $z^*$ is typically a factor of
$2$--$4$, depending on the diagnostic and fiducial cosmology. The complementary redshift
coverage is visible in Fig.~\ref{fig:redshift_distributions}. The
instrumental noise terms are defined in Eqs.~(\ref{eq:lisa_inst}) and
(\ref{eq:ET_inst}), and the resulting detector comparison is shown in
Fig.~\ref{fig:relative_precision}.

\item \textit{Model independence of the optimal reconstruction redshift.}
A central geometric inference result of this work is the near coincidence
of the optimal reconstruction redshift $z^*$ between ET and LISA across
all six fiducial cosmological models for the most stable diagnostics.
This convergence holds
despite the two detectors having entirely different noise floors, event
rates and redshift windows (see Appendix~\ref{app:precision}) and
crucially despite their event redshift distributions having qualitatively
different shapes and centroids. The coincidence arises because $z^*$ is
set mainly by the low-redshift anchoring region rather than by the bulk
of the source population. For LISA the boundary condition $d_C(0)=0$
stabilizes the reconstruction at the origin but it is not treated as an
observed bright siren event. The minimum in uncertainty appears only
after the first few nonzero redshift events fix the local GP
correlation scale. ET behaves in the same way. Since ET ($z_{\min}=0.07$) and LISA (effective
$z_{\min}\sim 0.1$--$0.2$) enter this anchored interpolation regime at
similar redshift their $z^*$ values are close. This establishes $z^*$
as a robust figure of merit determined mainly by the lower boundary of
the event redshift distribution and by the local GP correlation scale.
The signal variance $\sigma_f$ and the noise prescription mainly control
the amplitude of the precision curve and therefore the absolute precision
at $z^*$, while the lengthscale $\ell$ helps shape the curve near
$z_{\min}$ and therefore helps determine the \emph{location} of $z^*$.
In the mock catalogues studied here, changing the noise floor moves the
absolute precision level much more than it moves $z^*$. The stability of
$z^*_{\rm LISA}$ across cosmologies is expected
by construction since LISA events are drawn from a fixed empirical
redshift distribution with a cosmology-independent lower boundary. The
corresponding stability of $z^*_{\rm ET}$ is genuinely nontrivial since
ET events are generated self-consistently from each cosmology's $H(z)$
and $d_C(z)$. This stability arises because $z_{\min}=0.07$ is fixed by
the ET detector sensitivity floor regardless of cosmology and, since
$z^*$ is anchored near $z_{\min}$, the cosmology-to-cosmology variation
in $p(z)$ at higher redshift is subdominant for the location of $z^*$.
The definition of $z^*$ is given in Eq.~(\ref{eq:zstar}). Its near
coincidence between ET and LISA is visible in
Fig.~\ref{fig:relative_precision} and across the alternative cosmology
precision curves in Figs.~\ref{fig:precision_bestfit},
\ref{fig:precision_cplcsb}, \ref{fig:precision_cplrsh},
\ref{fig:precision_intdm}, and \ref{fig:precision_intde}. The role of the
event distribution is set by the source redshift density in
Eq.~(\ref{eq:ET_pz}) and by the distributions shown in
Fig.~\ref{fig:redshift_distributions}.

\item \textit{Quantitative model distinguishability.} The pointwise
marginal Hellinger distance analysis gives a diagnostic map of where
cosmological information enters the reconstruction. Denoting the
pointwise marginal Hellinger distance by $H_{\rm Hell}$, the largest
values in $H(z)$ occur at low redshift, with several pairs reaching
$H_{\rm Hell}\simeq 0.96$--$1.00$ near
$z\simeq 0.15$--$0.23$ for LISA and
$H_{\rm Hell}\simeq 0.73$--$0.98$ near
$z\simeq 0.17$--$0.21$ for ET. These peaks mostly trace different
fiducial $H_0$ values and should not be interpreted as independent
evidence for different late-time dynamics. The more physically
interesting separation appears in derivative-sensitive diagnostics.
For ET, $H'(z)$ reaches pairwise distances as large as
$H_{\rm Hell}\simeq 0.64$--$0.87$ around
$z\simeq 1.66$--$1.72$, while $\mathcal{O}_m(z)$ reaches
$H_{\rm Hell}\simeq 0.61$--$0.81$ around
$z\simeq 1.62$--$1.74$. For LISA, the same structures move to higher
redshift: $H'(z)$ reaches $H_{\rm Hell}\simeq 0.79$--$0.85$ near
$z\simeq 2.67$--$2.81$, and $\mathcal{O}_m(z)$ reaches
$H_{\rm Hell}\simeq 0.74$--$0.80$ near
$z\simeq 2.62$--$2.85$ for the most separated model pairs. The
$q(z)$ and $w_{\rm tot}(z)$ diagnostics are more modest, typically
peaking at $H_{\rm Hell}\simeq 0.52$ near $z\simeq 1.77$ for ET and
$H_{\rm Hell}\simeq 0.61$ near $z\simeq 0.19$--$0.75$ for LISA in the
most favorable pairs, while $\kappa(z)$ remains useful as a correlated
logarithmic-slope consistency check. Thus the result is not
featureless: the Hellinger curves identify specific redshift windows
where model discrimination begins to emerge, even though the propagated
one-point marginals remain overlapping enough to prevent robust
background-only model separation. The distance measure is defined in
Eqs.~(\ref{eq:hellinger_def}) and (\ref{eq:hellinger_gaussian}). The
summary result is shown in Fig.~\ref{fig:hellinger_summary}, with the
detector-specific panels collected in Figs.~\ref{fig:hellinger_H}
through \ref{fig:hellinger_kappa}.

\end{enumerate}


Taken together these results establish that Gaussian Process
reconstruction of gravitational wave standard sirens provides a powerful
and model-independent way of testing the concordance cosmological model.
Our comparison demonstrates that the median reconstructions successfully recover the
$\Lambda$CDM, CPL, CPL$+\Lambda$, IntDM, IntDE and AXI\_CLASS backgrounds well within their respective confidence regions.
The differences between cosmologies appear mainly as changes in the
fiducial trend and in the width of the high-redshift confidence bands,
while the ordering of diagnostic precision remains stable. Quantitatively,
for LISA with $N_{\rm ev}=160$, $H(z)$ reaches a best relative precision
of $0.394\%$--$0.421\%$ across the six models, whereas $H'(z)$ reaches
$5.09\%$--$6.55\%$ and $\mathcal{O}_m(z)$ reaches
$5.74\%$--$6.12\%$. The second-order diagnostics have best absolute
uncertainties of order $10^{-2}$, but their intervals broaden rapidly at
high redshift. Thus the reconstruction is faithful to each fiducial
cosmology and, equally importantly, it shows where the first signs of
model discrimination are expected to appear.

The Hellinger analysis sharpens this point. Low-redshift separation in
$H(z)$ can be very large, reaching near-unity values for some model
pairs, but this mainly reflects different fiducial $H_0$ choices. The
more informative windows occur in derivative-sensitive diagnostics: for
ET, the strongest nontrivial separation appears around
$z\simeq 1.6$--$1.8$ in $H'(z)$ and $\mathcal{O}_m(z)$, while for LISA
the corresponding leverage shifts to $z\simeq 2.6$--$2.9$. In these
windows the pointwise marginal Hellinger distance reaches
$H_{\rm Hell}\simeq 0.8$ for the most separated pairs, indicating that
the reconstructed distributions are beginning to pull apart even though
they have not reached decisive separation. This is the technical lesson
of the analysis: background-only GPR does not simply return a null
result, but identifies the redshift ranges and observables where future
standard-siren catalogues should gain the most discriminatory power.
$\kappa(z)$ provides a stable logarithmic-slope check of the second-order
reconstruction, although it should be interpreted together with
$q(z)$ and $w_{\rm tot}(z)$ rather than as an independent model
classifier, as follows from Eq.~(\ref{eq:kappa_q_relation}) and
Fig.~\ref{fig:hellinger_summary}. The remaining limitation is physical
and statistical: viable cosmologies have similar $d_L(z)$ shapes over
much of the observed range, so stronger separation requires either
tighter uncertainties in these redshift windows or additional
\subsection*{Future outlook}

Several natural extensions follow from this work. The present analysis
uses background standard siren distances. A first direction is to extend
the covariance learning beyond background expansion data. One can train
and validate GP kernels on perturbation sector observables or on joint
background and perturbation data. This would allow the reconstruction to
use information from both geometry and structure growth. It would also
test whether models with similar luminosity distance histories can be
separated through their perturbation dynamics.

A second direction is to relax the stationarity of the GP kernel. In
this work the hyperparameters $(\sigma_f,\ell)$ are sampled with
\texttt{emcee}. Each sampled pair is constant over the full redshift
range. This is consistent with the stationary Mat\'ern kernel used here
but it may be restrictive for catalogs whose information content changes
strongly with redshift. A natural extension is to train a neural network
that takes redshift as input and returns $\sigma_f(z)$ and $\ell(z)$.
This would define a nonstationary GP with redshift dependent amplitude
and correlation length. Such a model could adapt to dense low redshift
regions and sparse high redshift regions within a single reconstruction.
Other flexible parametric maps could also be explored but the covariance
construction must still preserve positive definiteness.

A third direction is to reduce the computational cost of repeated
reconstructions. Exact GP regression scales as $\mathcal{O}(N^3)$
because the covariance matrix must be factorized through a Cholesky
decomposition or an equivalent linear algebra operation. This cost
becomes substantial for large mock ensembles, dense catalogs and repeated
hyperparameter sampling. Neural operator surrogate models, including
Fourier neural operators, provide a possible acceleration strategy. Such
surrogates could learn the map from catalog realizations and noise
prescriptions to reconstructed observables or posterior summaries. If
trained and validated carefully, they would reduce wall clock time while
preserving the accuracy needed for forecasts and model comparison.

The motivation for these extensions is the same asymmetry identified
above. The location of a reconstructed feature and its amplitude are
controlled by different ingredients. For $z^*$ the location is set by
the low redshift anchoring scale and by the kernel lengthscale $\ell$.
The amplitude of the precision metric is set mainly by $\sigma_f$ and
the noise prescription. For the pointwise marginal Hellinger distance
the physical separation is controlled by the shapes of $d_L(z)$. The
posterior width is controlled by the noise. The stability of $z^*$
across cosmologies follows because $z_{\min}$ is fixed by the detector
sensitivity floor. Variations in $p(z)$ at higher redshift have little
leverage on $z^*$. Improvements in detector sensitivity therefore
address only one part of the problem. New discriminatory power is more
likely to come from perturbation observables and from a dedicated use of
LISA's high redshift reach where the background histories diverge more
strongly. These links between location and amplitude are displayed by
the precision curves in Fig.~\ref{fig:relative_precision}, the kernel
scaling in Eq.~(\ref{eq:kernel_factored}) and the Hellinger summary in
Fig.~\ref{fig:hellinger_summary}.
\section*{Acknowledgement}
Authors thank Rahul Shah, Tuhin Ghosh, William Giare and Shouvik Roychoudhury for helpful discussions. DFM thanks the Research Council of Norway for their support and the resources provided by UNINETT Sigma2-the National Infrastructure for High-Performance Computing and Data Storage in Norway. A.G. acknowledges the support from the Royal Society, UK, Funding Reference: NIF\ R1\ 253963. 
\clearpage
\onecolumn
\appendix

\section{Relative reconstruction precision for alternative cosmologies}
\label{app:precision}

This appendix presents the redshift-dependent precision curves
$\sigma/|\mathrm{median}|$ (or absolute $\sigma$ for zero-crossing
diagnostics) for all fiducial cosmological models considered in this
work, complementing Fig.~\ref{fig:relative_precision}, which shows the
$\Lambda$CDM case in the main text. For each model, we display both the
ET ($N_{\rm ev}=1000$, blue) and LISA ($N_{\rm ev}=160$, red)
configurations overlaid in a single panel per diagnostic, with vertical
dashed lines marking the optimal redshift $z^*$ for each detector. The
qualitative precision hierarchy---$H(z)$ most stable, $H'(z)$ least
stable---is preserved across all models and both detector configurations.

A central feature of these panels is the near-coincidence of the
$z^*_{\rm ET}$ (blue dashed) and $z^*_{\rm LISA}$ (red dashed) lines for
$H$, $q$, $w_{\rm tot}$ and $\kappa$ across all six cosmological models.
The exceptions are $\mathcal{O}_m$, whose denominator suppresses the
low-redshift minimum, and to a lesser extent $H'$, which is more
sensitive to derivative noise. This alignment holds despite the two
detectors having qualitatively different event redshift distributions
with different shapes and centroids as shown in
Fig.~\ref{fig:redshift_distributions}. The coincidence arises because
$z^*$ is anchored near the lower boundary
$z_{\min}$ of the event distribution for both detectors. Below $z_{\min}$
the GP has no anchoring data and derivative uncertainties are
extrapolation dominated. Just above $z_{\min}$ the first events provide
genuine interpolation anchoring and the precision metric reaches its
minimum. This transition is only weakly sensitive to the noise amplitude
and to the shape and centroid of the event distribution at higher
redshift for the mock catalogues considered here.
Since ET ($z_{\min}=0.07$) and LISA (effective $z_{\min}\sim 0.1$--$0.2$)
share a similar lower boundary, $z^*$ coincides between them for the
most stable diagnostics across all models. The main exception is
$\mathcal{O}_m$, where the denominator $(1+z)^3-1\to 0$ pushes the minimum to
somewhat higher redshift for both detectors; $H'$ also shows larger
scatter because it is derivative-noise dominated. In every case LISA
achieves the lower
(better) absolute precision metric value consistent with the discussion
in Sec.~\ref{sec:zstar}.

\begin{figure}[H]
  \centering
  \includegraphics[width=.95\linewidth]{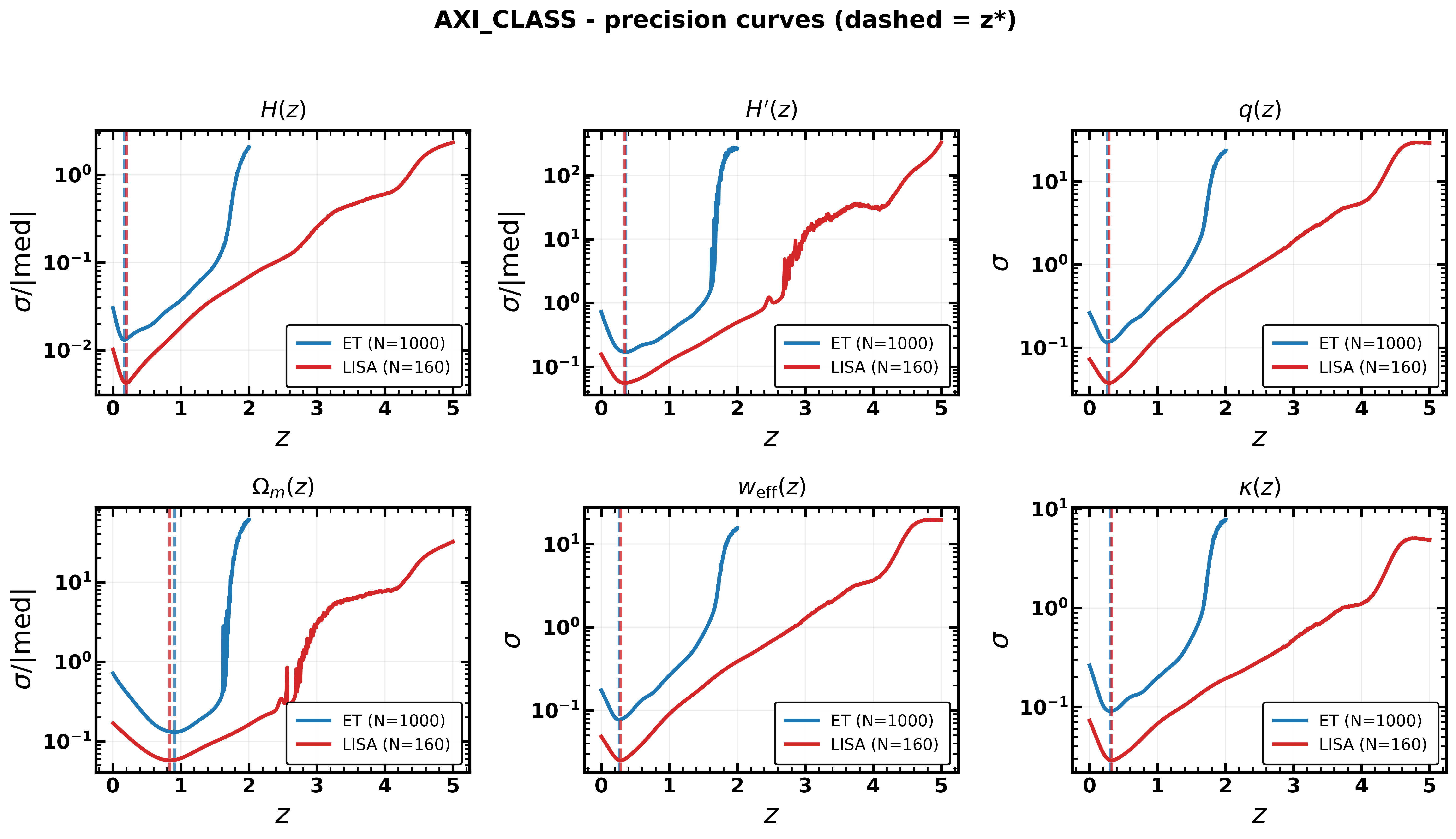}
  \caption{\it Precision curves for the AXI\_CLASS (axion/early dark energy)
    fiducial cosmology. Blue: ET ($N_{\rm ev}=1000$); red: LISA
    ($N_{\rm ev}=160$). Dashed vertical lines mark $z^*$ for each
    detector and diagnostic. Note the near-coincidence of $z^*_{\rm ET}$
    and $z^*_{\rm LISA}$ for $H$, $q$, $w_{\rm tot}$, and $\kappa$.}
  \label{fig:precision_bestfit}
\end{figure}

\begin{figure}[H]
  \centering
  \includegraphics[width=.95\linewidth]{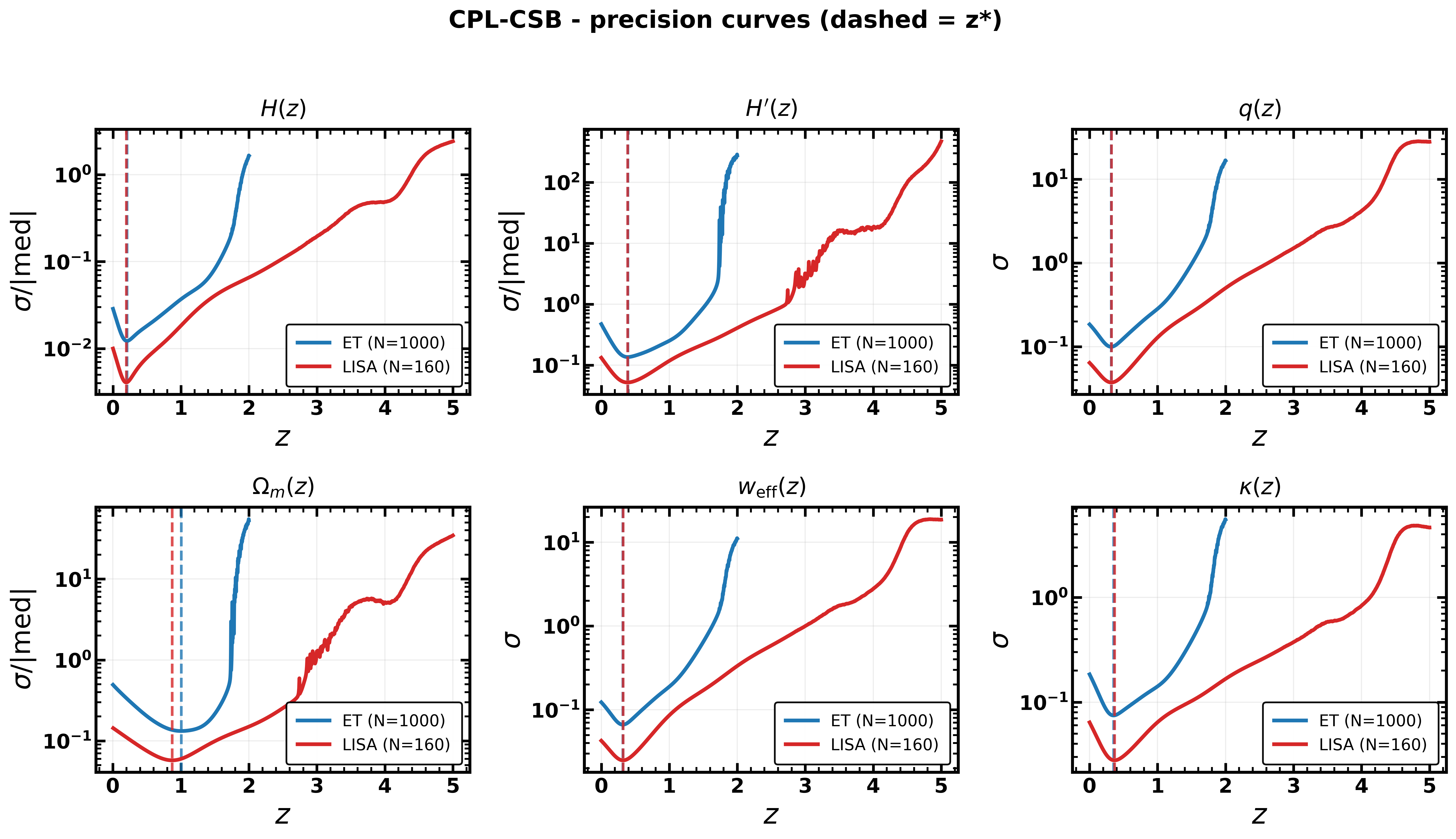}
  \caption{\it Same as Fig.~\ref{fig:precision_bestfit} for the CPL
    fiducial cosmology.}
  \label{fig:precision_cplcsb}
\end{figure}

\begin{figure}[H]
  \centering
  \includegraphics[width=.95\linewidth]{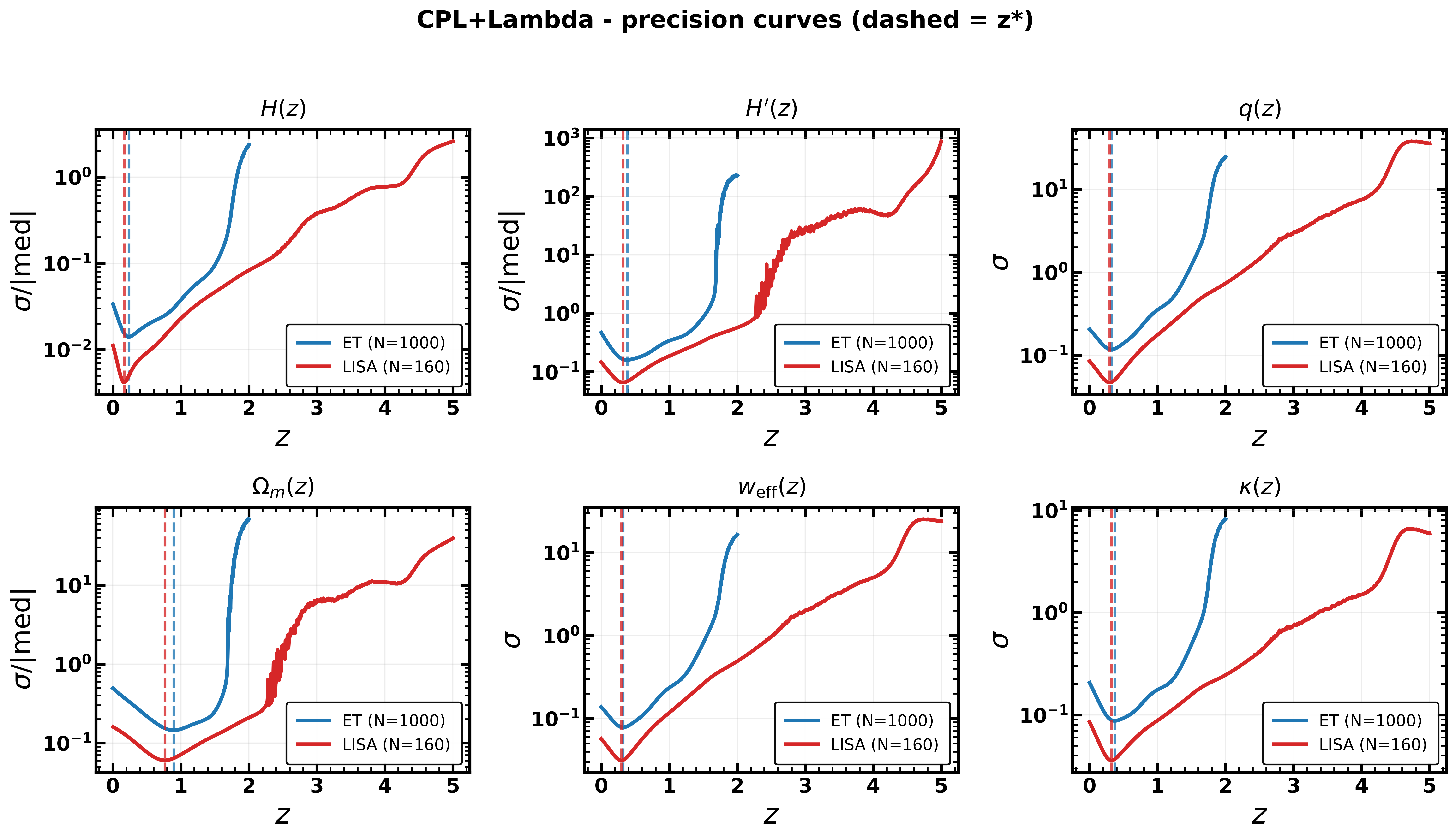}
  \caption{\it Same as Fig.~\ref{fig:precision_bestfit} for the
    CPL$+\Lambda$ fiducial cosmology.}
  \label{fig:precision_cplrsh}
\end{figure}

\begin{figure}[H]
  \centering
  \includegraphics[width=.95\linewidth]{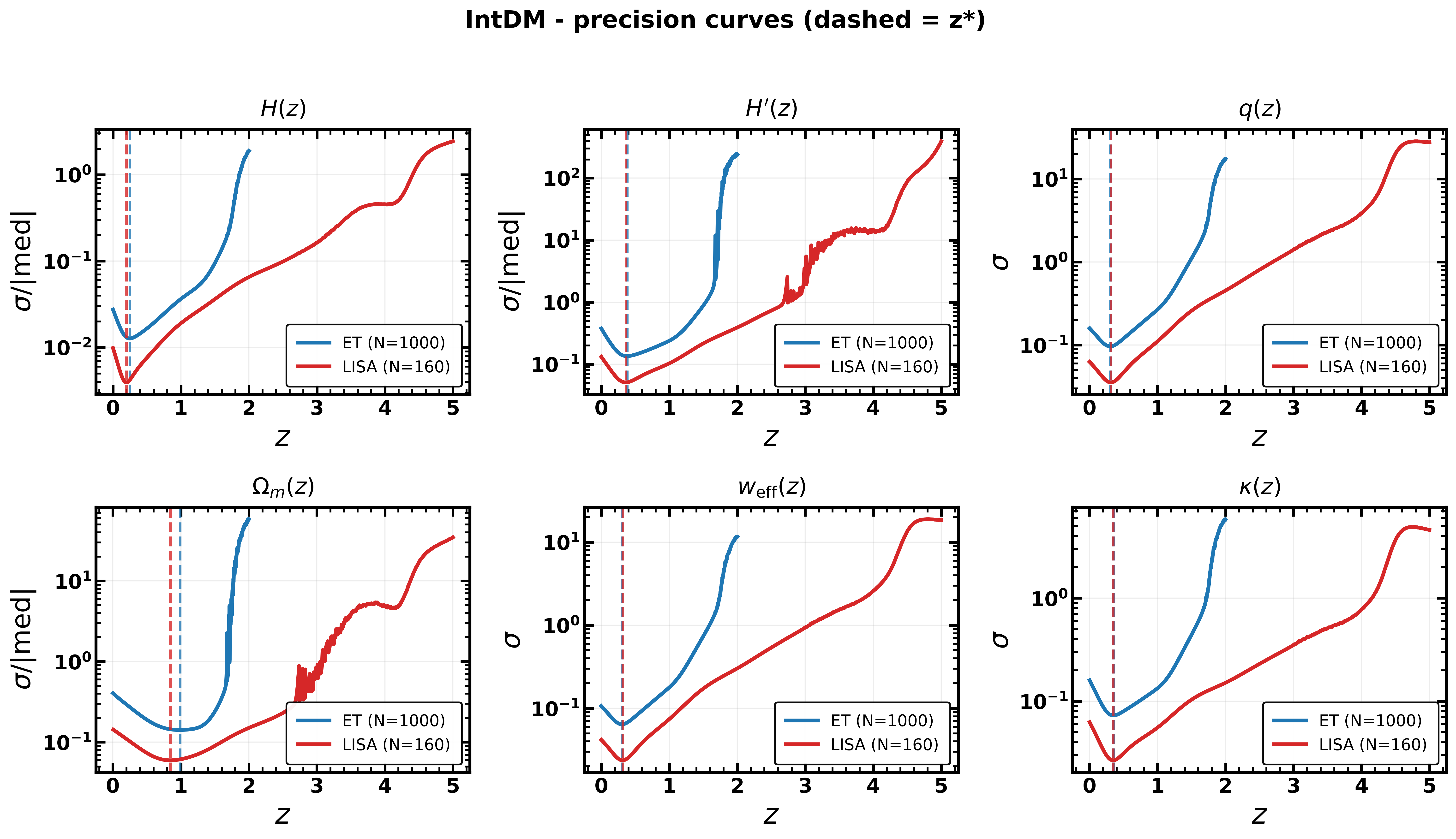}
  \caption{\it Same as Fig.~\ref{fig:precision_bestfit} for the interacting
    dark matter (IntDM) fiducial cosmology. Note the distinctive feature
    in the ET $H'$, $\mathcal{O}_m$, $w_{\rm tot}$, and $\kappa$ panels near
    $z\sim 1.5$, arising from the sign change in $d_C''$ associated with
    the matter--dark energy transition in this model.}
  \label{fig:precision_intdm}
\end{figure}

\begin{figure}[H]
  \centering
  \includegraphics[width=.95\linewidth]{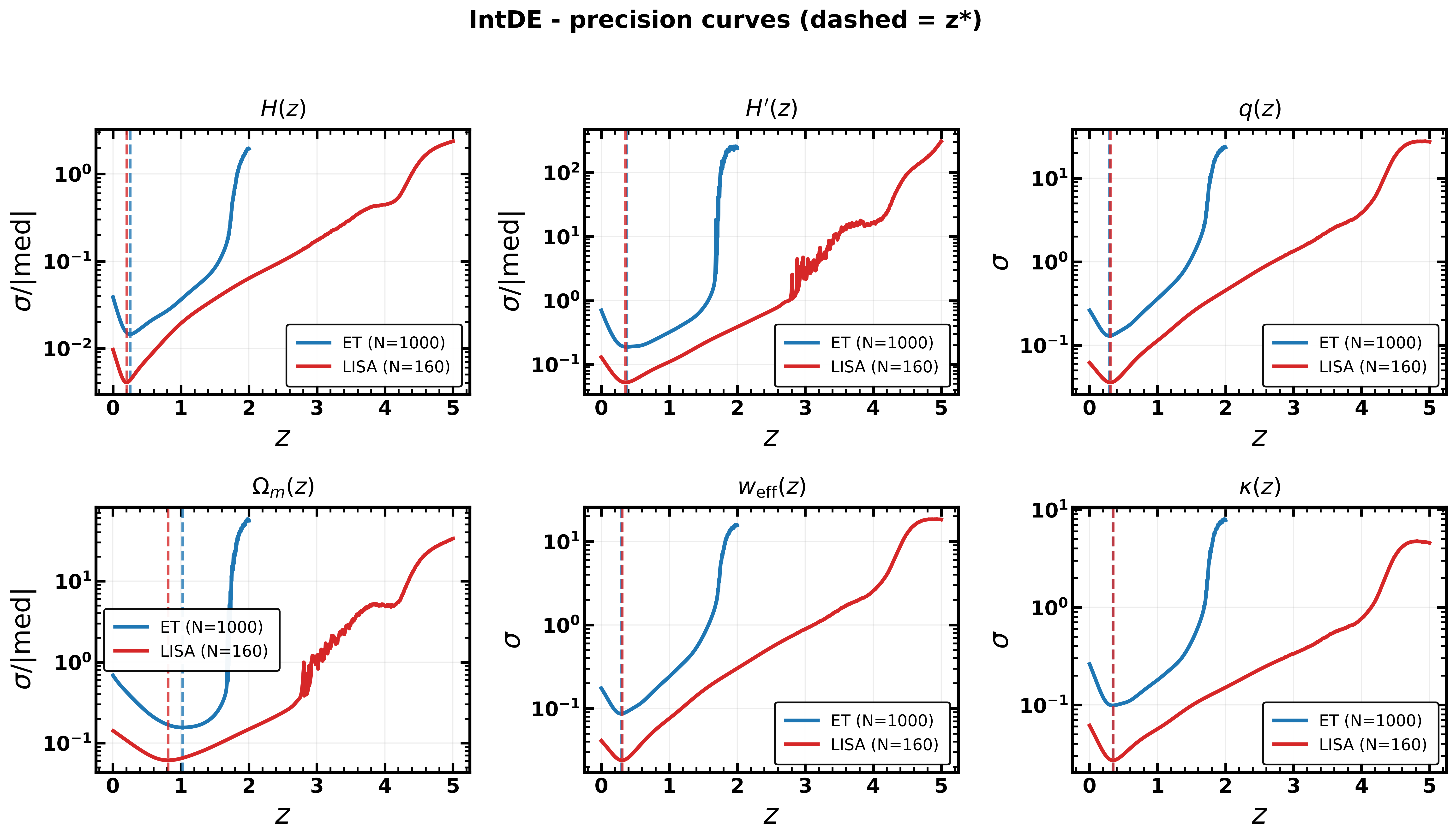}
  \caption{\it Same as Fig.~\ref{fig:precision_bestfit} for the interacting
    dark energy (IntDE) fiducial cosmology.}
  \label{fig:precision_intde}
\end{figure}

The precision curves for each model are shown in
Figs.~\ref{fig:precision_bestfit} through \ref{fig:precision_intde}.
Figure~\ref{fig:precision_bestfit} gives the AXI\_CLASS case.
Figures~\ref{fig:precision_cplcsb} and \ref{fig:precision_cplrsh}
show the two CPL variants. Figures~\ref{fig:precision_intdm} and
\ref{fig:precision_intde} show the interacting dark matter and
interacting dark energy cases. Across these figures the same hierarchy
is preserved. $H(z)$ remains the most stable diagnostic, $H'(z)$ remains
the least stable and LISA generally reaches a lower precision metric
than ET under the adopted noise prescriptions.

Table~\ref{tab:zstar_all} provides a complete summary of the optimal
reconstruction redshifts $z^*$ and precision metric values at $z^*$ for
all six cosmological models and both detector configurations, evaluated
at the maximum event counts of the scaling sweep, namely
$N_{\rm ev}=1000$ (ET) and
$N_{\rm ev}=160$ (LISA). LISA achieves the lower precision metric in
every case, consistent with the discussion in Sec.~\ref{sec:zstar}.
The row labelled CPL denotes the standard CPL fiducial based on the CSB
constraints discussed in Sec.~\ref{sec:fiducial_models}.
The near-coincidence of $z^*$ values between ET and LISA for $H$, $q$,
$w_{\rm tot}$, and $\kappa$ is evident across all models, confirming the
model-independence of this figure of merit. The bottom panel quantifies
this coincidence via $\Delta z^* \equiv |z^*_{\rm ET} - z^*_{\rm LISA}|$;
values $\leq 0.067$ for $H$, $q$, $w_{\rm tot}$, and $\kappa$ confirm
that the precision minimum is anchored near $z_{\min}$ for both
detectors and is therefore largely insensitive to detector configuration
and cosmological model.

\begin{table}[H]
\caption{\it Portrait summary of the optimal reconstruction redshift $z^*$
  and precision metric $\mathcal{P}(z^*)$ for all fiducial cosmological
  models and both detector configurations. In the ET and LISA blocks
  each entry is $z^*/\mathcal{P}(z^*)$. For $H$, $H'$ and
  $\mathcal{O}_m$ the metric is $\sigma/|\mathrm{median}|$; for $q$,
  $w_{\rm tot}$ and $\kappa$ it is the absolute $\sigma$. In the final
  block each entry is $\Delta z^*\equiv |z^*_{\rm ET}-z^*_{\rm LISA}|$.
  This portrait layout contains the same information as the full
  detector comparison while avoiding a sideways page.}
\label{tab:zstar_all}
\centering
\scriptsize
\setlength{\tabcolsep}{2.7pt}
\renewcommand{\arraystretch}{1.12}
\resizebox{\textwidth}{!}{%
\begin{tabular}{llcccccc}
\toprule
Block & Model & $H$ & $H'$ & $q$ & $\mathcal{O}_m$ & $w_{\rm tot}$ & $\kappa$ \\
\midrule
\multicolumn{8}{l}{\itshape Einstein Telescope ($N_{\rm ev}=1000$): entries are $z^*/\mathcal{P}(z^*)$} \\
\midrule
ET & AXI\_CLASS       & 0.166/0.01312 & 0.362/0.17072 & 0.260/0.11639 & 0.905/0.13036 & 0.260/0.07759 & 0.302/0.09101 \\
ET & CPL              & 0.206/0.01225 & 0.388/0.13581 & 0.316/0.09965 & 1.005/0.13218 & 0.316/0.06644 & 0.352/0.07470 \\
ET & CPL$+\Lambda$    & 0.232/0.01408 & 0.378/0.15924 & 0.320/0.11750 & 0.895/0.14532 & 0.320/0.07834 & 0.368/0.08749 \\
ET & IntDM            & 0.246/0.01270 & 0.378/0.13524 & 0.304/0.09640 & 0.987/0.14185 & 0.304/0.06427 & 0.350/0.07270 \\
ET & IntDE            & 0.250/0.01454 & 0.374/0.18883 & 0.292/0.12977 & 1.025/0.15567 & 0.292/0.08651 & 0.340/0.09877 \\
ET & $\Lambda$CDM     & 0.210/0.01287 & 0.400/0.14201 & 0.316/0.10599 & 0.993/0.13275 & 0.316/0.07066 & 0.364/0.07916 \\
\midrule
\multicolumn{8}{l}{\itshape LISA ($N_{\rm ev}=160$): entries are $z^*/\mathcal{P}(z^*)$} \\
\midrule
LISA & AXI\_CLASS     & 0.195/0.00421 & 0.345/0.05551 & 0.285/0.03766 & 0.836/0.05760 & 0.285/0.02511 & 0.325/0.02885 \\
LISA & CPL            & 0.195/0.00408 & 0.385/0.05190 & 0.320/0.03738 & 0.871/0.05738 & 0.320/0.02492 & 0.365/0.02783 \\
LISA & CPL$+\Lambda$  & 0.165/0.00417 & 0.320/0.06547 & 0.295/0.04705 & 0.766/0.06047 & 0.295/0.03137 & 0.325/0.03595 \\
LISA & IntDM          & 0.195/0.00394 & 0.360/0.05089 & 0.315/0.03549 & 0.846/0.05958 & 0.315/0.02366 & 0.345/0.02669 \\
LISA & IntDE          & 0.200/0.00406 & 0.355/0.05235 & 0.305/0.03587 & 0.811/0.06121 & 0.305/0.02391 & 0.345/0.02708 \\
LISA & $\Lambda$CDM   & 0.205/0.00414 & 0.375/0.05172 & 0.320/0.03753 & 0.871/0.05887 & 0.320/0.02502 & 0.360/0.02799 \\
\midrule
\multicolumn{8}{l}{\itshape Difference between detector optima: entries are $\Delta z^*$} \\
\midrule
$\Delta z^*$ & AXI\_CLASS      & 0.029 & 0.017 & 0.025 & 0.069 & 0.025 & 0.023 \\
$\Delta z^*$ & CPL             & 0.011 & 0.003 & 0.004 & 0.134 & 0.004 & 0.013 \\
$\Delta z^*$ & CPL$+\Lambda$   & 0.067 & 0.058 & 0.025 & 0.129 & 0.025 & 0.043 \\
$\Delta z^*$ & IntDM           & 0.051 & 0.018 & 0.011 & 0.141 & 0.011 & 0.005 \\
$\Delta z^*$ & IntDE           & 0.050 & 0.019 & 0.013 & 0.214 & 0.013 & 0.005 \\
$\Delta z^*$ & $\Lambda$CDM    & 0.005 & 0.025 & 0.004 & 0.122 & 0.004 & 0.004 \\
\bottomrule
\end{tabular}%
}
\renewcommand{\arraystretch}{1}
\end{table}

\section{Detector-specific pointwise Hellinger distance plots}
\label{app:hellinger}

This appendix collects the detector-specific pointwise marginal
Hellinger distance curves
for all reconstructed diagnostics. Each figure shows ET and LISA side
by side for the same quantity. The $q(z)$, $w_{\rm tot}(z)$ and
$\kappa(z)$ panels are nearly identical as discussed in
Sec.~\ref{sec:hellinger_results} because these diagnostics are related
by simple affine transformations at fixed redshift. The $H'(z)$ and
$\mathcal{O}_m(z)$ panels show the strongest detector-dependent structure at
high redshift, while the $H(z)$ panels are dominated at low redshift by
the differing fiducial $H_0$ values.

\begin{figure}[H]
  \centering
  \begin{subfigure}[t]{0.49\linewidth}
    \centering
    \includegraphics[width=\linewidth]{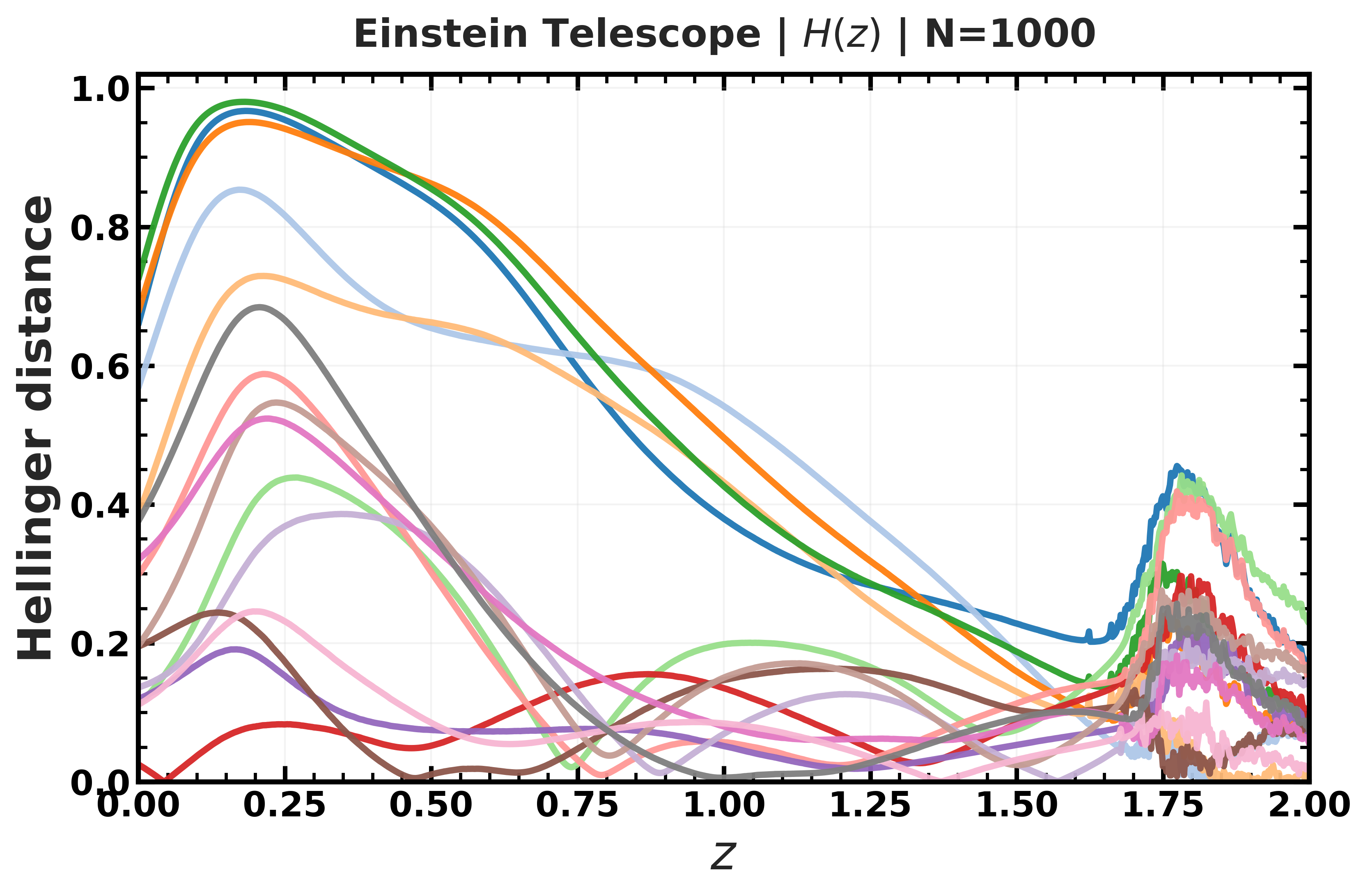}
    \caption{\it ET}
  \end{subfigure}
  \hfill
  \begin{subfigure}[t]{0.49\linewidth}
    \centering
    \includegraphics[width=\linewidth]{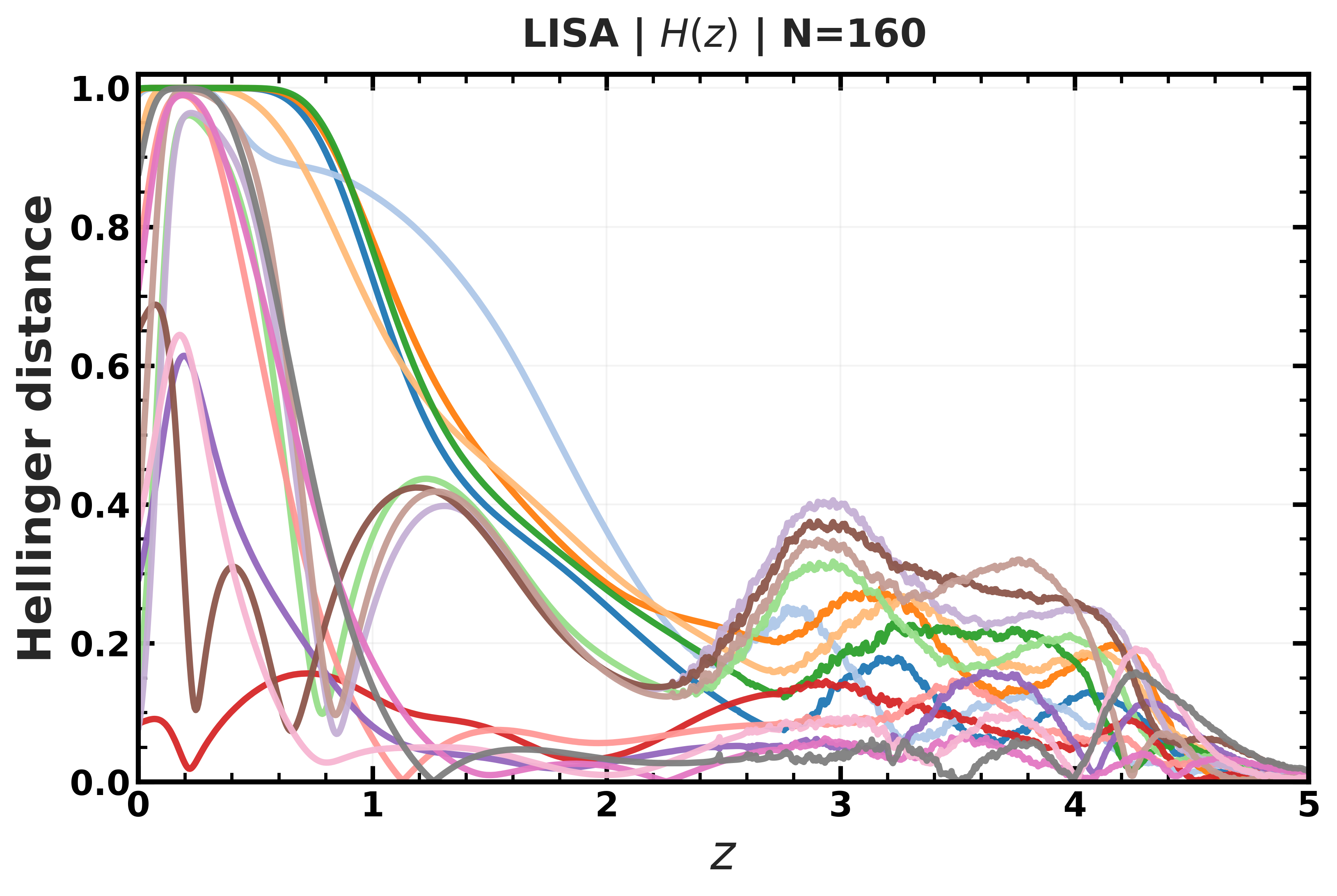}
    \caption{\it LISA}
  \end{subfigure}
  \caption{\it Pairwise pointwise marginal Hellinger distance curves for $H(z)$. The near-unity
    values at very low redshift reflect the differing fiducial $H_0$
    values across cosmologies rather than divergence in the full
    expansion history.}
  \label{fig:hellinger_H}
\end{figure}

\begin{figure}[H]
  \centering
  \begin{subfigure}[t]{0.49\linewidth}
    \centering
    \includegraphics[width=\linewidth]{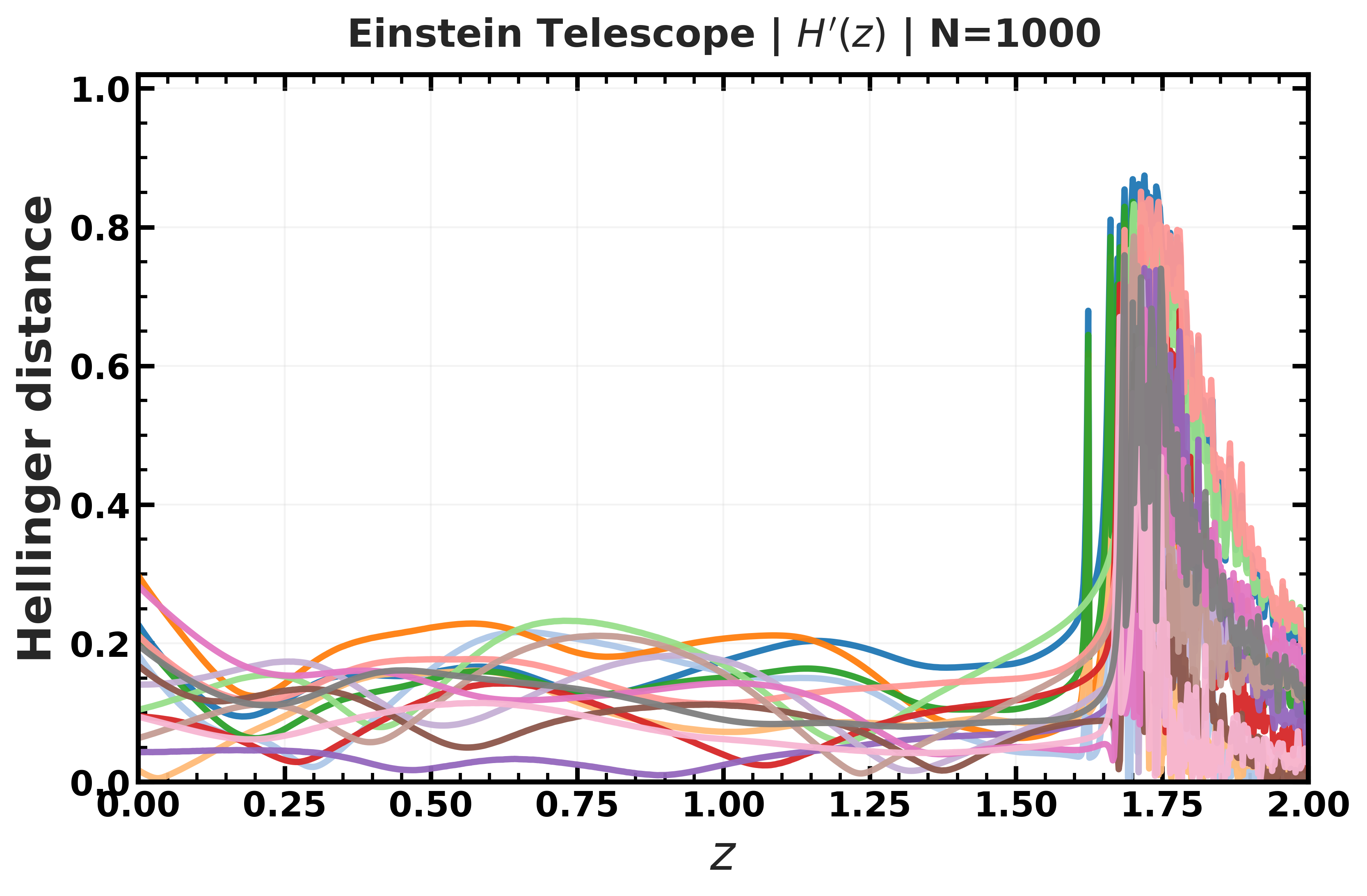}
    \caption{\it ET}
  \end{subfigure}
  \hfill
  \begin{subfigure}[t]{0.49\linewidth}
    \centering
    \includegraphics[width=\linewidth]{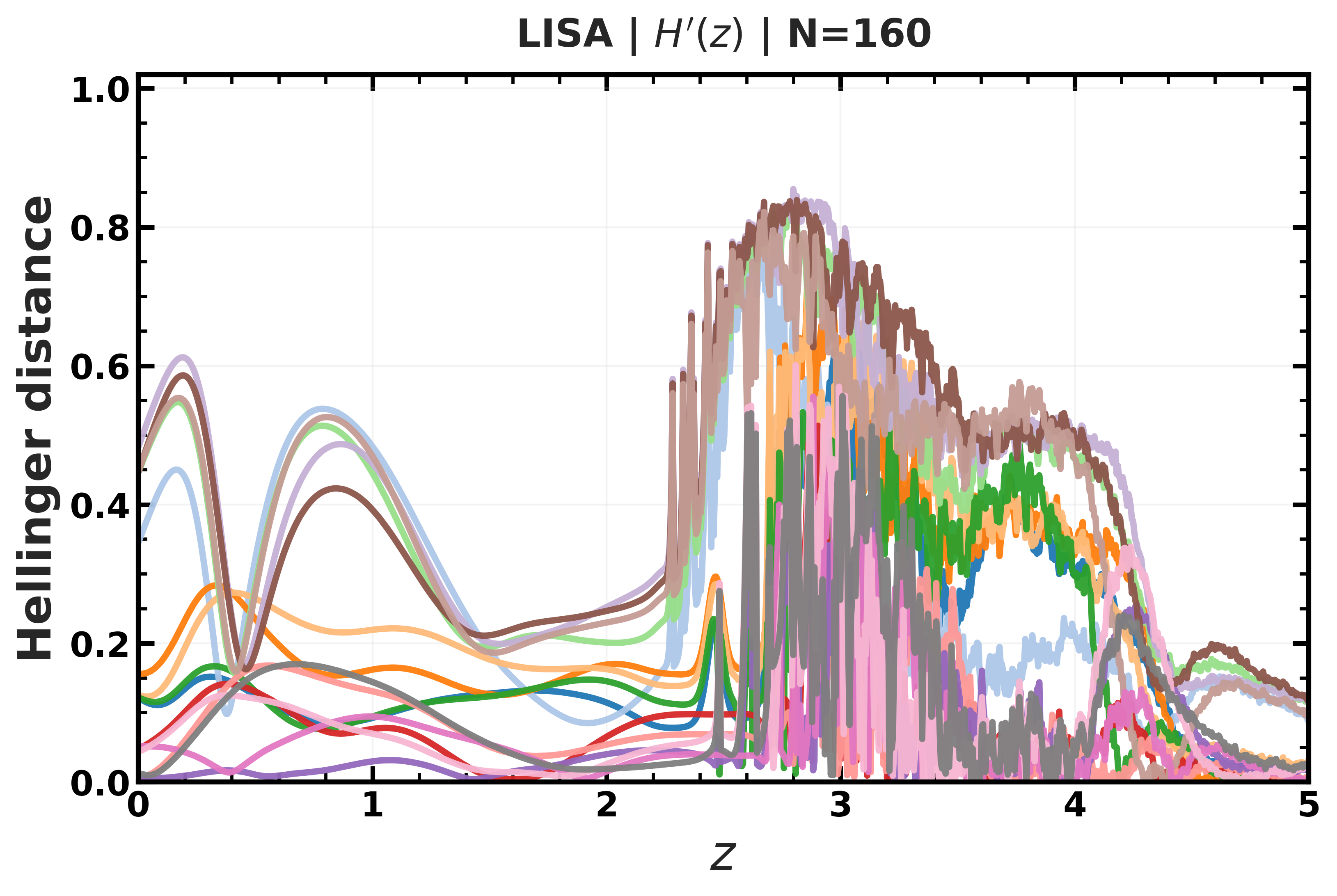}
    \caption{\it LISA}
  \end{subfigure}
  \caption{\it Pairwise pointwise marginal Hellinger distance curves for $H'(z)$. Both detectors
    show sharp structure near the upper end of their reconstruction
    window, reflecting the greater sensitivity of this quantity to
    fluctuations in $d_C''(z)$.}
  \label{fig:hellinger_Hp}
\end{figure}

\begin{figure}[H]
  \centering
  \begin{subfigure}[t]{0.49\linewidth}
    \centering
    \includegraphics[width=\linewidth]{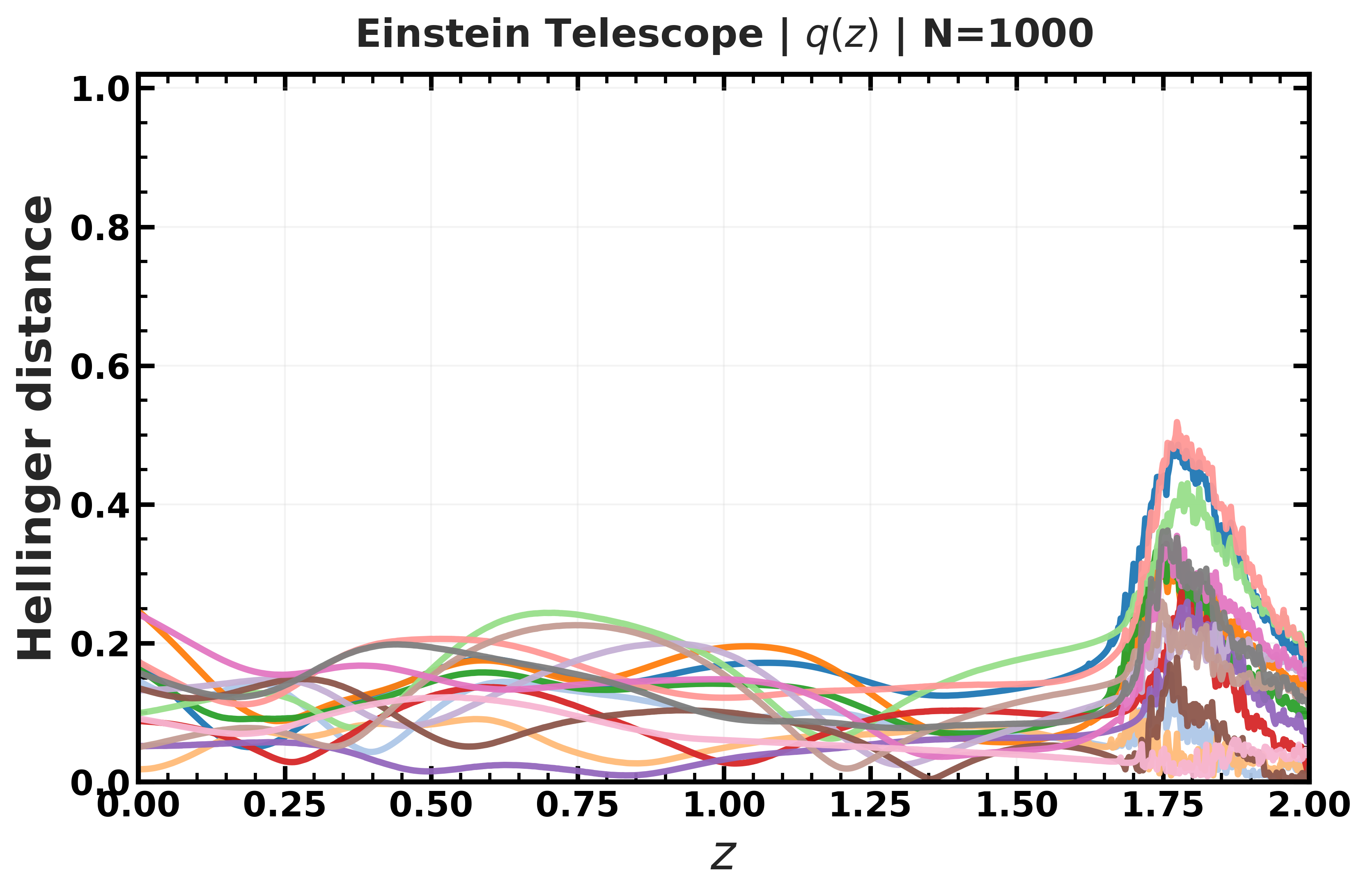}
    \caption{\it ET}
  \end{subfigure}
  \hfill
  \begin{subfigure}[t]{0.49\linewidth}
    \centering
    \includegraphics[width=\linewidth]{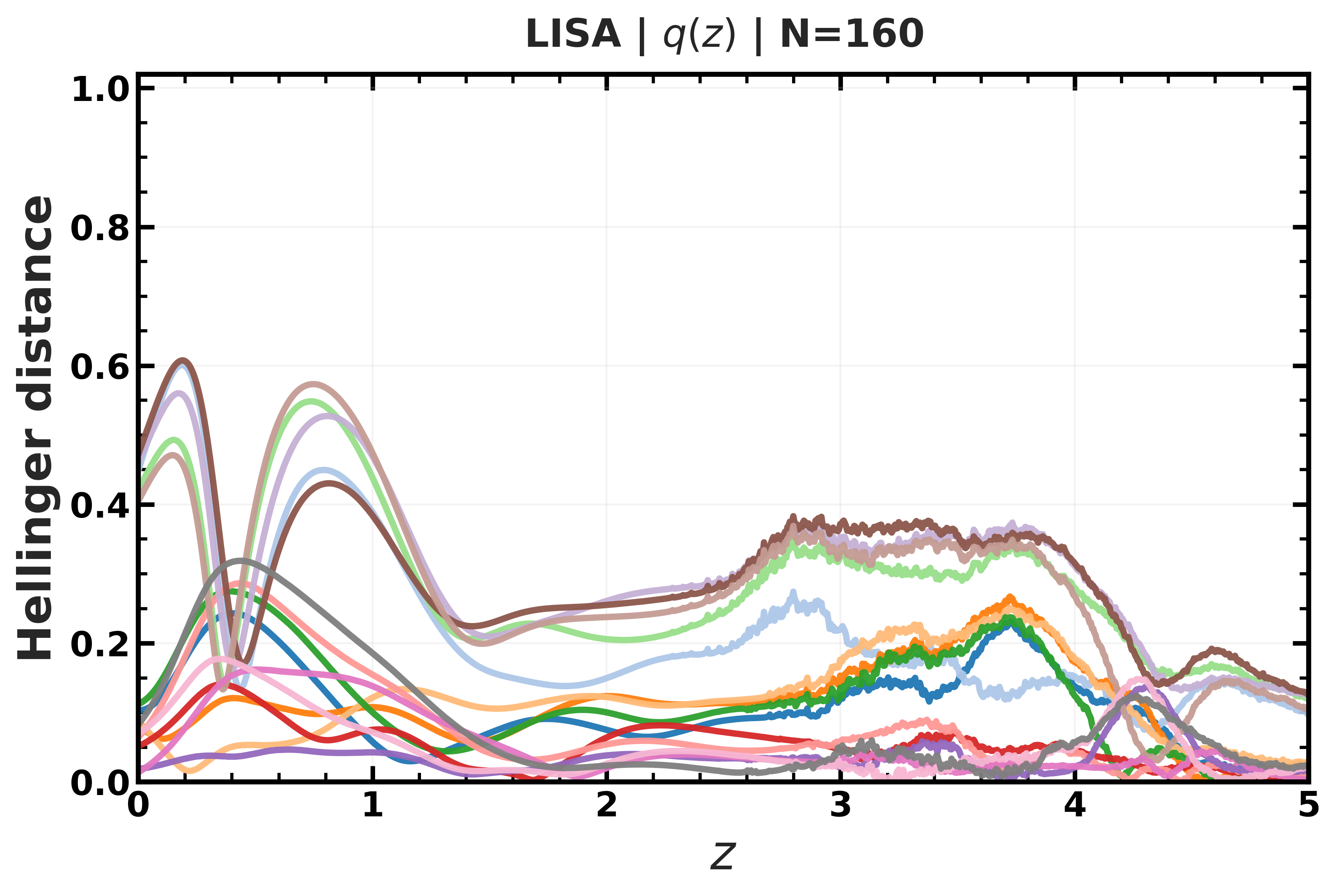}
    \caption{\it LISA}
  \end{subfigure}
  \caption{\it Pairwise pointwise marginal Hellinger distance curves for the deceleration
    parameter $q(z)$. For ET the main rise appears near $z\sim 1.8$.
    For LISA the structure is broader and extends over
    $z\sim 2.5$--$4$.}
  \label{fig:hellinger_q}
\end{figure}

\begin{figure}[H]
  \centering
  \begin{subfigure}[t]{0.49\linewidth}
    \centering
    \includegraphics[width=\linewidth]{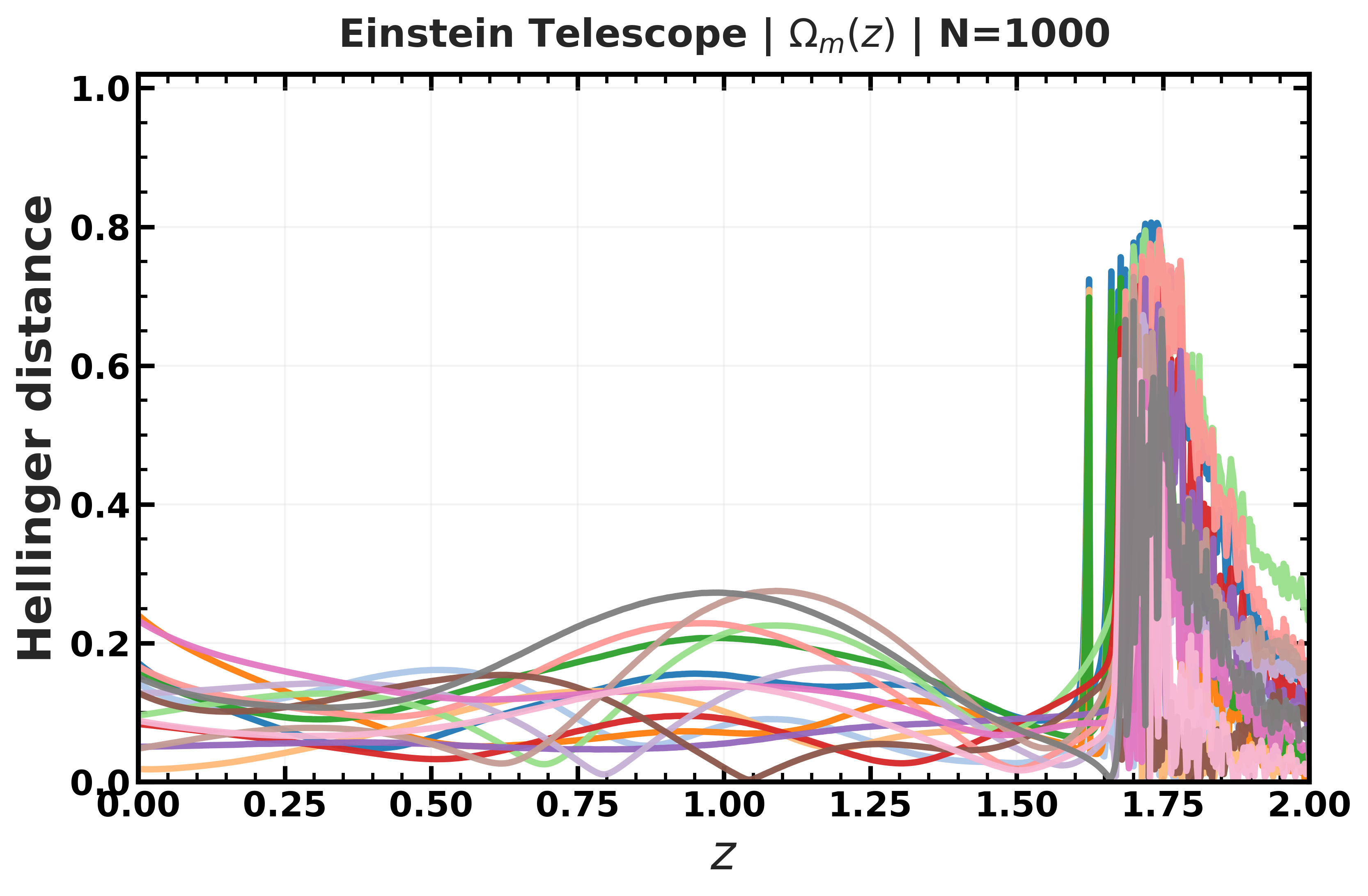}
    \caption{\it ET}
  \end{subfigure}
  \hfill
  \begin{subfigure}[t]{0.49\linewidth}
    \centering
    \includegraphics[width=\linewidth]{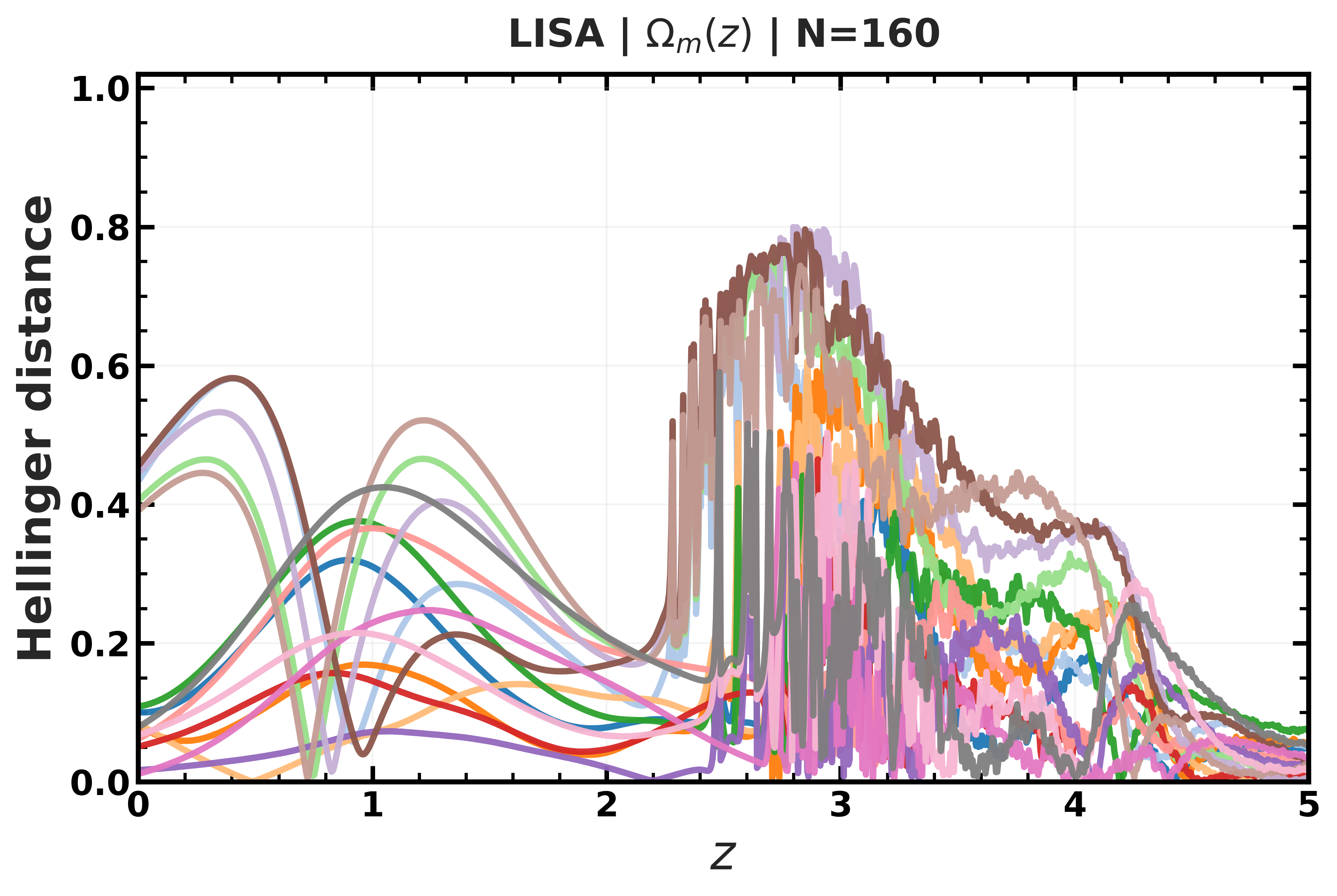}
    \caption{\it LISA}
  \end{subfigure}
  \caption{\it Pairwise pointwise marginal Hellinger distance curves for the
    Om diagnostic $\mathcal{O}_m(z)$. This quantity develops the most
    jagged high-redshift structure because the denominator
    $(1+z)^3-1$ and the reconstructed numerator can both enhance local
    fluctuations.}
  \label{fig:hellinger_Om}
\end{figure}

\begin{figure}[H]
  \centering
  \begin{subfigure}[t]{0.49\linewidth}
    \centering
    \includegraphics[width=\linewidth]{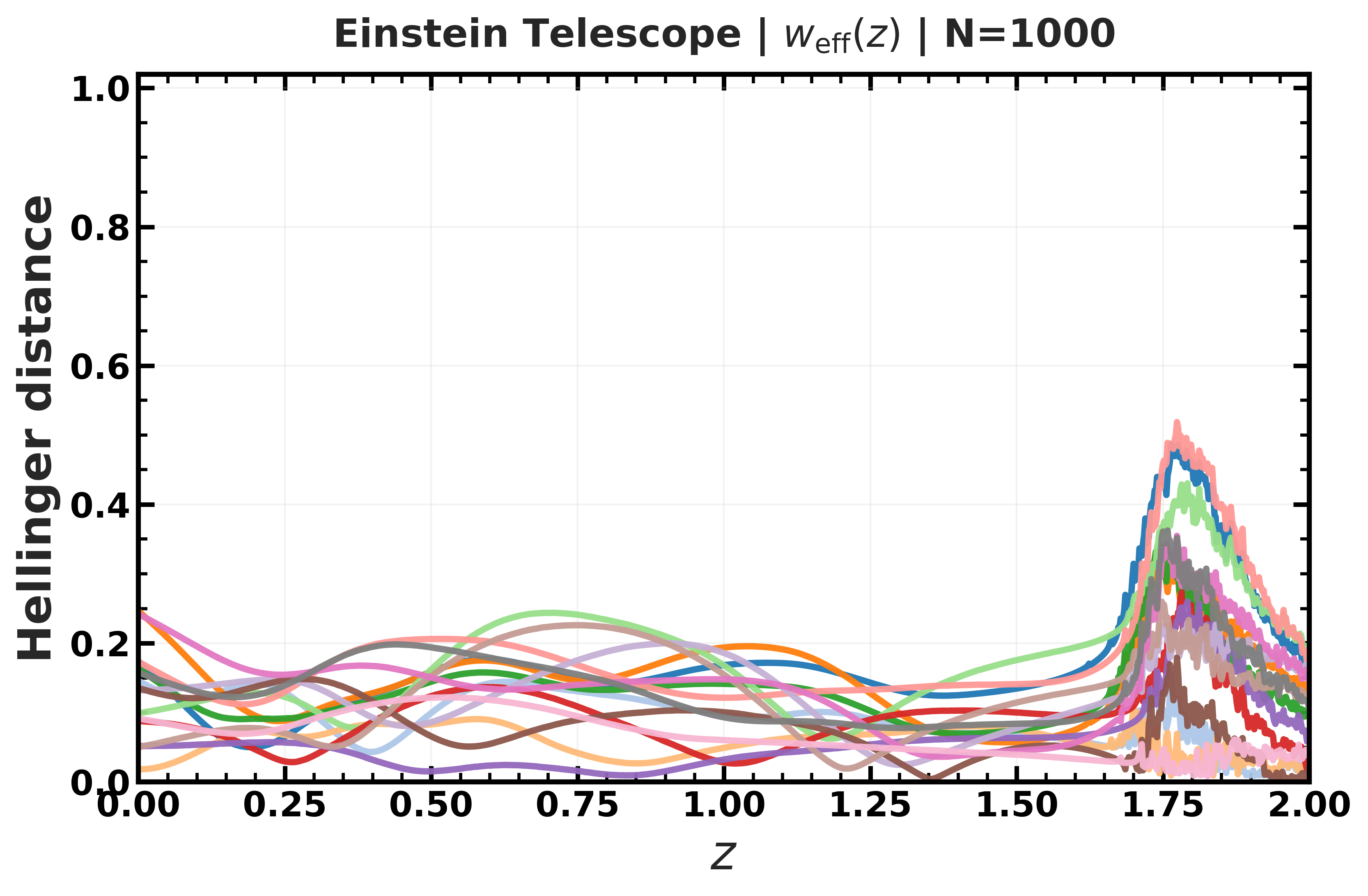}
    \caption{\it ET}
  \end{subfigure}
  \hfill
  \begin{subfigure}[t]{0.49\linewidth}
    \centering
    \includegraphics[width=\linewidth]{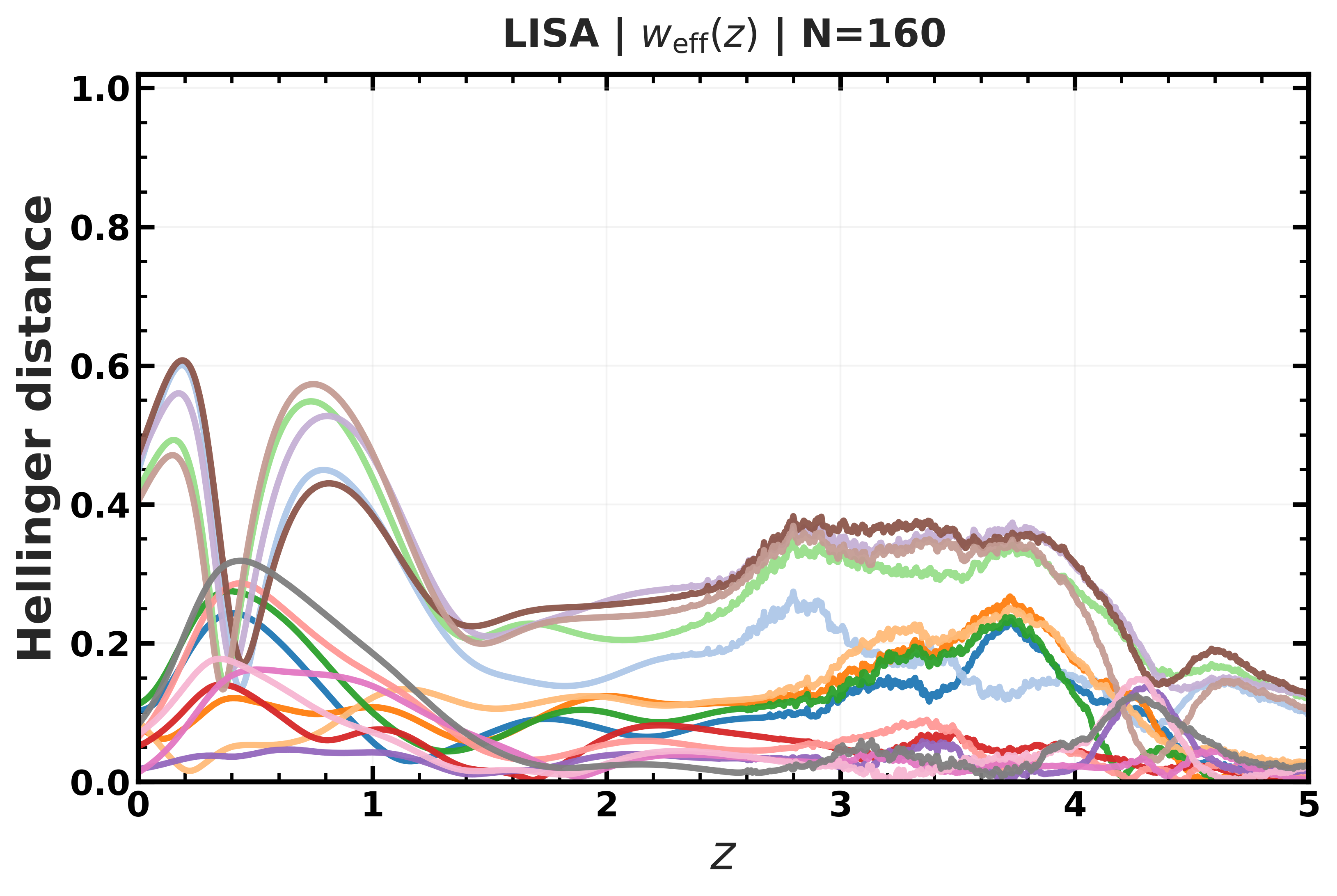}
    \caption{\it LISA}
  \end{subfigure}
  \caption{\it Pairwise pointwise marginal Hellinger distance curves for the
    total effective equation of state $w_{\rm tot}(z)$. The detector dependence matches that of
    $q(z)$ because the two diagnostics differ only by a redshift
    dependent affine transformation at fixed $z$.}
  \label{fig:hellinger_wtot}
\end{figure}

\begin{figure}[H]
  \centering
  \begin{subfigure}[t]{0.49\linewidth}
    \centering
    \includegraphics[width=\linewidth]{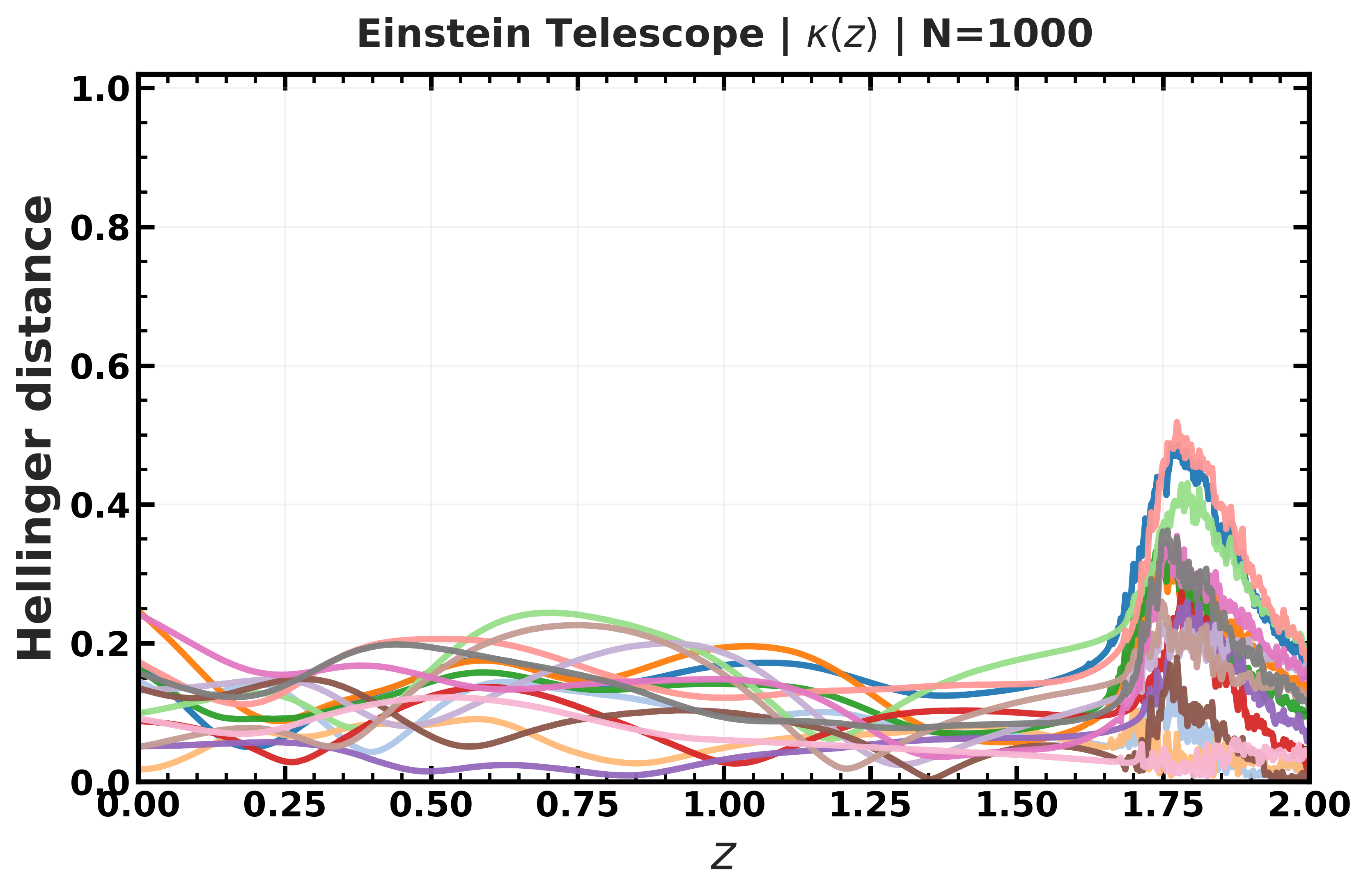}
    \caption{\it ET}
  \end{subfigure}
  \hfill
  \begin{subfigure}[t]{0.49\linewidth}
    \centering
    \includegraphics[width=\linewidth]{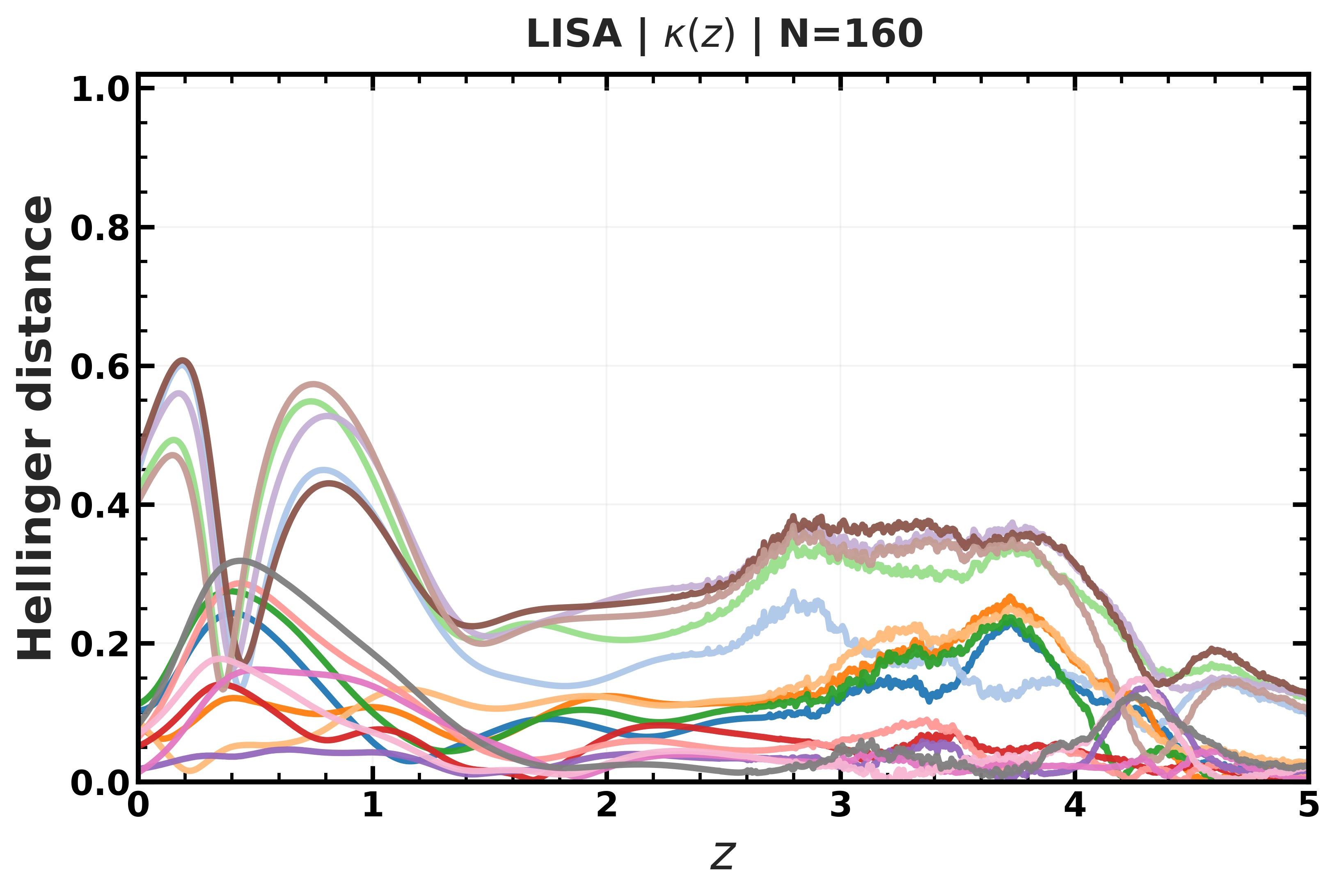}
    \caption{\it LISA}
  \end{subfigure}
  \caption{\it Pairwise pointwise marginal Hellinger distance curves for
    $\kappa(z)=E'(z)/E(z)$. The pattern is nearly identical in practice
    to that of $q(z)$ and $w_{\rm tot}(z)$ because
    $\kappa(z)=[1+q(z)]/(1+z)=3[1+w_{\rm tot}(z)]/[2(1+z)]$ at fixed
    redshift.}
  \label{fig:hellinger_kappa}
\end{figure}

Figure~\ref{fig:hellinger_H} shows the $H(z)$ comparison and isolates the
low redshift feature produced by the different fiducial $H_0$ values.
Figure~\ref{fig:hellinger_Hp} shows the stronger derivative sensitivity
of $H'(z)$. Figures~\ref{fig:hellinger_q},
\ref{fig:hellinger_wtot} and \ref{fig:hellinger_kappa} show the nearly
identical behavior of diagnostics related by affine maps.
Figure~\ref{fig:hellinger_Om} shows the Om diagnostic, whose high
redshift structure is enhanced by the denominator in Eq.~(\ref{eq:Omz}).

\section{Higher-order covariance structure}
\label{app:third_deriv}

For completeness, we present representative covariance blocks involving
third derivatives of the reconstructed comoving distance. These
quantities are not used directly in the analysis due to the rapid growth
of uncertainties, but they illustrate the strong oscillatory behavior
and amplified correlations arising from repeated kernel differentiation.

\begin{figure}[H]
  \centering
  \begin{subfigure}[t]{0.47\linewidth}
    \includegraphics[width=\linewidth]{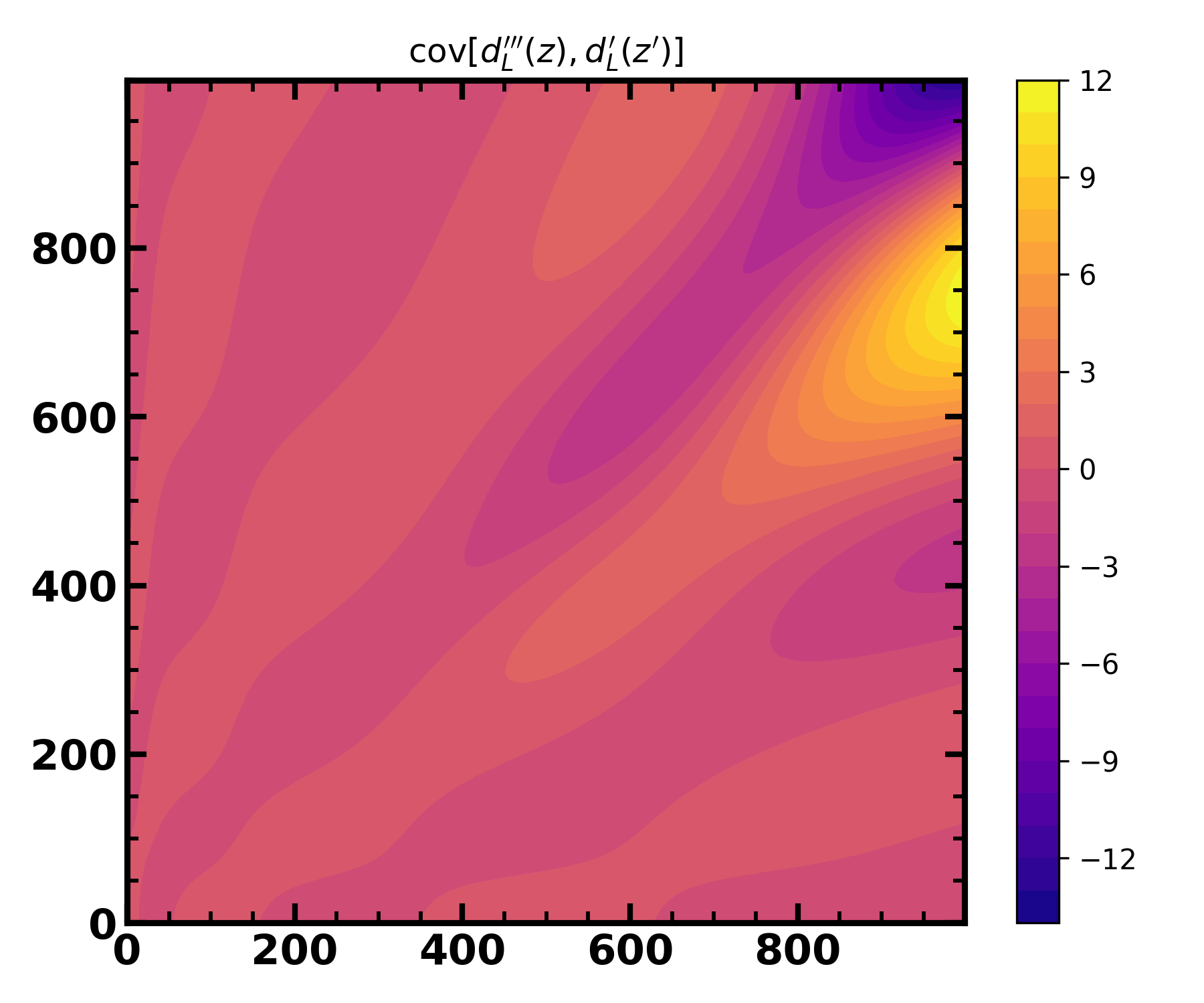}
    \caption{\it $\mathrm{Cov}[d_C'''(z),\,d_C'(z')]$}
  \end{subfigure}
  \hfill
  \begin{subfigure}[t]{0.47\linewidth}
    \includegraphics[width=\linewidth]{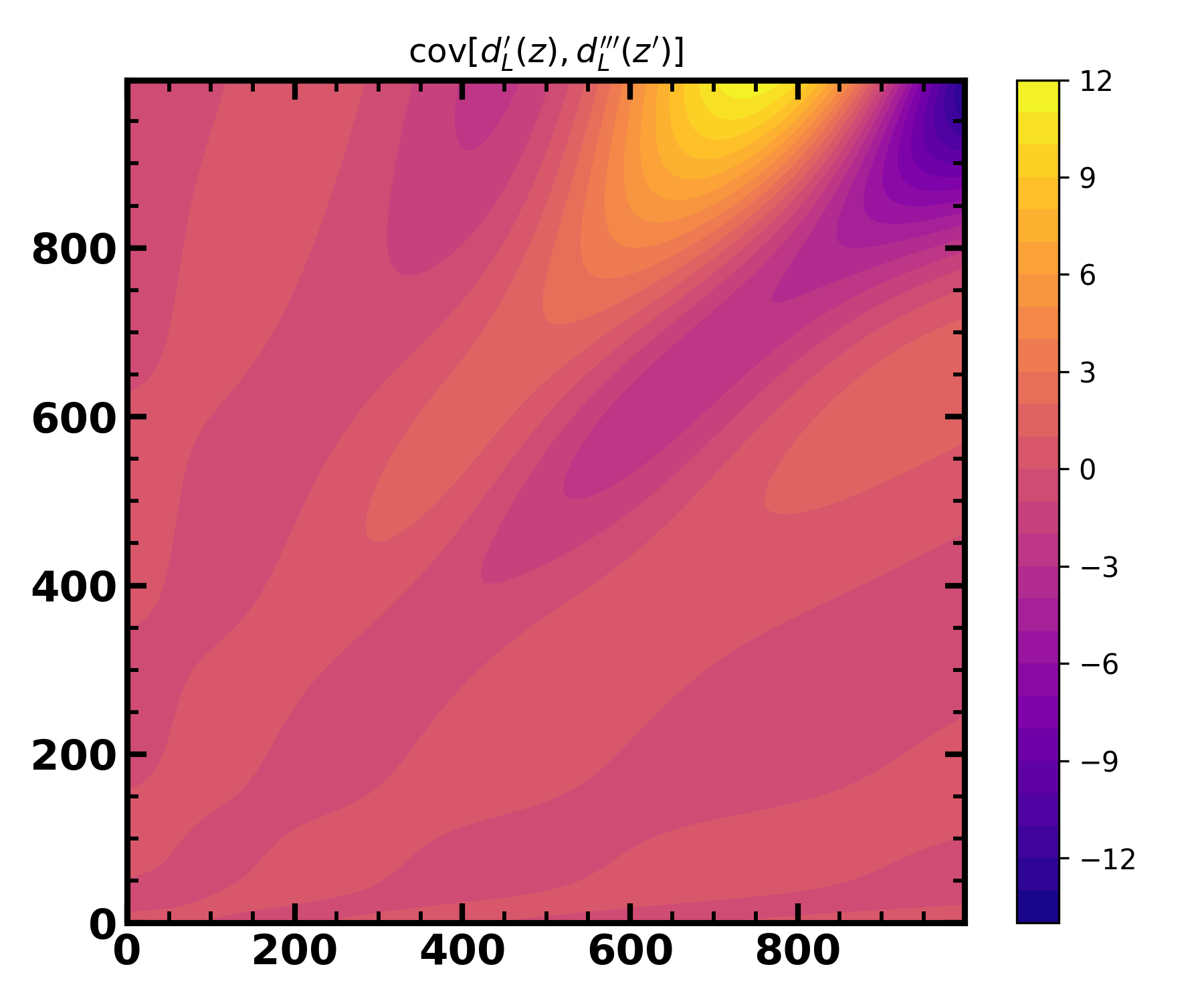}
    \caption{\it $\mathrm{Cov}[d_C'(z),\,d_C'''(z')]$}
  \end{subfigure}
  \caption{\it Predictive covariance blocks involving third derivatives of
    the reconstructed comoving distance. The pronounced oscillatory
    structure and large amplitudes toward high redshift explain the
    instability of third-order kinematical diagnostics and justify our
    preference for diagnostics requiring only up to second-order
    differentiation.}
  \label{fig:covariance_third_deriv}
\end{figure}

These are cross covariance blocks, so an individual panel is not
expected to be symmetric. The transpose relation is recovered only after
the derivative order is exchanged. The alternating positive and negative
regions at large redshift show that third derivative information is
highly sensitive to small changes in the reconstructed slope. This is
the visual reason why third order kinematical diagnostics are avoided in
the main analysis. Figure~\ref{fig:covariance_third_deriv} displays the
two representative cross covariance blocks used to illustrate this
instability.

\bibliographystyle{JHEP}
\bibliography{paper_bib}

\end{document}